\documentstyle[10pt,epsf,epsfig,dp_delphititle]{dp_delphi}
\makeindex
\def\DpPaperGroup{EP}
\def\DpPaperRef{2000-043}
\def\DpDate{08 February 2000}
\def\DpAuthors{DELPHI Collaboration}
\def\DpSubmit{(Eur. Phys. J. C16(2000)555)}
\def\DpTitle{{{Measurement of the \boldmath 
${\mathrm{B^0_s}}$ Lifetime and Study 
of ${\mathrm{B^0_s}}$-${\overline{\mathrm{B^0_s}}}$ 
Oscillations using ${{\mathrm{ D_s}}\ell}$ Events}}}
\begin{document}
\makeatletter
\newcount\@tempcntc
\def\@citex[#1]#2{\if@filesw\immediate\write\@auxout{\string\citation{#2}}\fi
  \@tempcnta\z@\@tempcntb\m@ne\def\@citea{}\@cite{\@for\@citeb:=#2\do
    {\@ifundefined
       {b@\@citeb}{\@citeo\@tempcntb\m@ne\@citea\def\@citea{,}{\bf ?}\@warning
       {Citation `\@citeb' on page \thepage \space undefined}}%
    {\setbox\z@\hbox{\global\@tempcntc0\csname b@\@citeb\endcsname\relax}%
     \ifnum\@tempcntc=\z@ \@citeo\@tempcntb\m@ne
       \@citea\def\@citea{,}\hbox{\csname b@\@citeb\endcsname}%
     \else
      \advance\@tempcntb\@ne
      \ifnum\@tempcntb=\@tempcntc
      \else\advance\@tempcntb\m@ne\@citeo
      \@tempcnta\@tempcntc\@tempcntb\@tempcntc\fi\fi}}\@citeo}{#1}}
\def\@citeo{\ifnum\@tempcnta>\@tempcntb\else\@citea\def\@citea{,}%
  \ifnum\@tempcnta=\@tempcntb\the\@tempcnta\else
   {\advance\@tempcnta\@ne\ifnum\@tempcnta=\@tempcntb \else \def\@citea{--}\fi
    \advance\@tempcnta\m@ne\the\@tempcnta\@citea\the\@tempcntb}\fi\fi}
 
\makeatother
\begin{titlepage}
\pagenumbering{roman}
\CERNpreprint{\DpPaperGroup}{\DpPaperRef} 
\date{{\small\DpDate}} 
\title{\DpTitle} 
\address{\DpAuthors} 
\begin{shortabs} 
\noindent
%
\newcommand{\Ds}{\mathrm{D_s}}
\newcommand{\Bd}{\mathrm{B^0_d}}
\newcommand{\Bs}{\mathrm{B^0_s}}
\newcommand{\dms}{\Delta m_{\Bs}}
\newcommand{\dgbs}{\Delta \Gamma_{\rm \Bs}/\Gamma_{\rm \Bs}}
\newcommand{\Zz}{\ifmmode {\mathrm Z^0} \else ${\mathrm Z^0 } $ \fi}

\noindent 
Lifetime and oscillations of $\Bs$ mesons have been studied in events
with a large transverse momentum lepton and a $\Ds$ of opposite electric charge in 
the same hemisphere, selected from about 3.6 million hadronic $\Zz$ 
decays accumulated by DELPHI between 1992 and 1995. \\
The $\Bs$ lifetime and the fractional width difference between the two physical $\Bs$ states have been found to be: 
$$
\tau_{\Bs}  = (1.42^{+0.14}_{-0.13}(stat.)\pm 0.03(syst.))~ps
$$
$$
\dgbs < 0.46 ~\mbox{at the 95\% C.L.}
$$
In the latter result it has been assumed that $\tau_{\Bs}=\tau_{\Bd}$.\\
Using the same sample, a limit on the mass difference between the physical 
$\mbox{B}^0_{\mathrm s}$ states has been set:
\begin{eqnarray*}
   \begin{array}{ll}
~~~~~~~~~~\dms > 7.4 ~ \mbox{ps}^{-1}~\mbox{at the 95\% C.L.} & \\
   \end{array}
 \end{eqnarray*}
with a corresponding sensitivity equal to  $8.1~ps^{-1}$.
\end{shortabs}
\vfill
\begin{center}
\DpSubmit \ \\ 
\end{center}
\vfill
\clearpage
\headsep 10.0pt
\addtolength{\textheight}{10mm}
\addtolength{\footskip}{-5mm}
\begingroup
%
\newcommand{\DpName}[2]{\hbox{#1$^{\ref{#2}}$},\hfill}
\newcommand{\DpNameTwo}[3]{\hbox{#1$^{\ref{#2},\ref{#3}}$},\hfill}
\newcommand{\DpNameThree}[4]{\hbox{#1$^{\ref{#2},\ref{#3},\ref{#4}}$},\hfill}
\newskip\Bigfill \Bigfill = 0pt plus 1000fill
\newcommand{\DpNameLast}[2]{\hbox{#1$^{\ref{#2}}$}\hspace{\Bigfill}}
%
\footnotesize
\noindent
\DpName{P.Abreu}{LIP}
\DpName{W.Adam}{VIENNA}
\DpName{T.Adye}{RAL}
\DpName{P.Adzic}{DEMOKRITOS}
\DpName{Z.Albrecht}{KARLSRUHE}
\DpName{T.Alderweireld}{AIM}
\DpName{G.D.Alekseev}{JINR}
\DpName{R.Alemany}{VALENCIA}
\DpName{T.Allmendinger}{KARLSRUHE}
\DpName{P.P.Allport}{LIVERPOOL}
% ...
\normalsize
\endgroup
\titlefoot{Department of Physics and Astronomy, Iowa State
     University, Ames IA 50011-3160, USA
    \label{AMES}}
\titlefoot{Physics Department, Univ. Instelling Antwerpen,
     Universiteitsplein 1, BE-2610 Wilrijk, Belgium \\
     \indent~~and IIHE, ULB-VUB,
     Pleinlaan 2, BE-1050 Brussels, Belgium \\
     \indent~~and Facult\'e des Sciences,
     Univ. de l'Etat Mons, Av. Maistriau 19, BE-7000 Mons, Belgium
    \label{AIM}}
\titlefoot{Physics Laboratory, University of Athens, Solonos Str.
     104, GR-10680 Athens, Greece
    \label{ATHENS}}
\titlefoot{Department of Physics, University of Bergen,
     All\'egaten 55, NO-5007 Bergen, Norway
    \label{BERGEN}}
\titlefoot{Dipartimento di Fisica, Universit\`a di Bologna and INFN,
     Via Irnerio 46, IT-40126 Bologna, Italy
    \label{BOLOGNA}}
\titlefoot{Centro Brasileiro de Pesquisas F\'{\i}sicas, rua Xavier Sigaud 150,
     BR-22290 Rio de Janeiro, Brazil \\
     \indent~~and Depto. de F\'{\i}sica, Pont. Univ. Cat\'olica,
     C.P. 38071 BR-22453 Rio de Janeiro, Brazil \\
     \indent~~and Inst. de F\'{\i}sica, Univ. Estadual do Rio de Janeiro,
     rua S\~{a}o Francisco Xavier 524, Rio de Janeiro, Brazil
    \label{BRASIL}}
\titlefoot{Comenius University, Faculty of Mathematics and Physics,
     Mlynska Dolina, SK-84215 Bratislava, Slovakia
    \label{BRATISLAVA}}
\titlefoot{Coll\`ege de France, Lab. de Physique Corpusculaire, IN2P3-CNRS,
     FR-75231 Paris Cedex 05, France
    \label{CDF}}
\titlefoot{CERN, CH-1211 Geneva 23, Switzerland
    \label{CERN}}
\titlefoot{Institut de Recherches Subatomiques, IN2P3 - CNRS/ULP - BP20,
     FR-67037 Strasbourg Cedex, France
    \label{CRN}}
\titlefoot{Institute of Nuclear Physics, N.C.S.R. Demokritos,
     P.O. Box 60228, GR-15310 Athens, Greece
    \label{DEMOKRITOS}}
\titlefoot{FZU, Inst. of Phys. of the C.A.S. High Energy Physics Division,
     Na Slovance 2, CZ-180 40, Praha 8, Czech Republic
    \label{FZU}}
\titlefoot{Dipartimento di Fisica, Universit\`a di Genova and INFN,
     Via Dodecaneso 33, IT-16146 Genova, Italy
    \label{GENOVA}}
\titlefoot{Institut des Sciences Nucl\'eaires, IN2P3-CNRS, Universit\'e
     de Grenoble 1, FR-38026 Grenoble Cedex, France
    \label{GRENOBLE}}
\titlefoot{Helsinki Institute of Physics, HIP,
     P.O. Box 9, FI-00014 Helsinki, Finland
    \label{HELSINKI}}
\titlefoot{Joint Institute for Nuclear Research, Dubna, Head Post
     Office, P.O. Box 79, RU-101 000 Moscow, Russian Federation
    \label{JINR}}
\titlefoot{Institut f\"ur Experimentelle Kernphysik,
     Universit\"at Karlsruhe, Postfach 6980, DE-76128 Karlsruhe,
     Germany
    \label{KARLSRUHE}}
\titlefoot{Institute of Nuclear Physics and University of Mining and Metalurgy,
     Ul. Kawiory 26a, PL-30055 Krakow, Poland
    \label{KRAKOW}}
\titlefoot{Universit\'e de Paris-Sud, Lab. de l'Acc\'el\'erateur
     Lin\'eaire, IN2P3-CNRS, B\^{a}t. 200, FR-91405 Orsay Cedex, France
    \label{LAL}}
\titlefoot{School of Physics and Chemistry, University of Lancaster,
     Lancaster LA1 4YB, UK
    \label{LANCASTER}}
\titlefoot{LIP, IST, FCUL - Av. Elias Garcia, 14-$1^{o}$,
     PT-1000 Lisboa Codex, Portugal
    \label{LIP}}
\titlefoot{Department of Physics, University of Liverpool, P.O.
     Box 147, Liverpool L69 3BX, UK
    \label{LIVERPOOL}}
\titlefoot{LPNHE, IN2P3-CNRS, Univ.~Paris VI et VII, Tour 33 (RdC),
     4 place Jussieu, FR-75252 Paris Cedex 05, France
    \label{LPNHE}}
\titlefoot{Department of Physics, University of Lund,
     S\"olvegatan 14, SE-223 63 Lund, Sweden
    \label{LUND}}
\titlefoot{Universit\'e Claude Bernard de Lyon, IPNL, IN2P3-CNRS,
     FR-69622 Villeurbanne Cedex, France
    \label{LYON}}
\titlefoot{Univ. d'Aix - Marseille II - CPP, IN2P3-CNRS,
     FR-13288 Marseille Cedex 09, France
    \label{MARSEILLE}}
\titlefoot{Dipartimento di Fisica, Universit\`a di Milano and INFN,
     Via Celoria 16, IT-20133 Milan, Italy
    \label{MILANO}}
\titlefoot{Dipartimento di Fisica, Univ. di Milano-Bicocca and
     INFN-MILANO, Piazza delle Scienze 2, IT-20126 Milan, Italy
    \label{MILANO2}}
\titlefoot{Niels Bohr Institute, Blegdamsvej 17,
     DK-2100 Copenhagen {\O}, Denmark
    \label{NBI}}
\titlefoot{NC, Nuclear Centre of MFF, Charles University, Areal MFF,
     V Holesovickach 2, CZ-180 00, Praha 8, Czech Republic
    \label{NC}}
\titlefoot{NIKHEF, Postbus 41882, NL-1009 DB
     Amsterdam, The Netherlands
    \label{NIKHEF}}
\titlefoot{National Technical University, Physics Department,
     Zografou Campus, GR-15773 Athens, Greece
    \label{NTU-ATHENS}}
\titlefoot{Physics Department, University of Oslo, Blindern,
     NO-1000 Oslo 3, Norway
    \label{OSLO}}
\titlefoot{Dpto. Fisica, Univ. Oviedo, Avda. Calvo Sotelo
     s/n, ES-33007 Oviedo, Spain
    \label{OVIEDO}}
\titlefoot{Department of Physics, University of Oxford,
     Keble Road, Oxford OX1 3RH, UK
    \label{OXFORD}}
\titlefoot{Dipartimento di Fisica, Universit\`a di Padova and
     INFN, Via Marzolo 8, IT-35131 Padua, Italy
    \label{PADOVA}}
\titlefoot{Rutherford Appleton Laboratory, Chilton, Didcot
     OX11 OQX, UK
    \label{RAL}}
\titlefoot{Dipartimento di Fisica, Universit\`a di Roma II and
     INFN, Tor Vergata, IT-00173 Rome, Italy
    \label{ROMA2}}
\titlefoot{Dipartimento di Fisica, Universit\`a di Roma III and
     INFN, Via della Vasca Navale 84, IT-00146 Rome, Italy
    \label{ROMA3}}
\titlefoot{DAPNIA/Service de Physique des Particules,
     CEA-Saclay, FR-91191 Gif-sur-Yvette Cedex, France
    \label{SACLAY}}
\titlefoot{Instituto de Fisica de Cantabria (CSIC-UC), Avda.
     los Castros s/n, ES-39006 Santander, Spain
    \label{SANTANDER}}
\titlefoot{Dipartimento di Fisica, Universit\`a degli Studi di Roma
     La Sapienza, Piazzale Aldo Moro 2, IT-00185 Rome, Italy
    \label{SAPIENZA}}
\titlefoot{Inst. for High Energy Physics, Serpukov
     P.O. Box 35, Protvino, (Moscow Region), Russian Federation
    \label{SERPUKHOV}}
\titlefoot{J. Stefan Institute, Jamova 39, SI-1000 Ljubljana, Slovenia
     and Laboratory for Astroparticle Physics,\\
     \indent~~Nova Gorica Polytechnic, Kostanjeviska 16a, SI-5000 Nova Gorica, Slovenia, \\
     \indent~~and Department of Physics, University of Ljubljana,
     SI-1000 Ljubljana, Slovenia
    \label{SLOVENIJA}}
\titlefoot{Fysikum, Stockholm University,
     Box 6730, SE-113 85 Stockholm, Sweden
    \label{STOCKHOLM}}
\titlefoot{Dipartimento di Fisica Sperimentale, Universit\`a di
     Torino and INFN, Via P. Giuria 1, IT-10125 Turin, Italy
    \label{TORINO}}
\titlefoot{Dipartimento di Fisica, Universit\`a di Trieste and
     INFN, Via A. Valerio 2, IT-34127 Trieste, Italy \\
     \indent~~and Istituto di Fisica, Universit\`a di Udine,
     IT-33100 Udine, Italy
    \label{TU}}
\titlefoot{Univ. Federal do Rio de Janeiro, C.P. 68528
     Cidade Univ., Ilha do Fund\~ao
     BR-21945-970 Rio de Janeiro, Brazil
    \label{UFRJ}}
\titlefoot{Department of Radiation Sciences, University of
     Uppsala, P.O. Box 535, SE-751 21 Uppsala, Sweden
    \label{UPPSALA}}
\titlefoot{IFIC, Valencia-CSIC, and D.F.A.M.N., U. de Valencia,
     Avda. Dr. Moliner 50, ES-46100 Burjassot (Valencia), Spain
    \label{VALENCIA}}
\titlefoot{Institut f\"ur Hochenergiephysik, \"Osterr. Akad.
     d. Wissensch., Nikolsdorfergasse 18, AT-1050 Vienna, Austria
    \label{VIENNA}}
\titlefoot{Inst. Nuclear Studies and University of Warsaw, Ul.
     Hoza 69, PL-00681 Warsaw, Poland
    \label{WARSZAWA}}
\titlefoot{Fachbereich Physik, University of Wuppertal, Postfach
     100 127, DE-42097 Wuppertal, Germany
    \label{WUPPERTAL}}
\titlefoot{On leave of absence from IHEP Serpukhov
    \label{MILAN-SERPOU}}
\addtolength{\textheight}{-10mm}
\addtolength{\footskip}{5mm}
\clearpage
\headsep 30.0pt
\end{titlepage}
%
\pagenumbering{arabic} 
\setcounter{footnote}{0} %
\large
%
\sloppy
\newcommand{\epst}{\epsilon_{tag}}
\newcommand{\Dstar}{\mbox{D}^{\ast \pm}}
\newcommand{\Dstarp}{\mbox{D}^{\ast +}}
\newcommand{\Dstarm}{\mbox{D}^{\ast -}}
\newcommand{\Dm}{\mbox{D}^{-}}
\newcommand{\Dsstar}{\mbox{D}^{\ast \ast}}
\newcommand{\Dsstarpm}{\mbox{D}^{\ast \ast \pm}}
\newcommand{\Dsstars}{\mbox{D}^{\ast \ast \pm}_s}
\newcommand{\Dob}{\overline{\mbox{D}^{0}}}
\newcommand{\Kp}{\mbox{K}^{+}}
\newcommand{\Km}{\mbox{K}^{-}}
\newcommand{\Ko}{\mbox{K}^{0}}
\newcommand{\Kob}{\overline{\mbox{K}^{0}}}
\newcommand{\dfrac}[2]{\frac{\displaystyle #1}{\displaystyle #2}}
\newcommand{\bptre}{\mathrm b^{+}_{3}}
\newcommand{\bp}{\mathrm b^{+}_{1}}
\newcommand{\bo}{\mathrm b^0}
\newcommand{\bos}{\mathrm B^0_s}
\newcommand{\bosbar}{\mathrm \overline{B^0_s}}
\newcommand{\bss}{\mathrm b^s_s}
\newcommand{\qq}{\mathrm q \overline{q}}
\newcommand{\cc}{\mathrm c \overline{c}}
\newcommand{\BsDmX}{{B_{s}^{0}} \rightarrow D \mu X}
\newcommand{\BsDsm}{{B_{s}^{0}} \rightarrow D_{s} \mu X}
\newcommand{\BsDsX}{{B_{s}^{0}} \rightarrow D_{s} X}
\newcommand{\BDsX}{B \rightarrow D_{s} X}
\newcommand{\BDomX}{B \rightarrow D^{0} \mu X}
\newcommand{\BDpmX}{B \rightarrow D^{+} \mu X}
\newcommand{\Dsfmn}{D_{s} \rightarrow \phi \mu \nu}
\newcommand{\Dsfipi}{D_{s} \rightarrow \phi \pi}
\newcommand{\DsfX}{D_{s} \rightarrow \phi X}
\newcommand{\DpfX}{D^{+} \rightarrow \phi X}
\newcommand{\DofX}{D^{0} \rightarrow \phi X}
\newcommand{\DfX}{D \rightarrow \phi X}
\newcommand{\DsD}{B \rightarrow D_{s} D}
\newcommand{\epsb}{\epsilon_b^{tag}}
\newcommand{\epsc}{\epsilon_c^{tag}}
\newcommand{\epsh}{\epsilon_h^{tag}}
\newcommand{\epsB}{\epsilon_{B_q}^{tag}}
\newcommand{\epsC}{\epsilon_C^{tag}}
\newcommand{\epsH}{\epsilon_H^{tag}}
\newcommand{\dmd}{\Delta m_{\Bd}}
\newcommand{\dms}{\Delta m_{\Bs}}
\newcommand{\dmq}{\Delta m_{\rm B^0_q}}
\newcommand{\dgs}{\Delta \Gamma_{\Bs}}
\newcommand{\dgbs}{\Delta \Gamma_{\rm \Bs}/\Gamma_{\rm \Bs}}
\newcommand{\tbs}{\tau_{\Bs}}
\newcommand{\tbd}{\tau_{\Bd}}
\newcommand{\gammas}{\Gamma_{\Bs}}
\newcommand{\gammad}{\Gamma_{\Bd}}
\newcommand{\mbs}{m_{\Bs}}
\newcommand{\mbd}{m_{\Bd}}
\newcommand{\DsmX}{D_{s} \rightarrow \mu X}
\newcommand{\DmX}{D \rightarrow \mu X}
\newcommand{\Zbb}{Z^{0} \rightarrow \mathrm b \overline{b}}
\newcommand{\Zcc}{Z^{0} \rightarrow \mathrm c \overline{c}}
\newcommand{\Rbb}{\frac{\Gamma_{Z^0 \rightarrow \mathrm b \overline{b}}}
{\Gamma_{Z^0 \rightarrow Hadrons}}}
\newcommand{\Rcc}{\frac{\Gamma_{Z^0 \rightarrow \mathrm c \overline{c}}}
{\Gamma_{Z^0 \rightarrow Hadrons}}}
\newcommand{\bb}{\mathrm b \overline{b}}
\newcommand{\str}{\mathrm s \overline{s}}
\newcommand{\Bd}{\mathrm{B^0_d}}
\newcommand{\Bdb}{\overline{\mathrm{B^0_d}}}
\newcommand{\Bo}{\mathrm{B^0}}
\newcommand{\Bob}{\overline{\mathrm{B^0}}}
\newcommand{\Bs}{\mathrm{B^0_s}}
\newcommand{\Bsb}{\overline{\mathrm{B^0_s}}}
\newcommand{\Bp}{\mathrm{B^{+}}}
\newcommand{\Kstar}{\mathrm{K^{\star 0}}}
\newcommand{\phim}{\mathrm{\phi}}
\newcommand{\Ds}{{\mathrm{ D_s}}}
\newcommand{\Dsl}{{\mathrm{ D_s}}\ell}
\newcommand{\Dsp}{{\mathrm{ D_s^+}}}
\newcommand{\Dp}{{\mathrm{ D^{+}}}}
\newcommand{\Do}{{\mathrm{ D^{0}}}}
\newcommand{\KKpi}{\mathrm{ K K \pi }}
\newcommand{\GeV}{\mathrm{GeV}}
\newcommand{\MeV}{\mathrm{MeV}}
\newcommand{\Ge}{{\mathrm{GeV}}}
\newcommand{\Gec}{{\mathrm{GeV/c}}}
\newcommand{\Gecqt}{{$\mathrm{GeV/c^2}$}}
\newcommand{\Gecqm}{{\mathrm{GeV/c^2}}}
\newcommand{\Me}{{\mathrm{MeV}}}
\newcommand{\Mec}{{\mathrm{MeV/c}}}
\newcommand{\Mecqt}{{$\mathrm{MeV/c^2}$}}
\newcommand{\Mecqm}{{\mathrm{MeV/c^2}}}
\newcommand{\nb}{\mathrm{nb}}
\newcommand{\Zzero}{{\mathrm Z}^0}
\newcommand{\MZ}{\mathrm{M_Z}}
\newcommand{\MW}{\mathrm{M_W}}
\newcommand{\GF}{\mathrm{G_F}}
\newcommand{\Gm}{\mathrm{G_{\mu}}}
\newcommand{\MH}{\mathrm{M_H}}
\newcommand{\MT}{\mathrm{m_{top}}}
\newcommand{\GZ}{\Gamma_{\mathrm Z}}
\newcommand{\Afb}{\mathrm{A_{FB}}}
\newcommand{\Afbs}{\mathrm{A_{FB}^{s}}}
\newcommand{\sigmaf}{\sigma_{\mathrm{F}}}
\newcommand{\sigmab}{\sigma_{\mathrm{B}}}
\newcommand{\NF}{\mathrm{N_{F}}}
\newcommand{\NB}{\mathrm{N_{B}}}
\newcommand{\Nnu}{\mathrm{N_{\nu}}}
\newcommand{\RZ}{\mathrm{R_Z}}
\newcommand{\rhob}{\rho_{eff}}
\newcommand{\Gammanz}{\mathrm{\Gamma_{Z}^{new}}}
\newcommand{\Gammani}{\mathrm{\Gamma_{inv}^{new}}}
\newcommand{\Gammasz}{\mathrm{\Gamma_{Z}^{SM}}}
\newcommand{\Gammasi}{\mathrm{\Gamma_{inv}^{SM}}}
\newcommand{\Gammaxz}{\mathrm{\Gamma_{Z}^{exp}}}
\newcommand{\Gammaxi}{\mathrm{\Gamma_{inv}^{exp}}}
\newcommand{\rhoZ}{\rho_{\mathrm Z}}
\newcommand{\thw}{\theta_{\mathrm W}}
\newcommand{\swsq}{\sin^2\!\thw}
\newcommand{\swsqmsb}{\sin^2\!\theta_{\mathrm W}^{\overline{\mathrm MS}}}
\newcommand{\swsqbar}{\sin^2\!\overline{\theta}_{\mathrm W}}
\newcommand{\cwsqbar}{\cos^2\!\overline{\theta}_{\mathrm W}}
\newcommand{\swsqb}{\sin^2\!\theta^{eff}_{\mathrm W}}
\newcommand{\ee}{{e^+e^-}}
\newcommand{\eeX}{{e^+e^-X}}
\newcommand{\gaga}{{\gamma\gamma}}
\newcommand{\mumu}{\ifmmode {\mu^+\mu^-} \else ${\mu^+\mu^-} $ \fi}
\newcommand{\eeg}{{e^+e^-\gamma}}
\newcommand{\mumug}{{\mu^+\mu^-\gamma}}
\newcommand{\tautau}{{\tau^+\tau^-}}
\newcommand{\qqb}{{q\overline{q}}}
\newcommand{\eegg}{e^+e^-\rightarrow \gamma\gamma}
\newcommand{\eeggg}{e^+e^-\rightarrow \gamma\gamma\gamma}
\newcommand{\eeee}{e^+e^-\rightarrow e^+e^-}
\newcommand{\eeeeee}{e^+e^-\rightarrow e^+e^-e^+e^-}
\newcommand{\eeeeg}{e^+e^-\rightarrow e^+e^-(\gamma)}
\newcommand{\eeeegg}{e^+e^-\rightarrow e^+e^-\gamma\gamma}
\newcommand{\eeeg}{e^+e^-\rightarrow (e^+)e^-\gamma}
\newcommand{\eemumu}{e^+e^-\rightarrow \mu^+\mu^-}
\newcommand{\eetautau}{e^+e^-\rightarrow \tau^+\tau^-}
\newcommand{\eehad}{e^+e^-\rightarrow {\mathrm hadrons}}
\newcommand{\eettg}{e^+e^-\rightarrow \tau^+\tau^-\gamma}
\newcommand{\eell}{e^+e^-\rightarrow l^+l^-}
\newcommand{\Ztopig}{{\mathrm Z}^0\rightarrow \pi^0\gamma}
\newcommand{\Ztogg}{{\mathrm Z}^0\rightarrow \gamma\gamma}
\newcommand{\Ztoee}{{\mathrm Z}^0\rightarrow e^+e^-}
\newcommand{\Ztoggg}{{\mathrm Z}^0\rightarrow \gamma\gamma\gamma}
\newcommand{\Ztomumu}{{\mathrm Z}^0\rightarrow \mu^+\mu^-}
\newcommand{\Ztotautau}{{\mathrm Z}^0\rightarrow \tau^+\tau^-}
\newcommand{\Ztoll}{{\mathrm Z}^0\rightarrow l^+l^-}
\newcommand{\Ztocc}{{\mathrm Z^0\rightarrow c \overline c}}
\newcommand{\Lamp}{\Lambda_{+}}
\newcommand{\Lamm}{\Lambda_{-}}
\newcommand{\Pt}{\mathrm P_{t}}
\newcommand{\Gee}{\Gamma_{ee}}
\newcommand{\Gpig}{\Gamma_{\pi^0\gamma}}
\newcommand{\Ggg}{\Gamma_{\gamma\gamma}}
\newcommand{\Gggg}{\Gamma_{\gamma\gamma\gamma}}
\newcommand{\Gmumu}{\Gamma_{\mu\mu}}
\newcommand{\Gtautau}{\Gamma_{\tau\tau}}
\newcommand{\Ginv}{\Gamma_{\mathrm inv}}
\newcommand{\Ghad}{\Gamma_{\mathrm had}}
\newcommand{\Gnu}{\Gamma_{\nu}}
\newcommand{\GnuSM}{\Gamma_{\nu}^{\mathrm SM}}
\newcommand{\Gll}{\Gamma_{l^+l^-}}
\newcommand{\Gff}{\Gamma_{f\overline{f}}}
\newcommand{\Gtot}{\Gamma_{\mathrm tot}}
\newcommand{\al}{a_l}
\newcommand{\vl}{v_l}
\newcommand{\af}{a_f}
\newcommand{\vf}{v_f}
\newcommand{\ael}{a_e}
\newcommand{\ve}{v_e}
\newcommand{\amu}{a_\mu}
\newcommand{\vmu}{v_\mu}
\newcommand{\atau}{a_\tau}
\newcommand{\vtau}{v_\tau}
\newcommand{\ahatl}{\hat{a}_l}
\newcommand{\vhatl}{\hat{v}_l}
\newcommand{\ahate}{\hat{a}_e}
\newcommand{\vhate}{\hat{v}_e}
\newcommand{\ahatmu}{\hat{a}_\mu}
\newcommand{\vhatmu}{\hat{v}_\mu}
\newcommand{\ahattau}{\hat{a}_\tau}
\newcommand{\vhattau}{\hat{v}_\tau}
\newcommand{\vtildel}{\tilde{\mathrm v}_l}
\newcommand{\avsq}{\ahatl^2\vhatl^2}
\newcommand{\Ahatl}{\hat{A}_l}
\newcommand{\Vhatl}{\hat{V}_l}
\newcommand{\Afer}{A_f}
\newcommand{\Ael}{A_e}
\newcommand{\Aferb}{\overline{A_f}}
\newcommand{\Aelb}{\overline{A_e}}
\newcommand{\AVsq}{\Ahatl^2\Vhatl^2}
\newcommand{\Iwk}{I_{3l}}
\newcommand{\Qch}{|Q_{l}|}
\newcommand{\roots}{\sqrt{s}}
\newcommand{\pT}{p_{\mathrm T}}
\newcommand{\mt}{m_t}
\newcommand{\Rechi}{{\mathrm Re} \left\{ \chi (s) \right\}}
\newcommand{\up}{^}
\newcommand{\abscosthe}{|cos\theta|}
\newcommand{\dsum}{\Sigma |d_\circ|}
\newcommand{\zsum}{\Sigma z_\circ}
\newcommand{\sint}{\mbox{$\sin\theta$}}
\newcommand{\cost}{\mbox{$\cos\theta$}}
\newcommand{\mcost}{|\cos\theta|}
\newcommand{\epair}{\mbox{$e^{+}e^{-}$}}
\newcommand{\mupair}{\mbox{$\mu^{+}\mu^{-}$}}
\newcommand{\taupair}{\mbox{$\tau^{+}\tau^{-}$}}
\newcommand{\gamgam}{\mbox{$e^{+}e^{-}\rightarrow e^{+}e^{-}\mu^{+}\mu^{-}$}}
\newcommand{\fullskip}{\vskip 16cm}
\newcommand{\halfskip}{\vskip  8cm}
\newcommand{\quarskip}{\vskip  6cm}
\newcommand{\abitskip}{\vskip 0.5cm}
\newcommand{\ba}{\begin{array}}
\newcommand{\ea}{\end{array}}
\newcommand{\bc}{\begin{center}}
\newcommand{\ec}{\end{center}}
\newcommand{\bt}{\begin{tabular}}
\newcommand{\et}{\end{tabular}}
\newcommand{\beq}{\begin{eqnarray}}
\newcommand{\eeq}{\end{eqnarray}}
\newcommand{\bes}{\begin{eqnarray*}}
\newcommand{\ees}{\end{eqnarray*}}
\newcommand{\Kz}{\ifmmode {\mathrm K^0_S} \else ${\mathrm K^0_S} $ \fi}
\newcommand{\Zz}{\ifmmode {\mathrm Z^0} \else ${\mathrm Z^0 } $ \fi}
\newcommand{\qqbar}{\ifmmode {\mathrm q\overline{q}} \else ${\mathrm q\overline{q}} $ \fi}
\newcommand{\ccbar}{\ifmmode {\mathrm c\overline{c}} \else ${\mathrm c\overline{c}} $ \fi}
\newcommand{\bbbar}{\ifmmode {\mathrm b\overline{b}} \else ${\mathrm b\overline{b}} $ \fi}
\newcommand{\xxbar}{\ifmmode {\mathrm x\overline{x}} \else ${\mathrm x\overline{x}} $ \fi}
\newcommand{\rphi}{\ifmmode {\mathrm R\phi} \else ${\mathrm R\phi} $ \fi}
\renewcommand{\arraystretch}{1.2}

\section{Introduction}
\label{sec:intro}

In this paper, the average lifetime of the $\Bs$ meson has been measured
and limits have been derived on the oscillation frequency of the 
$\Bs$-$\Bsb$ system, $\dms$, and on the decay width difference, $\dgs$, 
between mass eigenstates of this system.\\
Starting with a $\mbox{B}^0_s$ meson produced at time $t$=0, 
the probability, ${\cal P}$, to observe a $\Bs$ or a $\Bsb$ decaying at 
the proper time $t$ can be written, neglecting effects from CP violation:
\begin{equation}
{\cal P}[\Bs \rightarrow \Bs ({\Bsb})]~=~\frac{\gammas}{2} e^{- \gammas t}
[\cosh (\frac{\dgs}{2} t)~\pm~\cos (\dms t) ]
\label{eq:tot}
\end{equation}
\noindent 
where $\gammas~=~(\gammas^H~+~\gammas^L)/2$, $\dgs~=~\gammas^L-\gammas^H$ 
and $\dms~=~\mbs^H~-~\mbs^L$. L and H denote the light and heavy physical 
states, respectively; $\dgs$ and $\dms$ are defined to be positive \cite{ref:bbd}
and the plus (minus) signs refer to $\Bs$ ($\Bsb$) decays. 
The oscillation period gives a direct measurement of the mass difference 
between the two physical states. 
The Standard Model predicts that $\dgs \ll \dms$, for which
the previous expression simplifies to :
 \begin{equation}
{\cal P}_{\rm B^0_s}^{unmix.}~=~{\cal P}(\Bs \rightarrow \Bs)~
=~\gammas e^{- \gammas t} \cos^2 (\frac{\dms t}{2} )
\label{eq:osc1}
\end{equation}
and similarly:
\begin{equation}
{\cal P}_{\rm B^0_s}^{mix.}~={\cal P}(\Bs \rightarrow \Bsb)~
=~\gammas e^{- \gammas t} \sin^2 (\frac{\dms t}{2} )
\label{eq:osc2}
\end{equation}
The oscillation frequency, proportional to $\dms$, can be obtained from the 
fit of the  time distributions given in relations (\ref{eq:osc1}) and 
(\ref{eq:osc2}), whereas expression (\ref{eq:tot}), without distinguishing 
between the $\Bs$ and the $\Bsb$, can be used to determine the average 
lifetime and the difference between the lifetimes of the heavy and light 
mass eigenstates.\\


\noindent B physics allows a precise determination of 
some of the parameters of the Cabibbo Kobayashi
Maskawa (CKM) matrix. 
All the nine elements can be expressed in term of four parameters that are, 
in Wolfenstein parametrization 
\cite{ref:wolf}, $\lambda$, $\rm A$, $\rho$ and $\eta$.
The values of $\rho$ and $\eta$ are the most uncertain.\\
Several quantities which depend on $\rho$ and $\eta$ can be measured
and, if the Standard Model is correct, they must define compatible values for 
the two parameters, inside measurement errors and theoretical uncertainties.\\
These quantities are $\epsilon_K$, the parameter introduced to measure CP 
violation in the K system, $|V_{ub}|/|V_{cb}|$, the ratio between the modulus 
of the CKM matrix elements corresponding to $b \rightarrow u$ and 
$b \rightarrow c$ transitions and the mass difference $\dmd$. 

\vskip 0.1cm

In the Standard Model, $\rm B^0_q$-$\overline{\rm B^0_q}$ 
($q = \mathrm{d,s}$) mixing is a direct consequence of second order weak 
interactions. Having kept only the dominant top quark contribution, $\dmq$ 
can be expressed in terms of Standard Model parameters \cite{ref:dmd_teo}:
\begin{equation}
\dmq= \frac{G_F^2}{6 \pi^2}|V_{tb}|^2|V_{tq}|^2 m_t^2 m_{\rm B_q} 
      f_{\rm B_q}^2 B_{\rm B_q} 
      \eta_{\rm B} F(\frac{m_t^2}{m_W^2}).
\label{eq:dmth}
\end{equation}
In this expression $G_F$ is the Fermi coupling constant; $F(x_t)$, with 
$x_t=\frac{m_t^2}{m_W^2}$, results from the evaluation of 
the second order weak ``box'' diagram responsible for the mixing
and has a smooth dependence on $x_t$;
$\eta_{\rm B}$ is a QCD correction factor obtained at next to leading order 
in perturbative QCD \cite{ref:bubu}.
The dominant uncertainties in Equation (\ref{eq:dmth}) come from
the evaluation of the B meson decay constant $f_{\rm B_q}$ and of the ``bag'' 
parameter $B_{\rm B_q}$. \\
The mass differences $\dmd$ and $\dms$ involve the CKM elements $V_{td}$ 
and $V_{ts}$. 
Neglecting terms of order $\lambda^4$, these are given by:
\begin{equation}
 |V_{td}| = A \lambda^3 \sqrt{( 1- \rho )^2 +  \eta^2}
  ~~~~~ ;~~~~~        |V_{ts}| = A \lambda^2.
\end{equation}

In the Wolfenstein parametrization, $|V_{ts}|$ 
is independent of $\rho$ and $\eta$. A measurement of $\dms$ is  
thus a way to measure the value of the non perturbative QCD
parameters.\\
Direct information on $V_{td}$ can be inferred by measuring $\dmd$.\\
Several experiments have accurately measured $\dmd$, nevertheless this 
precision cannot be fully exploited to extract information on $\rho$ and 
$\eta$ because of the large uncertainty which originates in the 
evaluation of the non-perturbative QCD parameters.\\
An efficient constraint is the ratio between the Standard Model expectations 
for $\dmd$ and $\dms$, given by:
\begin{equation}
\frac{\dmd}{\dms}~=~\frac{ m_{\Bd} f^2_{\Bd} B_{\Bd} \eta_{\Bd}}
{ m_{\Bs} f^2_{\Bs} B_{\Bs} \eta_{\Bs}} \frac{\left | V_{td} \right |^2}
{\left | V_{ts} \right |^2}
\label{eq:ratiodms}
\end{equation}
A measurement of the ratio ${\dmd}/{\dms}$ gives the same type of constraint, 
in the ${\rho}-{\eta}$ plane, as a measurement of $\dmd$, 
but because only ratio  $f_{\Bd}/f_{\Bs}$ and $B_{\Bd}/B_{\Bs}$ are involved, 
some of the theoretical uncertainties cancel \cite{ref:bbag1}.\\
Using existing measurements which constrain $\rho$ and $\eta$, 
except those on $\dms$, the distribution for
the expected values of
$\dms$ can be obtained. It has been shown, in the context of Standard Model
and QCD assumptions, that $\dms$ has to lie, at 68$\%$ C.L., 
between 12 and 17.6 $ps^{-1}$ and is expected to be smaller than $20ps^{-1}$
at 95$\%$ C.L. \cite{ref:bello}.\\

The $\Bs$ meson lifetime is expected to be equal to the 
$\Bd$  lifetime \cite{ref:bbd} within one percent. In the Standard Model, the ratio between the 
mass difference and decay width in the $\Bo$-$\Bob$ system is of the order 
$(m_b/m_t)^2$, although large QCD corrections are expected.\\
Explicit calculations to leading order in QCD correction, in the HQE
(Heavy Quark Expansion) 
formalism \cite{ref:bbd}, predict:
\bes
\dgbs  = 0.16 ^{+0.11}_{-0.09} 
\ees
where the quoted error is dominated by the uncertainty related to
hadronic matrix elements.\\
Recent calculations \cite{ref:bbd_new} at next-to-leading order predict a 
lower value:
\bes
 \dgbs = 0.054^{+0.016}_{-0.032}
\ees
An interesting approach consists in using the ratio between
$\dgs$ and $\dms$ \cite{ref:bbd_new}:
\begin{equation}
\frac{\dgs}{\dms} = (2.63 ^{+0.67}_{-1.36}) 10^{-3}
\end{equation}
to constrain the upper part of the $\dms$ spectrum with an upper
limit on $\dgbs$.\\
If, in future, the theoretical uncertainty can be reduced, this method 
can give an alternative approach in determining $\dms$ via 
$\dgs$ and, in conjunction with the determination of $\dmd$,
 can provide an extra constraint on the $\rho$ and 
$\eta$ parameters.\\

The results presented in the following have been obtained from data accumulated
by DELPHI experiment at LEP between 1992 and 1995, corresponding 
to about 3.6 million hadronic $\Zz$ decays. 
The main features of these analyses are:
\begin{itemize}
\item
a precise measurement of the B decay proper time; 
\item
a determination of the charge of the $b$ quark at the B-meson decay time 
(decay tag);
\item
a determination of the sign of the $b$ quark at production time 
(production tag).
\end{itemize}
The first item is common to the three studies on  $\dms$, $\tau_{\Bs}$ and 
$\dgs$ while the others are specific to the oscillation analyses. 
For these last, the principle of the measurement 
is as follows. Each of the charged and 
neutral particles measured in an event is assigned to one of the two 
hemispheres defined by the plane transverse to the sphericity axis.
A ``production tag'' is used to estimate the $\rm b/\overline{\rm b}$ sign of 
the initial quark at the production point.
The decay time of the B hadron is evaluated and a ``decay tag'' is defined,
correlated with the $\rm b/\overline{\rm b}$ content of the decaying hadron.
The analysis is performed using events containing a lepton emitted at large 
transverse momentum, $p_T$, relative to its jet axis  accompanied by an 
exclusively (or partially) reconstructed $\Ds$ in the same hemisphere and of opposite 
electric charge.
The lepton charge defines the ``decay tag''.
Different variables defined in the same and in the opposite hemisphere, 
are used to determine the ``production tag''.\\
Similar analyses have been performed by the ALEPH, CDF and OPAL Collaborations
\cite{aledsl,cdfdsl,opdsl}.
\vskip 0.2cm
\noindent
Section~\ref{sec:delphi_det}  describes the main features of the 
DELPHI detector, the event selection and the event simulation. 
Section \ref{sec:dsl_sel} describes the selection of the $\Dsl$ sample.
Section \ref{sec:taubs} presents the $\Bs$ lifetime measurement. 
Section \ref{sec:dg} presents the result on the lifetime difference.
Section \ref{sec:os} is devoted to the study of 
$\Bs$-$\Bsb$ oscillations with the $\Ds \ell$ sample: the first part 
describes the ``production tag'' algorithm while the second part presents the fitting 
procedure and the result on $\dms$.

\section{The DELPHI detector}
\label{sec:delphi_det}
 
The events used in this analysis have been recorded with the DELPHI detector 
at LEP operating at energies close to the  $\Zz$ peak. 
The DELPHI detector and its performance 
have been described in detail elsewhere \cite{ref:perfo}.
In this section are summarized the most relevant characteristics for this 
analysis.

\subsection{Global event reconstruction}

\subsubsection{Charged particles reconstruction}

The detector elements used for tracking are the Vertex
Detector (VD), the Inner Detector (ID), the Time Projection Chamber
(TPC) and the Outer Detector (OD).

The VD provided the high precision needed near the primary vertex. 
For the data taken from 1991 to 1993,
the VD consisted of three cylindrical layers of silicon detectors 
(radii 6.3, 9.0 and 10.9~cm)
measuring points in the plane transverse to the beam direction 
($r\phi$ coordinate)
in the polar angle range $ 43^\circ < \theta < 137^\circ$.
In 1994, two layers have been equipped with detector modules with double sided 
readout, providing a single hit precision of 7.6~$\mu$m in the $r\phi$ 
coordinate, similar to that obtained previously, and 9~$\mu$m in the 
coordinate parallel to the beam ($z$)~\cite{Delphi:VD}.
For high momentum particles with associated hits in the VD, the extrapolation 
precision close to the interaction region is 20~$\mu$m in the $r\phi$ plane 
and 34~$\mu$m in the $rz$ plane.

Charged particle tracks have been reconstructed with $95\%$ efficiency and with a 
momentum resolution $\sigma_p/p < 2.0 \times 10^{-3} p$ ($p$ in $GeV/c$) in 
the polar angle region $ 25^\circ < \theta < 155^\circ$.

\subsubsection{Energy reconstruction}

The total energy in the event is determined by using all information 
available from the tracking detectors and the calorimeters. For charged 
particles, the momentum measured in the tracking detector is used. 
Photons are detected and their energy measured in the electromagnetic 
calorimeters, whereas the hadron calorimeter detects long lived neutral 
hadrons such as neutrons and $K^0_L$'s. 

The electromagnetic calorimetry system of DELPHI is composed of a barrel 
calorimeter,
the HPC, covering the polar angle region \mbox{$46^\circ<\theta<134^\circ$},
and a forward calorimeter, the FEMC, for polar angles 
$8^\circ < \theta < 35 ^\circ$ and $145^\circ < \theta < 172^\circ$. 
The relative precision on the measured energy $E$ has been parametrized as 
$\sigma_E/E = 0.32/\sqrt E \oplus 0.043 $ \mbox{($E$ in $GeV$)} in the barrel,
and \mbox{$\sigma_E/E = 0.12/\sqrt E \oplus 0.03 $} \mbox{($E$ in $GeV$)} in 
the forward region.

The hadronic calorimeter, HCAL, has been installed in the return yoke of the 
DELPHI solenoid.
In the barrel region, the energy has been reconstructed with a precision of
\mbox{$\sigma_E/E = 1.12/\sqrt E \oplus 0.21 $} \mbox{($E$ in $GeV$)}.

\subsubsection{\boldmath Hadronic $Z^0$ selection}

Hadronic events from $Z^0$  decays have been selected by requiring a charged 
multiplicity greater than four and a total energy of charged particles 
greater than 0.12$\sqrt s$, where
$\sqrt s$ is the centre-of-mass energy and all particles have been assumed to
be pions; charged particles have been required to have a momentum greater than 
\mbox{0.4~$GeV/c$} and a polar angle between $20^\circ$ and $160^\circ$. 
The overall trigger and selection efficiency is 
(95.0$\pm$0.1)\%~\cite{delphi:Zshape}. 
A total of about 3.6 million hadronic events has been obtained from the 1992-1995 data.

\subsection{Particle identification}

\subsubsection{Lepton identification}
\label{lepid}

Lepton identification in the DELPHI detector is 
based on the barrel electromagnetic calorimeter and the muon chambers.
Only particles with momentum
larger than \mbox{2~$GeV/c$} have been considered as possible lepton candidates.

Muon chambers consisted, in the barrel region,
of three layers covering the polar regions
$53^\circ<\theta<88.5^\circ$ and
$91.5^\circ<\theta<127^\circ$.
The first layer contained three planes of chambers
and was inside the return yoke of the magnet after 90~cm
of iron, while the other two, with two chamber planes each,
were mounted outside the yoke behind a further 20~cm of iron.
In the end-caps there were two layers of muon chambers mounted
one outside and one inside the return yoke of the magnet.
Each consisted of two planes of active chambers covering
the polar angle regions
$20^\circ<\theta<42^\circ$ and
$138^\circ<\theta<160^\circ$ where the charged particle
tracking was efficient.

The probability of a particle being 
a muon has been calculated from a global $\chi^2$ of the match between
the track extrapolation to the muon chambers and the hits observed there.
Four identification flags are given as output of the muon identification
in decreasing order of efficiency: very loose, loose, standard and tight.
In this analysis the loose selection has been applied 
corresponding to an efficiency of $(94.8\pm 0.1 )\%$ with  a hadron 
misidentification probability of $(1.5\pm 0.1)\%$.

Electron identification has been performed using two independent and complementary
measurements, the $dE/dx$ measurement of the TPC (described in Section 2.2.2) 
and the energy deposition in the HPC. Probabilities from calorimetric 
measurements 
and tracking are combined to produce an overall probability for the electron 
hypothesis. Three levels of identification are given: loose, standard 
and tight.\\
The loose selection has been applied for this analysis corresponding to
an efficiency of 80~\% with an hadron misidentifation probability
of $\simeq$ 1.6 \%.

\subsubsection{Hadron identification}
\label{hadid}

Hadron identification relied on the RICH detector and on the
specific ionization measurement performed by the TPC.

The RICH detector~\cite{RICH:detector} used two radiators.
A gas radiator separated kaons from pions between \mbox{3 and 9 $GeV/c$},
where kaons gave no Cherenkov light whereas pions did, and between \mbox{9 
and 16 $GeV/c$}, using the measured Cherenkov angle. It also provided 
kaon/proton separation from \mbox{8 to 20 $GeV/c$}. 
A liquid radiator, which has been fully operational for 1994 and 1995 data, provided
$p/K/\pi$ separation in the momentum range \mbox{1.5--7 $GeV/c$}. 

The specific energy loss per unit length ($dE/dx$) is measured 
in the TPC by using up to 192 sense wires.
At least 30 contributing measurements have been required to compute the truncated 
mean. In the momentum range \mbox{$3 < p < 25$~$GeV/c$}, this is fulfilled
for 55\% of the tracks, and the $dE/dx$ measurement has a precision of 
$\pm7\%$.

The combination of the two measurements, $dE/dx$ and RICH angles, provides
three levels of pion, kaon and proton tag (loose, standard, tight) 
corresponding
to different purities. A tag for ``Heavy Particle'' is also given in order
to separate pions from heavier hadrons with high efficiency. \\
The Standard ``Heavy Particle'' flag has an efficiency of about 70 \%
with a pion misidentification probability of 10 \% for charged particle
with momentum greater than 0.7 $GeV/c$.

\subsubsection{$\Lambda^0$ and $K^0$ reconstruction}

The $\Lambda^0 \rightarrow p\pi^-$  and $K^0 \rightarrow \pi^+\pi^-$ decays 
have been reconstructed if the distance in the $r\phi$~plane between the 
$V^0$ decay point and the primary vertex is less than 90~cm. This 
condition meant that the decay products have track segments at least 20~cm 
long in the TPC. 
The reconstruction of the V$^0$ vertex and selection cuts are
described in detail in reference~\cite{ref:perfo}.\\
Only $K^0$ candidates passing the ``tight'' selection criteria
have been retained for this analysis.

\subsubsection{$\pi^0$ reconstruction}
\label{sec:pi0}
The $\pi^0 \rightarrow \gamma \gamma$ decays 
are reconstructed by fitting all $\gamma \gamma$ pairs
whose invariant mass is within 20 MeV of the nominal $\pi^0$
mass, using the nominal $\pi^0$ mass as a constraint. The
fit probability has to be larger than 1\%.

\subsection{Primary vertex reconstruction and event topology}

The location of the $e^+e^-$ interaction has been reconstructed on an 
event-by-event
basis using the beam spot position as a constraint~\cite{ref:perfo}.
In 1994 and 1995 data, the position of the primary vertex
transverse to the beam
has been determined with a precision of about 40~$\mu$m in the horizontal direction,
and about 10~$\mu$m in the vertical direction. For 1992 and 1993 data, 
the uncertainties are larger by about 50\%.

Each selected event has been divided into two hemispheres separated by 
the plane transverse to the sphericity axis. A clustering analysis based on 
the JETSET algorithm LUCLUS~\cite{ref:luclus} with default parameters has 
been used to define the jets, using both charged and neutral particles. 
These jets have been used to measure the $P_T$ of each particle in the 
event, defined as its momentum transverse to the axis of the rest 
of the jet it belonged to, after removing the particle itself.

The different detector configurations, both for hadron identification
and vertex resolution, implies, in the rest of the analysis, a separate 
treatment of the data taken before and after 1994.

\subsection{$b$-tagging}
\label{sec:btag}

The $b$-tagging package developed by the DELPHI collaboration has been 
described in reference~\cite{btag}.
The impact parameters of the charged particle tracks, with respect to 
the primary vertex, have been used to build the probability 
that all tracks come from this vertex. Due to the long 
$\rm B$-hadrons lifetime, the probability distribution is peaked at zero 
for events which contained beauty whereas it is flat for events containing 
light quarks.\\
The $b$-tagging algorithm has been used in this analysis to 
select control samples with low $b$ purity.

\subsection{\boldmath Event simulation}
\label{sec:ev_sim}

Simulated events have been generated using the JETSET 7.3 program 
\cite{ref:luclus}
with parameters tuned as in~\cite{ref:tuning} and 
using an updated description of B decays.
B hadron semileptonic decays have been simulated using the ISGW 
model \cite{ref:isgw}.
Generated events have been followed through the full 
simulation of the DELPHI
detector (DELSIM) \cite{ref:perfo}, and the resulting simulated raw data 
have been processed through the same reconstruction and analysis programs 
as the real data.

\section{ The \boldmath $\mathrm{D_s^{\pm}}\ell^\mp$ sample selection}
\label{sec:dsl_sel}
  
$\Bsb$
semileptonic decays\footnote{Charge conjugation is always implied.} 
have been selected requiring
the presence of a $\Dsp$ meson correlated with a 
high $p_T$ lepton of opposite electric charge
in the same hemisphere:
\begin{eqnarray*}
\overline{\Bs} \rightarrow {\mathrm {D_s^+}} \ell^- 
\overline{\nu_{\ell}} X.
\end{eqnarray*}
\noindent The $\Ds$ mesons have been reconstructed in six non-leptonic 
and two semileptonic decay channels:
\bes
\ba{ll}
\Dsp \rightarrow \phi \pi^{+} 
& \phi \rightarrow \mathrm{K}^{+}\mathrm{K}^{-};
\\
\Dsp \rightarrow {\overline {\mathrm{K}}}^{\star 0} \mathrm{K}^{+} &
{\overline {\mathrm{K}}}^{\star 0}
\rightarrow {\mathrm{K}}^{-}\pi^{+};
\\
\Dsp \rightarrow {\mathrm{K}^{0}_S} \mathrm{K}^{+}                 &
 {\mathrm{K}^{0}_S} \rightarrow \pi^{+}
\pi^{-} ;
\\
\Dsp \rightarrow    \mathrm \phi \pi^+\pi^-\pi^+
& \phi \rightarrow \mathrm{K}^{+}\mathrm{K}^{-};
\\
\Dsp \rightarrow    \mathrm \phi \pi^+\pi^0
& \phi \rightarrow \mathrm{K}^{+}\mathrm{K}^{-};
\\
\Dsp \rightarrow {\overline {\mathrm{K}}}^{\star 0} 
{\mathrm{K}}^{\star +}    &
{\overline {\mathrm{K}}}^{\star 0} \rightarrow {\mathrm{K}^{-}}\pi^{+},~~
{\mathrm{K}}^{\star +} \rightarrow {\mathrm{K}^{0}_S}\pi^{+} ;
\ea
\ees
\bes
\ba{ll}
\Dsp \rightarrow \phi e^{+} \nu_{e}  &
\phi \rightarrow \mathrm{K}^{+}\mathrm{K}^{-}; 
\\
\Dsp \rightarrow \phi \mu^{+} \nu_{\mu}    &
\phi \rightarrow \mathrm{K}^{+}\mathrm{K}^{-}. 
\ea
\ees
\noindent In addition, partially reconstructed $\Dsp$ have been selected 
requiring the presence of 
a $\phi$ meson (reconstructed in the $\rm K^+K^-$ decay channel)
accompanied by an hadron $h^+$ in the same hemisphere.
\bes
\ba{ll}
\Dsp \rightarrow \phi h^+ X
\ea
\ees
In the following the first eight decay modes will be referred as the 
$\Dsl$ sample and the last one as the $\phi \ell h$ sample.

\subsection{Selection of the $\phi \pi^+$, $\rm {\overline K}^{*0} K^+$, 
$\rm K^0_S K^+$ and $\phi \ell^+ \nu$ decay modes}
\label{dsl_sel_0}
Each $\Ds$ decay mode has been reconstructed by making all 
possible combinations of particles in the same hemisphere.
In $\Dsp$ semileptonic decays, the ambiguity between the two leptons
has been removed by assigning the lepton to the $\Dsp$ ($\Bsb$) if the
mass of the $\phi \ell$ system, $M(\phi \ell)$, is below (above) the 
nominal $\Dsp$ mass. 
If the two leptons 
both gave a $M(\phi \ell)$ above or below the $\Ds$ mass, 
the event was rejected.\\
The measured position of the $\Dsp$ decay vertex and momentum together with 
their measurement errors, have been used to form a new track
(called pseudo-track) that contains the measured parameters
of the $\Dsp$ particle.\\
A candidate $\Bsb$ decay vertex has been obtained by intercepting 
the $\Dsp$ pseudo-track with the one of a lepton. To guarantee 
a precise determination of the position of this secondary vertex, 
at least one VD hit has been required to be associated to the lepton and to
at least two tracks from the $\Dsp$ decay products. The $\chi^2$ of the
reconstructed $\Dsp$ and $\Bsb$ vertices have been required to be smaller than
40 and 20 respectively.\\
In order to suppress fake leptons and B hadron
cascade decays ($b \rightarrow c \rightarrow \ell^+$),
additional selection criteria have been applied to the $\Dsl$ pairs, which are 
summarized in Table~\ref{tab:sel}. \\
For the channel $\Dsp \rightarrow \phi \ell^+ \nu$ 
requirements on the $\phi \ell \ell$ mass and momentum have been 
reduced as compared to the other channels to account for the additional escaping neutrino.\\
Due to the smaller combinatorial background 
under the $\Ds$ signal, in the $\Ds \rightarrow \phi \pi^+$ and $\Ds \rightarrow
 \phi \ell^+ \nu$ decay channels,
the $p_T$ cut has been lowered to 
$1~GeV/c$.

\begin{table}[hbt]
\bc
\bt{|c|c|c|c|}
\hline
 & $\phi\pi^+$ & $\phi \ell^+$ & Others \\
\hline \hline
 $p_T$($\ell$)($GeV/c$)  & $>1$     & $>1$      & $>1.2$  \\
 M($\Dsl$)($GeV/c^2$)  & $\in[3,5.5]$ & $\in[2.5,5.5]$    & $\in[3,5.5]$    \\
 P($\Dsl$)($GeV/c$)  & $>14$    & $>12$     & $>14$  \\
\hline
\et
\caption []{\it Selection criteria applied to the lepton and 
$\Ds$ candidates.}
\label{tab:sel}
\ec
\end{table}

A tighter selection was then applied, separately for each decay mode,
using a discriminant variable built with the variables listed in 
Table~\ref{tab:discri_ds}.\\
These variables are:
\begin{itemize}
 \item the momenta, $P$, and masses, $M$, of the decay products;
 \item the cosine of the helicity angle, $\psi$, for the $\phi \pi^+$ and 
 $\rm {\overline K}^{\star 0} \rm K^{+}$
       decay modes;
 \item $H_{ID}$, defining whether the hadron identification from Section \ref{hadid}
               favours the $\pi$, K or proton hypothesis;
 \item $L_{ID}$, defining whether the lepton identification
               from Section \ref{lepid} identifies a particle from the 
               $\rm D^+_s$ semileptonic decay as an electron or a muon 
               (used only for leptons coming from the $\Dsp$ semileptonic decays).
\end{itemize}

\noindent For each quantity the probability densities for the signal (S) 
($\Dsl$ from $\Bs$ semileptonic decays) 
 and  for the combinatorial background (B) (fake $\Dsl$ candidates in $q\overline{q}$ events) 
have been parametrized using the simulation; the discriminant  variable 
$X_{\Ds}$ is then defined as
\bes
   R = \prod_i R_i = \prod_i {S_i(x_i)\over B_i(x_i)} \hspace{1cm} 
   X_{\Ds} = {R\over {R+1}}
\ees
where $i$ runs over the number of variables (which actual values are $x_i$). 
The combinatorial background is concentrated close to $X_{\Ds}=0$ while the 
$\Ds$ signal accumulates close to $X_{\Ds}=1$.
The definition of $X_{\Ds}$
provides an optimal separation between the signal and the combinatorial
background if the individual discriminant variables $x_i$ are independent;
in case of correlations the separation power decreases but no bias
is introduced.\\

\begin{figure}[h]
\begin{center}
\epsfig{figure=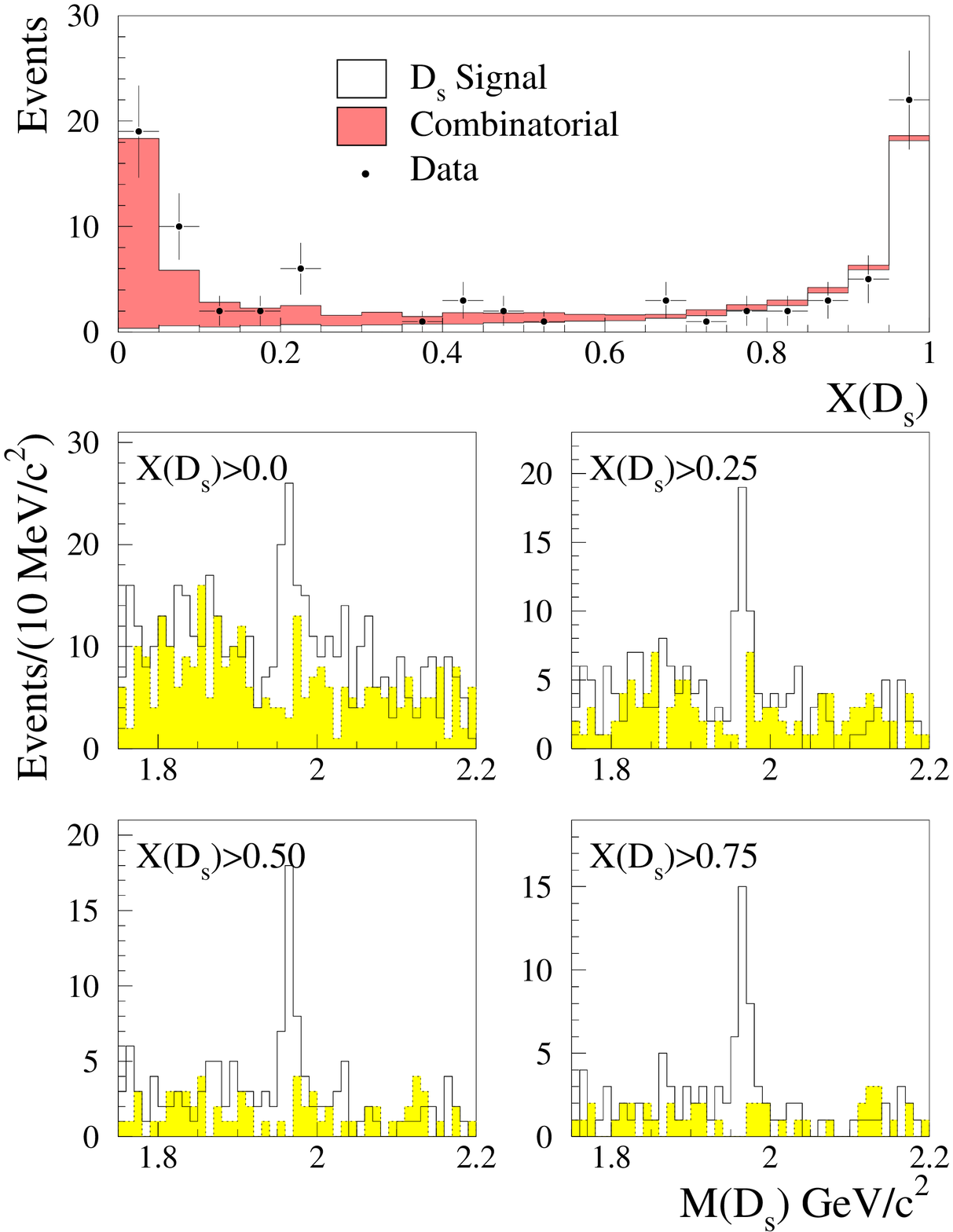,height=18cm}
\caption []{ \it The plot on the top shows the distribution of the $X_{\Ds}$ 
discriminant variable for the $\phi \pi$ channel in 94-95 data. 
The points with error bars represent the data, the white
histogram shows the contribution from the simulated signal and the shaded
histogram shows the contribution 
coming from simulated background events.\\
It could be seen that the $X_{\Ds}$ is able to discriminate the
signal ($\Ds$) from the combinatorial background.\\
The four figures on the bottom show the effect, on the $\phi \pi$ 
signal in the 94-95 data, of a cut on the discriminant variable
(white histograms represent ``right-sign'' events while shaded histogram 
show ``wrong-sign'' events). }
\label{fig:plot_chk_xdis_ds}
\end{center}
\end{figure}

\begin{table}[h]
\bc
\bt{|c|c|c|c|}
\hline 
  $\phi\pi$ & $\overline{\rm K^{0*}}\rm K$  & $\rm K^0_S K$         & 
  $\phi \ell^+$                     \\
\hline \hline
   $P(\Ds)$ & $P(\Ds)$ & $P(\Ds)$        & $P(\phi)$                     \\
   $P(\phi)/P(\Ds)$ & $P(K^{*0})/P(\Ds)$   & $P({\rm K^0_S})/P(\Ds)$ &   \\
   $H_{ID}$ $\rm K_1$ & $H_{ID}$ $\rm K_1$ & $H_{ID}$ $K$ & $H_{ID}$ $\rm K_1$       \\
   $H_{ID}$ $\rm K_2$ & $H_{ID}$ $\rm K_2$ &            & $H_{ID}$ $\rm K_2$       \\
   $H_{ID}$ $\rm \pi$ & $H_{ID}$ $\rm \pi$ &            & $L_{ID}$ $\rm \ell(\Ds)$ \\
   $cos(\psi)$ & $cos(\psi)$ &           &  \\
   $M(\phi)$   & $M(\rm K^{*0})$   &           &  \\
\hline
\et
\caption []{\it List of the quantities which are used, in the different
decay channels, to construct a discriminant variable between $\Bs$ 
semileptonic decays and background events.}
\label{tab:discri_ds}
\ec
\end{table}
\noindent
The distributions of this variable obtained in data and in the simulation
are shown in Figure~\ref{fig:plot_chk_xdis_ds} for the $\phi\pi^+$ decay 
channel.

The optimal value of the cut on the discriminant variable has been studied on 
simulated events, separately for each channel and for each detector 
configuration, in order to keep high efficiency (Table~\ref{tab:cuts}).
A very loose cut  has been applied on the 
$\phi \pi^+$ channel because of its small combinatorial background.

The individual event purity has been evaluated, in the following, from the
distribution of the discriminant variable for signal and combinatorial
background.
\begin{table}[h]
\bc
\bt{|c|c|c|c|c|}
\hline
   & $\phi\pi^+$ & $\rm \overline{K^{0*}}K$  & $\rm K^0_S K$  
   & $\phi \ell^+$        \\
\hline \hline
  92-93 & $>0.05$   & $>0.75$  & $>0.80$  & $>0.75$            \\
  94-95 & $>0.03$   & $>0.85$  & $>0.90$  & $>0.90$            \\
\hline
\et
\caption []{\it Values of the cuts applied on the discriminant 
             variable $\rm X_{\Ds}$
to select $\Bs$ semileptonic decay candidates.}
\label{tab:cuts}
\ec
\end{table}

In addition, for the two channels ($\rm \overline{K^{*0}}K$ and 
$\rm K^0_SK$), which receive contributions from kinematic reflections of 
non strange B decays, the bachelor kaon has been required to be incompatible 
with the pion hypothesis.\\
Further background suppression has been obtained by placing
a requirement on the $\Ds$ flight distance $L(\Ds)$. The small 
effect induced on the decay time acceptance has
been taken into account in the following. This requirement
has been applied, depending on the resolution on the decay distance
observed in the different $\Ds$ decay channels and on the level
of the combinatorial background:
$L(\Ds)>0$ for $\phi\pi$ and $\rm \overline{K^{*0}}K^+$,
$L(\Ds)/\sigma(L(\Ds))>-3$ for $\rm K^0_SK^+$ and
$L(\Ds)/\sigma(L(\Ds))>-1$ for $\phi \ell^+$.\\
Finally, for the semileptonic decay modes (with two neutrinos in the final 
state) an algorithm has been developed to estimate the missing energy,
$E_{miss}$, defined as:
\begin{eqnarray*}
E_{miss}=E_{tot}-E_{vis} 
\end{eqnarray*}
where the visible energy ($E_{vis}$) is the sum of the energies of charged 
particles and photons in the same hemisphere as the $\Ds\ell$ candidate. 
Using four-momentum conservation, the total energy ($E_{tot}$) in that 
hemisphere is: 
\begin{eqnarray*}
{E_{tot}}=E_{beam}+{{M_{same}^{2} - M_{opp}^{2}}\over{4 E_{beam}}}
\end{eqnarray*}
where $M_{same}$ and $M_{opp}$ are the hemisphere invariant masses of the 
same and opposite hemispheres respectively.
A positive missing energy $E_{miss}$ has been required.

\subsection{Selection of the
$\phi\pi^+\pi^+\pi^-$, $\phi\pi^+\pi^0$ and
$\rm {\overline K}^{*0}K^{*+}$ decay modes}

These three decay modes 
have been searched for in the 94 and 95 data only.\\
$\Dsl$ pairs have been selected by requiring 
$M(\Dsl)>3.0~GeV/c^2$, $p_T(\ell)>1.2~GeV/c$ and 
$\chi^2$ ($\Dsl$ vertex)$<20$
(except for the $\phi\pi^+\pi^+\pi^-$ decay mode
in which no $\chi^2$ cut has been applied).\\
In each event only one candidate is kept. The procedure is the following:
if more than one candidate passed all the selection criteria
only the one with the highest lepton transverse momentum and, if the 
same lepton candidate is attached to several $\rm D^{+}_{s}$ 
candidates the highest $\rm D_{s}^{+}$ momentum,  is kept.

\noindent It has been verified that this requirement 
keeps the signal with high efficiency and removes some of the combinatorial 
background. 

\subsubsection{$\rm D_{s}^{+} \rightarrow \overline{K}^{*0} K^{*+}$}

$\rm D_{s}^{+}$ candidates have been 
selected by reconstructing 
$\rm \overline{K}^{*0} \rightarrow K^{-} \pi^{+}$ and 
$\rm K^{*+} \rightarrow \rm K^{0}_{s} \pi^{+}$ decays. 
$\rm K^{0}_{s}$ candidates have been reconstructed in the mode 
$\rm K^{0}_{s} \rightarrow \pi^{+} \pi^{-}$ by combining all pairs of 
oppositely charged particles and applying the ``tight'' selection criteria 
described in \cite{ref:perfo}.
The $\rm K^{0}_{s}$ has been then combined with two charged particles of the same 
sign, and a third with opposite charge. If more than one $\rm D_{s}^{+}$ 
candidate could be reconstructed by the same four particles (by swapping 
the two pion candidates for example) the $\rm D_{s}^{+}$ candidate minimizing 
the squared mass difference 
$(M({\rm K^{-} \pi^{+}}) - M({\rm \overline{K}^{*0}}))^{2}  + 
(M({\rm K^{0}_{s} \pi^{+}}) - 
M({\rm K^{*+}}))^{2}$
has been chosen, where $M({\rm \overline{K}^{*0}})$
and $M({\rm K^{*+}})$ are the nominal $\rm K^{*}$ masses \cite{ref:book}.
The three charged particle tracks have been fitted to a common vertex
and the $\chi^2$ of this vertex has been required to be smaller than 30.
To improve the resolution on the vertex position, all three tracks have been 
required to have at least one VD hit. \\
$\rm K^-\pi^+$ and $\rm K^0_S \pi^+$ mass combinations have been selected if
their effective masses are within $\pm 75$ and $\pm 95$ $MeV/c^2$ of the nominal neutral 
and charged $\rm \overline{K^{\ast}}$ mass respectively.\\
The charged pion and kaon from $\rm \overline{K^{\ast}}$ decays must have a 
momentum larger than 1 and 1.5 $GeV/c$ respectively. The charged and neutral
$\rm \overline{K^{\ast}}$ mesons must have a momentum larger than 4 and 
3.5 $GeV/c$ respectively and $\Dsp$ mesons have a momentum larger than 
11 $GeV/c$.

\subsubsection{$\Dsp \rightarrow \phi \pi \pi \pi$}

The $\phi$ is reconstructed in the  decay channel
$\rm \phi \rightarrow K^{+} K^{-}$ by taking all 
possible pairs of oppositely charged particle tracks that have an invariant 
mass within 13 $MeV/c^{2}$ of the nominal $\phi$ meson mass \cite{ref:book}. 
Neither kaon candidate should be tagged by the combined RICH and dE/dX 
measurements as pions (``tight'' selection). Three tracks, each compatible 
with the pion hypothesis as given by the combined RICH and dE/dX 
measurements, have been then added to the $\phi$ candidate to make a 
$\rm D_{s}^{+}$. The five tracks have been required to be compatible with a 
single vertex, but no requirement 
has been applied on the $\chi^{2}$ of the vertex fit. 
Three of the five tracks have been required to have at 
least one VD hit and two of the three pion candidates have been required to 
have a momentum above 1.2 
GeV/c. 

In addition, kaons from the $\phi$ decays must have a momentum larger
than 1.8 GeV/c. Individual pion momenta must be larger than 700 $MeV/c$
and the $\Ds$ candidate momentum must be larger than 9 $GeV/c$.

\subsubsection{$\rm D^{+}_{s} \rightarrow \phi \pi \pi^{0}$}
The $\phi$ is  reconstructed using the 
same selection criteria as for the previous channel.
A third track, which has been required not to be tagged as a kaon by the combined 
RICH and $dE/dx$, and a reconstructed $\pi^{0}$ (Section.~\ref{sec:pi0}) have been added to the 
$\phi$ candidate.
\noindent The three charged tracks have been fitted to a common vertex.
To improve the resolution on the vertex position, each of the 
three tracks has been required to be associated to at least one VD 
hit each.\\
In addition, kaons from the $\phi$ decay must have a momentum larger than 
2.5 $GeV/c$. The momentum of the charged pion and of the $\Ds$ must be 
larger than 1 and 10 $GeV/c$ respectively.

\subsection{Summary for the $\Dsl$ selected 
events }
\label{sec:dsl_sel_1}

\subsubsection{Non leptonic $\Ds$ modes}

\begin{figure}[ph]
\begin{center}
\epsfig{figure=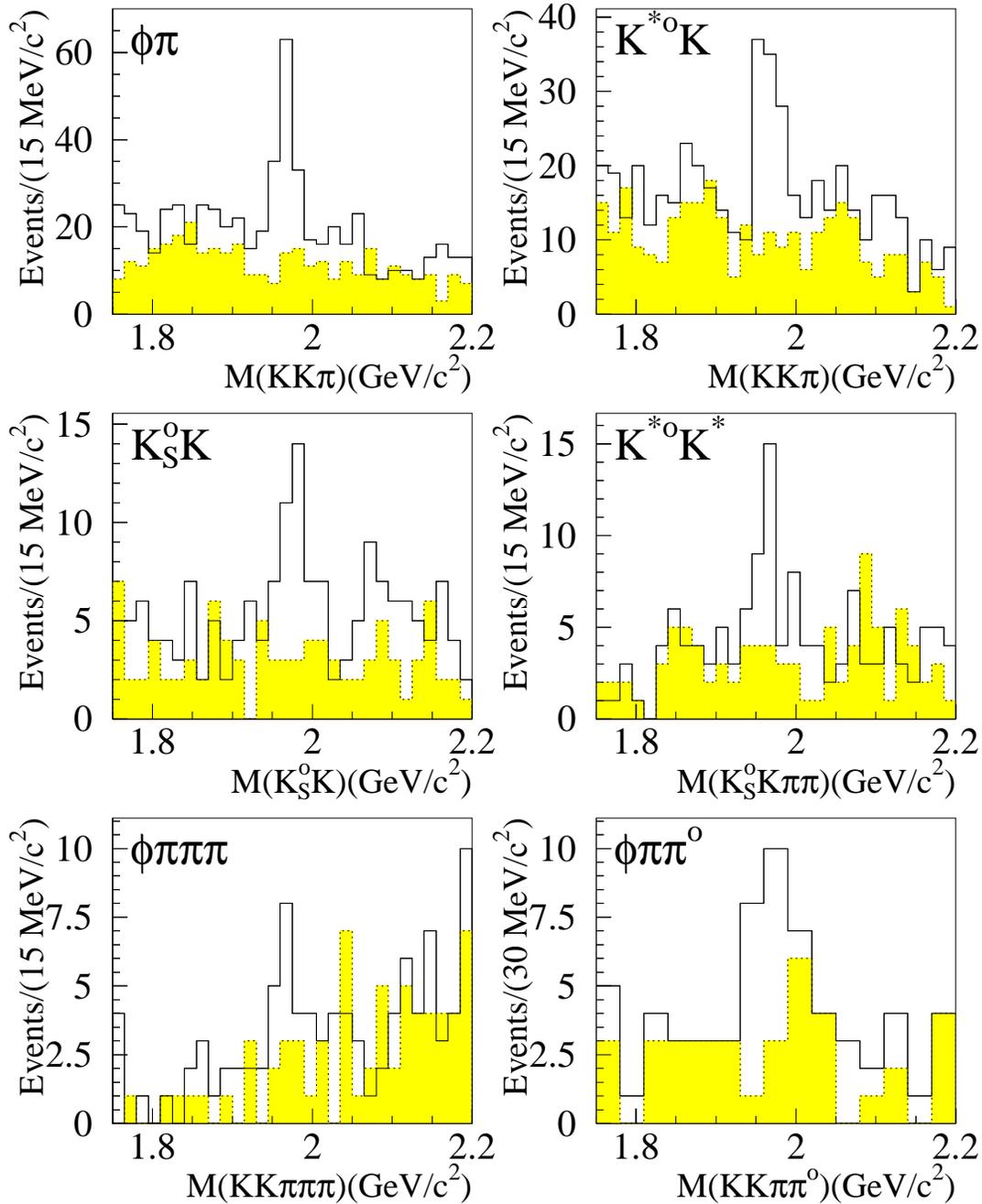,height=18cm}
\caption []{\it Invariant mass distributions for $\Ds$ candidates in 
            six non-leptonic decay modes 
($\phi \pi^+$, $\rm {\overline K}^{\star 0}K^+$, $\rm K^0 K^+$, $\rm 
{\overline K}^{\star 0} K^{\star +}$, $\phi \pi^+\pi^-\pi^+$ 
            and $\phi \pi^+\pi^0$).
The last three decay modes have been reconstructed using only the 
94-95 statistics. The corresponding distribution for wrong-sign 
combinations are given by the shaded histograms}
\label{fig:pl_sig_dsl_1}
\end{center}
\end{figure}

\begin{figure}[ph]
\begin{center}
\epsfig{figure=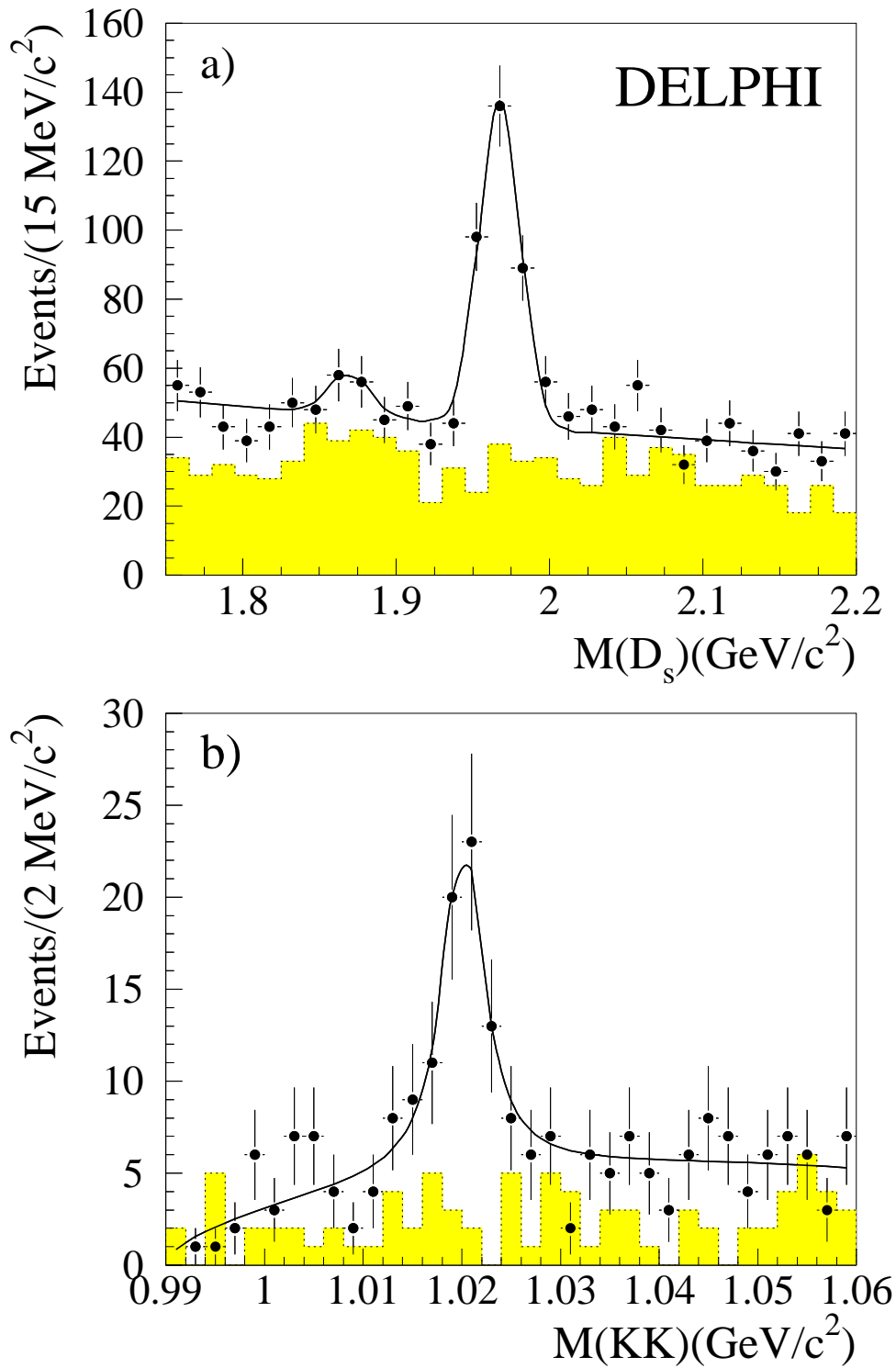,height=20cm}
\caption []{\it a) Invariant mass distributions 
          for $\Ds$ candidates in non-leptonic decay modes 
($\phi \pi^+$, $\rm {\overline K}^{\star 0}K^+$, $\rm K^0 K^+$, $\rm 
{\overline K}^{\star 0} K^{\star +}$, $\phi \pi^+\pi^-\pi^+$ and 
$\phi \pi^+\pi^0$).
b) $\rm \mbox{K}^+\mbox{K}^-$ invariant mass 
distribution for $\Ds$ candidates selected in the two semileptonic decay 
modes. The corresponding distribution for wrong-sign combinations are 
given by the shaded histograms. The curves show the result of fits 
described in the text.}
\label{fig:pl_sig_dsl_2}
\end{center}
\end{figure}

In the $\Dsp$ mass region, an excess of ``right-sign'' 
($\mathrm{D_s^{\pm}}\ell^\mp$) over ``wrong-sign'' 
($\mathrm{D_s^{\pm}}\ell^\pm$) combinations 
is  observed in each channel (Figure ~\ref{fig:pl_sig_dsl_1}). 
The estimated number of signal events and the yields
for the combinatorial background in all the studied 
modes are summarized in Table~\ref{tab1}. 
The mass distribution for non-leptonic decays has been
fitted with two Gaussian distributions of equal widths to account for the 
$\Dsp$ and $\Dp$ signals and a polynomial function  for the combinatorial 
background. The $\Dp$ mass has been fixed at the nominal value  of 
$1.869$ \Gecqt~\cite{ref:book}.
The overall mass distribution for non-leptonic decays is shown in 
(Figure~\ref{fig:pl_sig_dsl_2}a). The fit yields a signal of 
$(206\pm 21)$ $\Ds$ decays in ``right-sign'' combinations, centred at a mass 
of $(1.9680 \pm 0.0016)$ \Gecqt~with a width of $(14 \pm 1)$ \Mecqt .
\begin{table}[ht]
\bc
  \bt{|l|c|c|} \hline
  $\Ds$ decay modes &Estimated signal  & Combinatorial background / Total \\ 
\hline\hline 
   \hbox{$\Ds\rightarrow \phi\pi^+$} & 
   $83 \pm 11$ & $0.38 \pm 0.06$  \\ \hline
 \hbox{$\Ds \rightarrow {\overline {\mathrm{K}}}^{\star0} \mathrm{K}^+$} & 
   $60 \pm 11$ & $0.45 \pm 0.06$  \\ \hline 
  \hbox{$\Ds \rightarrow {\mathrm{K}^{0}_S} \mathrm{K}^+$} & 
   $22 \pm 7 $ & $0.48 \pm 0.10$  \\  \hline               
  \hbox{$ \Ds \rightarrow {\overline {\mathrm{K}}}^{\star 0}
{\mathrm{K}}^{\star +}$}  & 
   $21 \pm 5 $ & $0.31 \pm 0.07$  \\  \hline    
\hbox{$ \Ds \rightarrow     \mathrm \phi \pi^{+} \pi^{+} \pi^{-}$}& 
   $10 \pm 4 $ & $0.39 \pm 0.10$  \\ \hline
\hbox{$ \Ds \rightarrow     \mathrm \phi \pi^{+} \pi^{0}$}& 
   $18 \pm 6 $ & $0.39 \pm 0.10$  \\  \hline \hline
\hbox{$\Ds \rightarrow \phi \ell^{+} \nu$}  &
   $80 \pm 16$ & $0.38 \pm 0.06$  \\ \hline  
\et
\ec
\caption{ \it Numbers of $\Ds$ signal events and fractions of
 combinatorial background events measured in the different
$\Ds$ decay channels. The level of the combinatorial  background has
been  evaluated inside
a mass interval of $\pm 2\sigma$ ($\pm 1.5\Gamma$) centred on the measured 
$\Ds$ ($\phi$) mass.}
\label{tab1}
\end{table}

\subsubsection{Semileptonic $\Ds$ modes}
\label{sec:dsl_sel_2}

Selected events show an excess of ``right-sign'' with respect 
to ``wrong-sign'' combinations (Figure~\ref{fig:pl_sig_dsl_2}b). 
The $\mathrm {K^+K^-}$ invariant mass distribution for ``right sign'' events
has been fitted with a Breit--Wigner distribution to account for the signal
and a
polynomial function to describe  the combinatorial background. 
The fit gives $(80 \pm 16)$ events (see Table~\ref{tab1}) centred at a 
mass of $(1.020 \pm 0.001)$ \Gecqt~with a total width ($\Gamma$) of 
$(5 \pm 1)$ $\MeV /c^2$.  

\subsection{Selection of the $\phi\ell h$ inclusive channel}
\label{sec:phil_sel}

\begin{figure}[ph]
  \begin{center}
   \epsfig{figure=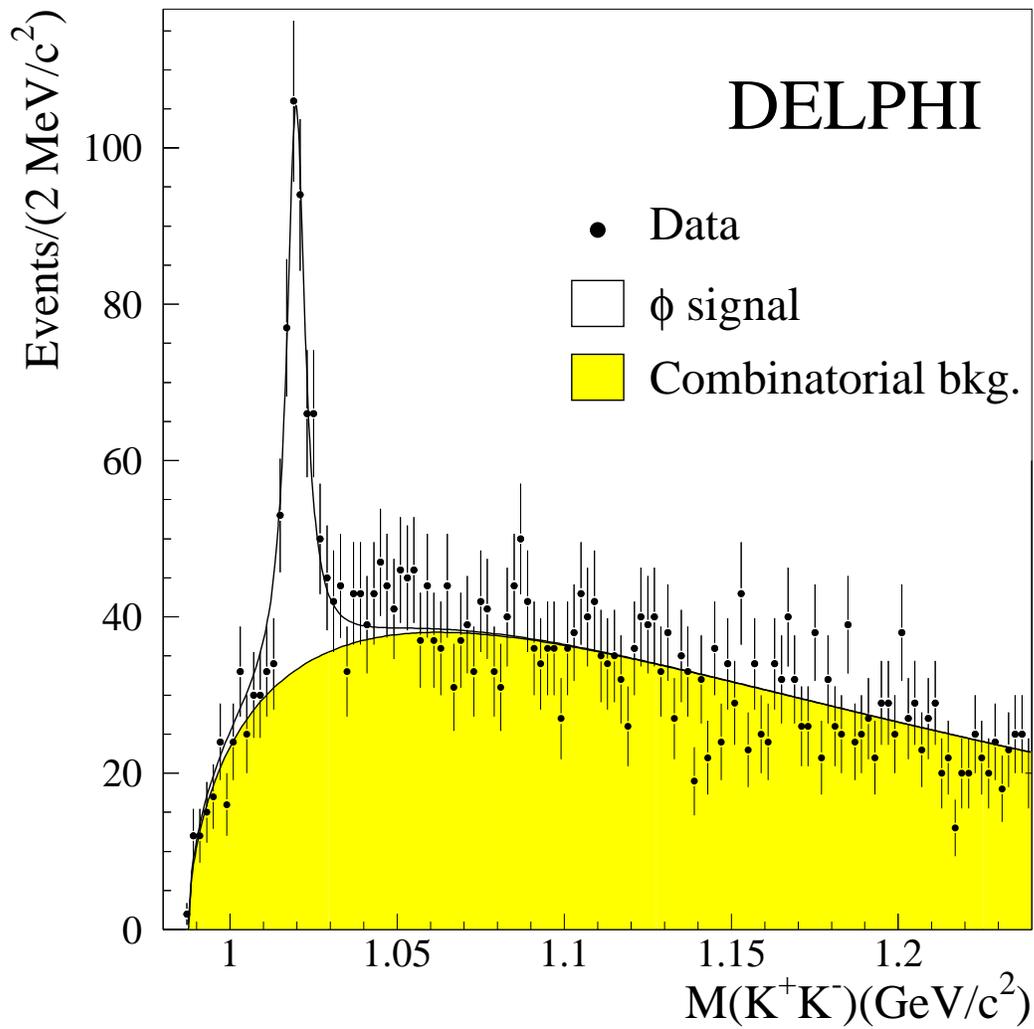,height=14cm}
  \end{center}
  \caption[]{
   \it $K^+ K^-$ invariant mass distribution of the $\phi \ell h$ candidates.
    The curves show the result of the fit described in the text
    (the signal and the combinatorial background components are represented 
     by the white and shaded histograms respectively).}
  \label{fig:phimas}
\end{figure}

Inclusive $\Bs$ semileptonic decays are reconstructed by
requiring, in the same hemisphere, a high $p_T$ lepton and a reconstructed
$\phi\rightarrow\mathrm{K}^+\mathrm{K}^-$. 
This analysis is expected to be more efficient than analyses based on
completely reconstructed $\Dsp$, at the cost of a higher
background. The extra contamination comes mainly from combinatorial
$\mathrm{K}^+\mathrm{K}^-$ pairs and from non-strange $\rm B$-decays.\\
In order to avoid a statistical overlap with the $\Dsl$ sample
considered previously, all $\rm K^+K^- \ell$ triplets 
selected in the $\Dsl$ channels containing
a $\phi$ in the final state have been excluded from the
present sample.\\
The analysis of the $\phi\ell h$ channel has been performed using 
94-95 data only.\\ 
Leptons are required to have a momentum 
and a transverse momentum larger than $3.0\mbox{ GeV/c}$ and 
$1.0\mbox{ GeV/c}$ respectively. A pair of oppositely
charged identified kaons is considered as a $\phi$ candidate provided
their combined momentum is above $3.0\mbox{ GeV/c}$. Considering the
remaining particles of charge opposite to the lepton, the hadron $h$ with the
highest momentum projected along the $\phi$ direction is associated to
the $\Dsp$ decay vertex. The $\mathrm{K}^+\mathrm{K}^-h^+$
vertex is fitted, and the $\Dsp$ pseudo-track is
reconstructed and fitted with the lepton track to estimate 
the $\rm B$ decay vertex.
The mass distribution of the $\mathrm{K}^+\mathrm{K}^-$ pairs has been
fitted with a Breit-Wigner function to account for true $\phi$ mesons
and a polynomial function for the combinatorial background (Figure~\ref{fig:phimas}).\\
Accepting events within $\pm 1\Gamma$ of the fitted $\phi$ mass,
where $\Gamma$ corresponds to the fitted width of the signal, 441
events are retained, including a combinatorial background of
$(45.2\pm4.5)\%$.

\subsection{Sample composition}
\label{sec:dsl_sel_3}

The lifetime and the oscillations of $\Bs$ mesons have been studied  
selecting, in the $\Dsl$ sample, right-sign events lying in a mass
interval of $\pm 2 \sigma$  ($\pm 1.5 \Gamma$) centered on the measured 
$\mathrm{D_s}$ ($\phi$) mass and, in the $\phi \ell h$ sample, events with the 
candidate $\phi$ meson in a mass interval of $\pm 1 \Gamma$ centered
on the measured $\phi$ mass.

\noindent The following components, entering into the selected sample, have 
to be considered:
\begin{itemize}
 \item $f_{bkg}$: fraction of candidates from the combinatorial background:
       it has been evaluated from the fit of the mass distributions on
       $\Dsl$ and $\phi \ell h$ events;
 \item $f_{f\ell}$: fraction of candidates coming from events 
       having a fake lepton and a real $\Ds$ or $\phi$ meson
       (in the $\phi \ell h$ analysis this category includes also events
        containing true leptons and $\phi$ mesons coming from charm decays or
        light quark hadronization);
 \item $f_{bc\ell}$: fraction of candidates
       in which the high $p_T$ lepton originates from a ``cascade'' decay
       ($b \rightarrow c \rightarrow \bar{\ell}$);
 \item $f_{b\ell}^{\rm B}$: fraction of semileptonic  decays 
       of non-strange $\rm B$ mesons
 \item $f_{b\ell}^{\Bs}$: fraction of semileptonic  decays
       of the $\Bs$ meson.
\end{itemize} 
Only the last four components (i.e. background and signal coming
from physical processes) will be detailed in the following: the 
estimation of the combinatorial background has been already reported 
in previous sections.

\subsubsection{Composition of the $\Dsl$ sample}
\label{sec:dsl_comp}

In the $\Dsl$ sample the $\Ds$ signal of the ``right'' sign correlation
is dominated by $\Bs$ semileptonic decays; other minor sources of $\Dsl$ candidates are:
\begin{itemize}
\item $f_{f\ell}$:\\
      a possible contribution from this source ($\Dsp$-fake $\ell$)
      would give the same contribution in right and wrong sign candidates.
      Since no excess has been observed in wrong sign candidates this component
      has been neglected.
\item $f_{bc\ell}$:\\
 it is the expected fraction of ``cascade'' decays 
         ($\mathrm B \rightarrow {{ \overline {\mathrm D}}^{(\star)}} 
         {\mathrm {D_s}}^{(\star)+} \mathrm X$) 
         followed by the semileptonic decay
         $\mathrm {\overline D} \rightarrow \ell^- \overline{\nu} X$
         yielding right-sign $\mathrm{D_s^{\pm}}\ell^\mp$ pairs
         (referred also as $f_{\mathrm D_s \mathrm D}$).
         This background corresponds approximately to the same number 
         of events as the signal \cite{ref:bspaper}, but the selection
         efficiency is lower because of the requirement of a high $p_T$ lepton and of
         a high mass of the$(\mbox{D}_{\mathrm s}\ell)$ system. 
         These selection criteria reduce the $\rm D_s \overline{D}$ background
         fractions to the values reported in Table~\ref{tab:dsd}.
         \begin{table}
         \bc 
         \bt{|l|c|c|c|} 
            \hline
            & $\phi \pi$ & $\phi \ell$ & Others \\ \hline \hline
           $f_{bcl}/f^{\Bs}_{bl}$  
           & $0.151 \pm 0.018$ & $0.148\pm0.025$    & $0.114\pm0.020$  \\ \hline
         \et
         \ec
         \caption{\it Ratio between $\rm D_s \overline{D}$ and signal yields
                  in the three $\Dsl$ classes.}
         \label{tab:dsd}
         \end{table}
         Quoted errors on these fractions result from the uncertainties
         on the branching
         fractions of the contributing processes and from the errors
         on the respective experimental selection efficiencies.
\item $f_{b\ell}^{\rm B}$:\\
      two contributions to this fraction have been considered:
      \begin{itemize}

  \item  $f_{refl}$: the fraction of events from 
         $\Dp\rightarrow\mathrm{K}^- \pi^+ \pi^+$ and 
         $\mathrm D^+\rightarrow{ {\mathrm{K}}^{ 0}_S} \pi^{+}$  decays 
         in which a $\pi^{+}$ has been misidentified as a $\mathrm{K}^{+}$ 
         which give candidates in the $\Ds$ mass region. 
         If the $\mathrm D^+$ is accompanied by an oppositely charged lepton 
         in the decay ${ \overline{\mathrm B}_{\mathrm {u,d}}}\rightarrow 
         \mathrm D^{+} \ell^{-} {\overline \nu} X$, it looks like a $\Bsb$ 
         semileptonic decay.
            The fractions 
            $f_{refl}/f_{{\mathrm B}_{\mathrm s}} = 0.054\pm 0.015$ and 
            $f_{refl}/f_{{\mathrm B}_{\mathrm s}} = 0.069\pm 0.025$ 
           have been estimated for
           the $\overline{\mathrm{K}}^{\star 0} \mathrm{K}^+$ and 
           $\mathrm{K}^0_S \mathrm{K}^+$ decay channels, respectively.
 \item A $\rm{D_s^{\pm}} \ell^{\mp}$ pair from a non-strange B meson decay, 
         with the lepton
         emitted from a direct B semileptonic decay, may come from the decay
         ${\rm \overline{B}} \rightarrow \Ds
          \rm{KX} \ell^{-}\overline{\nu}$.
         The production of $\Ds$ in B decays not originating from $W^{+} 
         \rightarrow c \overline{s}$, has been measured by CLEO 
         \cite{ref:lowervtx}, but no measurement of this production in 
         semileptonic decays exists yet. This process implies the production 
         of a $\rm D^{\star \star}$ followed by its decay into
       $\rm{D_{s}} \rm{K}$. This decay is suppressed by  phase space
      (the $\rm{D_s}$ $\rm{K}$ system has a large mass) and by the 
       required additional $s \overline{s}$ pair. A detailed
      calculation shows that the contribution of this process is \cite{teor1}:
      $$ \frac{{\rm Br}(b \rightarrow
      {\rm {\overline B}}\rightarrow \Ds \rm{K} \rm{X} \ell^{-}
      {\overline \nu})   }{
      {\rm Br}(b \rightarrow {\rm \overline{\Bs}} 
      \rightarrow \Ds \ell^{-}{\overline \nu})    } < 10 \%.
      $$
      Assuming a selection efficiency similar to the one for the $\rm D_s \overline{D}$
      component the contribution of this decay channel is below 2\% and,
      for this reason, has been neglected in the following.
      \end{itemize}

\end{itemize}

\noindent Taking into account the above components, the estimated number of $\Bs$ semileptonic 
decays in the sample of 436 candidates is $230\pm18$.\\
The signal composition for each $\Ds$ decay mode is given in Table~\ref{tab:comdsl}.
\begin{table}
\bc
\bt{|l|c|c|c|c|c|}
\hline
& $\phi \pi$ & $\phi\ell$        & $\rm K^0_S K^+$ & $\rm K^{*0} K^+$ & \mbox{Others} \\
\hline
\hline
$f_{\Bs}$          & $0.869\pm 0.014$ & $0.871\pm0.019$  & $0.845\pm0.023$  & $0.856\pm 0.018$ & $0.898\pm0.016$   \\ 
\hline
$f_{\rm \Ds \overline{D}}$    & $0.131\pm 0.016$ & $0.129\pm0.022$  & $0.096\pm0.021$  & $0.098\pm 0.020$ & $0.102\pm0.018$  \\ 
\hline
$f_{refl.}$       &        -          &       -          & $0.059\pm0.022$  & $0.046\pm 0.014$ & -  \\ 
\hline
\et
\caption{Estimated composition of the $\Ds$ signal in the $\Dsl$ sample}
\label{tab:comdsl}
\ec
\end{table}

In order to increase the effective $\Bs$ purity of the selected sample,
signal and background fractions have been calculated on an event by event 
basis using the probability density functions of $p_T(l)$ and $X_{\Ds}$
(defined in Section~\ref{dsl_sel_0}):
\bes
\ba{lll}
  f^{eff}_{bkg}            & = &   
  f_{bkg}{\cal F}_{Comb}(X_{\Ds}){\cal F}_{Comb}(p_T)/Tot          \\
  f^{eff}_{\mathrm {\Bs}}  & = &   
  f_{\Bs}{\cal F}_{\rm D_s}(X_{\Ds}){\cal F}_{\rm \Bs}(p_T)/Tot    \\
  f^{eff}_{D_sD}           & = &   
  f_{D_sD}{\cal F}_{\rm D_s}(X_{\Ds}){\cal F}_{\rm D_sD}(p_T)/Tot  \\
  f^{eff}_{refl}           & = & 
  f_{refl}{\cal F}_{\rm D_s}(X_{\Ds}){\cal F}_{\rm \Bs}(p_T)/Tot    
\ea
\ees
where ${\cal F}_{\rm D_s}$,
${\cal F}_{Comb}$,
${\cal F}_{\rm D_s D}$,
${\cal F}_{\Bs}$ are 
the probability densities for the $\rm \Ds$ mesons,
the combinatorial background,
the $\rm D_s D$ background and
the $\rm \Bs$ signal events, respectively.\\
In these expressions, $Tot$ is a normalisation factor such that:
$$
f^{eff}_{bkg}+f^{eff}_{\mathrm {\Bs}} +f^{eff}_{D_sD}+f^{eff}_{refl}~ =~ 1.
$$
The distributions of the values of the 
$X_{\Ds}$ and $p_T$ variables are shown in Figure \ref{fig:plot_chk_bcl}. 
The use of this procedure is equivalent to increasing the statistics 
by a factor 1.2.

\subsubsection{Composition of the $\phi\ell h$ sample}
\label{phi_comp}
The different contributions to $\phi\ell h$ candidates contained 
in the selected $\rm K^+ K^-$ mass interval of $\pm 6.6~MeV/c^2$
around the $\phi$ nominal mass and corresponding to a real $\phi$ meson
are shown in Table~\ref{tab:comp} and have been estimated using
simulated events and measured branching fractions. 
Quoted uncertainties originate from the finite Monte Carlo statistics, except for 
one attached to the signal fraction which is dominated 
by $f_{\mathrm{B}^0_s}\times\mathrm{Br}(\mathrm{B}^0_s\rightarrow
\Dsp \ell^-\bar{\nu}_\ell\mathrm{X})=(0.86\pm0.09^{+0.29}_{-0.20})\%$ 
\cite{ref:wav}.

\begin{table}[h]
  \begin{center}
    \bt{|l|c|} 
      \hline
      \mbox{Source}              & $(\%)$                   \\ \hline \hline
      $f_{f\ell}$                & $22.5\pm2.4$             \\ \hline
      $f_{b\ell}^{\rm B}$        & $49.4\pm2.1$             \\ \hline
      $f_{bc\ell}$               & $11.3\pm1.2$             \\ \hline
      $f_{b\ell}^{\rm \Bs}$      & $16.9^{+6.0}_{-4.3}$     \\ \hline
    \et
  \end{center}
  \caption{\it Estimated composition of the $\phi$ signal
  in the $\phi \ell h$ sample}
  \label{tab:comp}
\end{table}
The number of $\Bs$ semileptonic decays contained in this sample has
been evaluated to be $41^{+15}_{-10}$.

\subsection{Measurement of the B meson decay time}
\label{sec:dsl_sel_x}

\begin{figure}[ph]
\begin{center}
\epsfig{figure=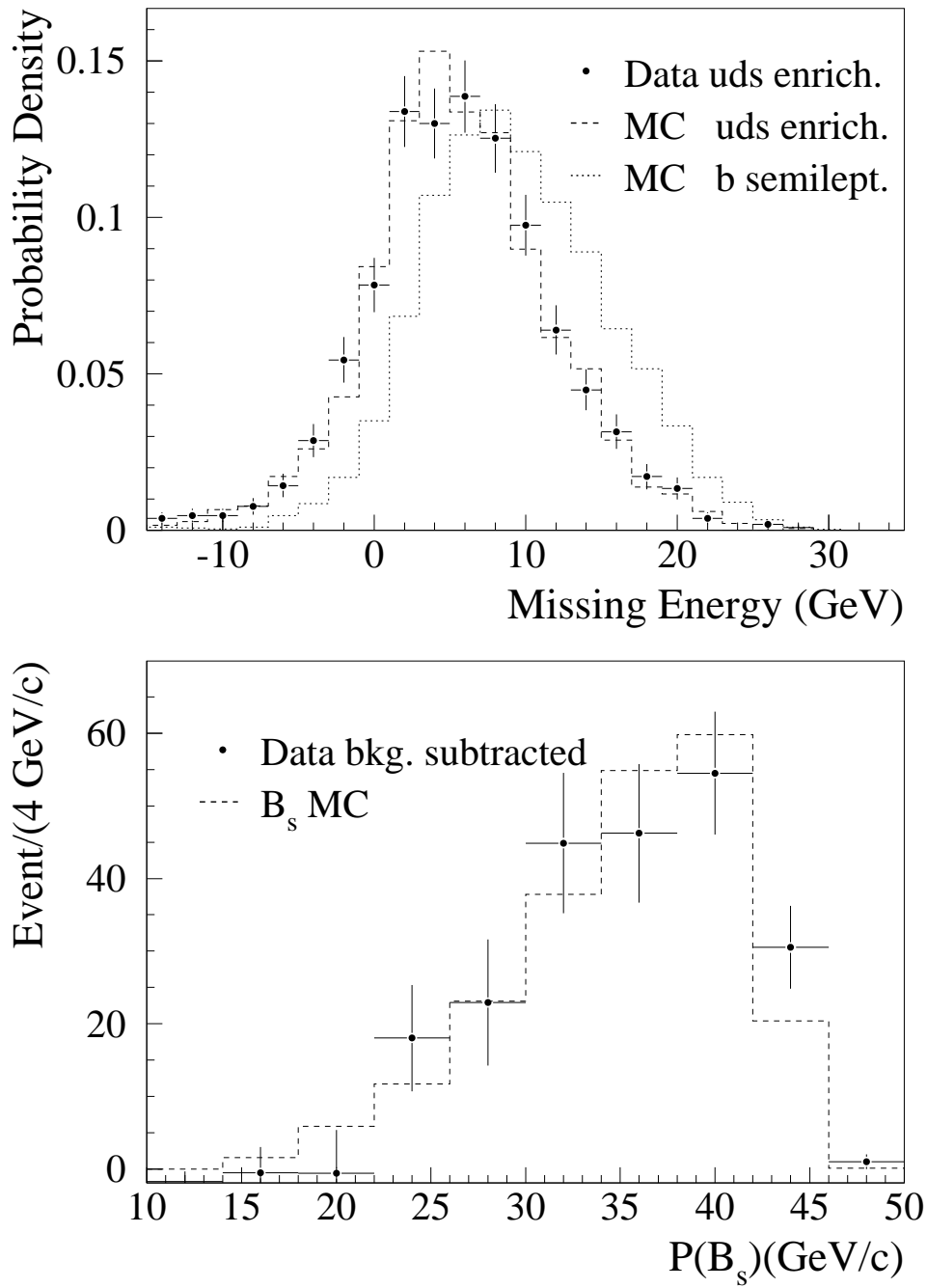,height=18cm}
\caption []{\it
The plot on the top shows the comparison between data and the simulation 
for the missing energy distribution after correction (see Section~\ref{sec:dsl_sel_x}). 
The data (dots with error bars) and the simulation (dashed 
histogram) are enriched in light quark events. 
The dotted histogram shows the missing energy distribution in simulated b semileptonic 
decays.
The plot on the bottom shows the comparison between the $\Bs$ momentum distribution 
for simulated events and the one estimated from  data in the signal region by 
subtracting the $\Bs$ momentum distribution of events in the $\Ds$ side 
bands from that of the events in the signal region.}
\label{fig:plot_chk_ene}
\end{center}
\end{figure}

For each event, the $\Bs$ decay time is obtained from the measured
decay length ($L_{\Bs}$) and the estimate of the $\Bs$ momentum ($p_{\Bs}$).
The $\Bs$ momentum is estimated using the measured energies:
\begin{eqnarray*}
{p_{\Bs}}^{2}=(E(\Ds\ell)+ E_{\nu})^{2}-{m_{\Bs}}^{2}.
\end{eqnarray*}
The neutrino energy $E_{\nu}$ is obtained from the measured value of
$E_{miss}$ (see Section \ref{dsl_sel_0}).\\
The agreement between data and simulation on $E_{miss}$ has 
been verified using the side bands of the $\Dsl$ sample. 
In order  to have enough statistics to perform this test,
cuts not correlated with the missing energy have been relaxed. 
In addition, to verify the resolution on the energy estimate, the 
studied sample has been enriched in light quark events by applying an 
anti--b-tagging cut (Section \ref{sec:btag}).
A relative shift of $\Delta(MC-Data)~=~500~MeV$ has been measured and the 
simulation has been
corrected. Figure~\ref{fig:plot_chk_ene} shows the agreement
between data and simulation after having applied this correction.\\
The neutrino $E_{\nu}$ resolution has been improved by correcting
$E_{miss}$ by a function of $(\Ds\ell)$ 
energy\footnote{here $\Ds$ means ``observed decay products of
$\Ds$'', including also the decays where the $\Ds$ is not fully 
reconstructed: specifically  $\Ds^+ \rightarrow \phi \ell^+\nu_{\ell}$
and $\Ds^+ \rightarrow \phi h^+X$}
and  which has been determined from simulated signal events:
\begin{eqnarray*}
{E_{\nu}}=E_{miss}+F(E(\Ds\ell)).  
\end{eqnarray*}
The data-simulation 
agreement on $p_{\Bs}$ has been verified  on the selected signal events after 
subtraction of the combinatorial background (estimated from 
events selected in side-bands
of $\Ds$ and $\phi$ signals) 
(Figure~\ref{fig:plot_chk_ene}). 

\subsection{Proper time resolution and acceptance}
\label{sec:propt}
\begin{figure}[ph]
\begin{center}
\epsfig{figure=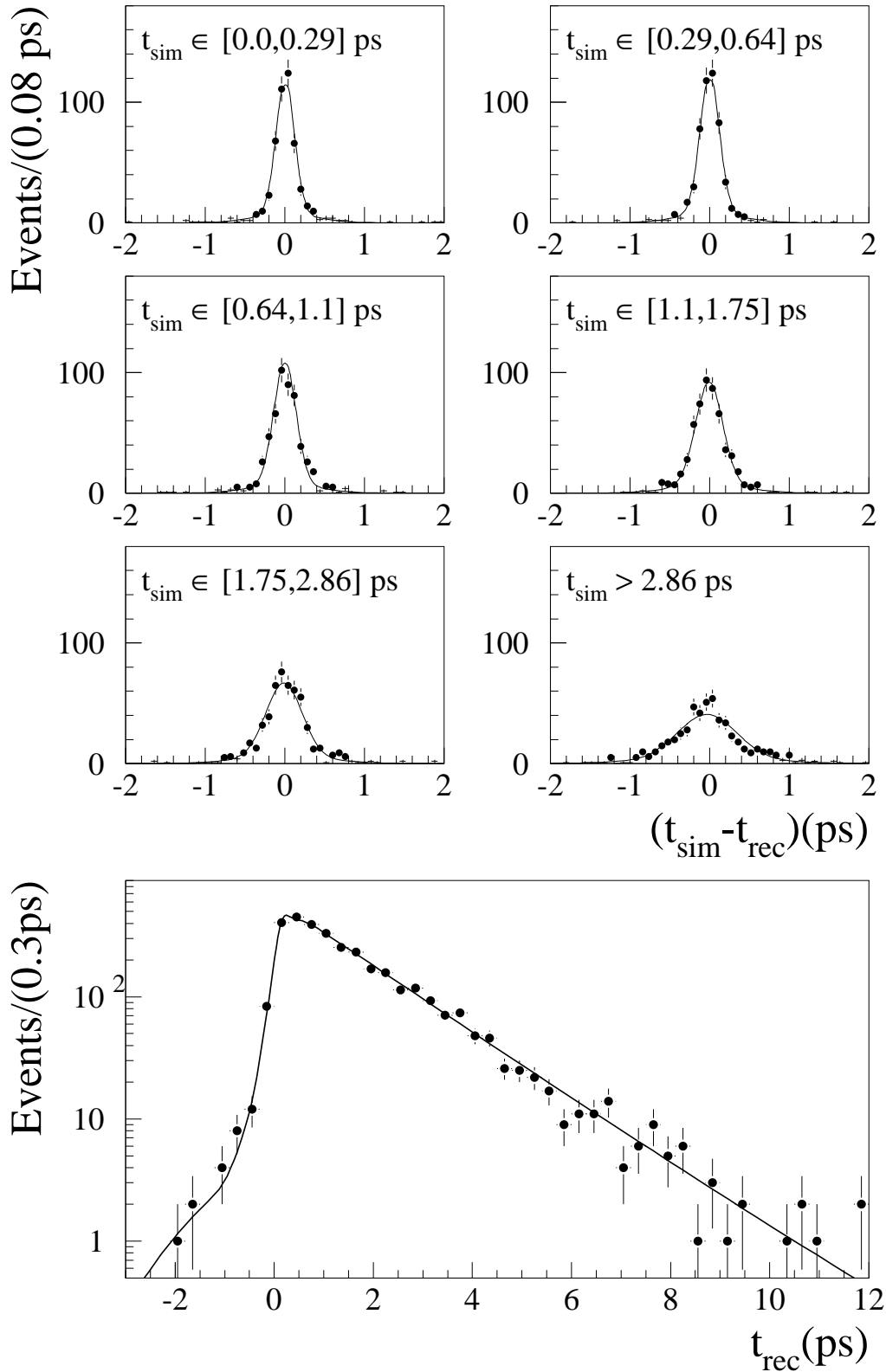,height=22cm}
\caption []{\it The six plots on the top show the proper time resolution, 
defined as the difference between the generated
time ($t_{sim}$) and the reconstructed time ($t_{rec}$), in bins of generated time
on a sample of $\phi \pi$ events simulated with the 94-95 Vertex Detector configuration.  
The plot on the bottom shows the distribution of the
reconstructed time with the fit superimposed.}
\label{fig:plot_res}
\end{center}
\end{figure}

The predicted decay
time distributions have been obtained by convoluting the theoretical 
distributions with resolution functions evaluated from simulated events.
Due to different resolutions on the decay length, different parametrizations of 
the proper time resolution have been used for
three different classes in the $\Dsl$ sample: $\mathrm{K}^{0}_S \mathrm{K}^{+}$ decays, 
other non-leptonic decays and semileptonic $\mathrm D_{s}$ decays.
Different parametrizations have been also used for the two Vertex
Detector configurations installed in 91-93 and in 94-95.
The proper time resolution is obtained from the distribution of the difference 
between the generated ($t$) and the reconstructed ($t_{i}$) time.
The following distributions have been considered:
\begin{itemize}
\item ${\cal R}_{bl}(t-t_{i})$ is the resolution function for direct
semileptonic $\rm B$ decays.\\
${\cal R}_{bl}$ is parametrized, for the $\Dsl$ sample, as the sum of three
Gaussian distributions. The width of the third Gaussian is taken
to be proportional to the width of the second Gaussian.
\bes
\ba{l}
  {\cal R}_{\rm b\ell}(t-t_{i}) = (1-f_2-f_3)G(t-t_i,\sigma_1)+ 
               f_2G(t-t_i,\sigma_2)+ 
               f_3G(t-t_i,\sigma_3)                          \\
  \mbox{with~}   \sigma_1 = \sqrt{\sigma_{L1}^2+\sigma_{P1}^2t^2}  
  \hspace{0.2cm}            \\
  \hspace{0.9cm}  \sigma_2 = \sqrt{\sigma_{L2}^2+\sigma_{P2}^2t^2} 
  \hspace{0.2cm}     \\
  \hspace{0.9cm}  \sigma_3 = s_3\sigma_2
\ea
\ees
In the $\phi \ell h$ analysis a fourth Gaussian distribution has been added.\\
The parameters related to the decay length and proper time resolutions,
$\sigma_{L_i}$ and $\sigma_{P_i}$ respectively, 
and the relative fractions $f_i$ are listed in Table~\ref{tab:b}.
A typical parametrization of the resolution, for the $\Dsl$ sample,
is shown in Figure~\ref{fig:plot_res} for the $\phi\pi^+$ decay mode obtained with the 94-95 
Vertex-Detector configuration.\\
\item ${\cal R}_{bcl}$ is the resolution function applied to
``cascade`` events. \\
Since the charm decay products have been only partially  reconstructed in 
these events, the momentum of the $\Bsb$ candidate is underestimated
giving a long positive tail in the proper time resolution 
function.\\
The function,
${\cal R}_{bcl}(t-t_{i})$, is well described by a Gaussian 
distribution convoluted with an exponential distribution. The variation of 
the shape of this distribution with the 
generated proper time has been neglected.
\end{itemize}
\begin{table}[hbt]
\bc
\bt{|l|l|l|l|l|l|l|l|} \hline 
  \multicolumn{8}{|c|}{$\Dsl$ sample} \\ \hline
  \mbox{$\mathrm D_s$ decay mode} &$\sigma_{L1}(ps)$&$\sigma_{P1}$
  &$\sigma_{L2}(ps)$&$\sigma_{P2}$&$s_3$&$f_2$&$f_3$ 
   \\ \hline \hline
   $\mathrm K^0_S K^+$ (92-93)  & 0.16 & 0.08  & 1.04 & 0.16 & - & 0.50 
   & 0    \\  \hline
   $\mathrm K^0_S K^+$ (94-95)  & 0.16 & 0.08  & 0.98 & 0.16 & - & 0.28 
   & 0    \\  \hline
   other non-leptonic (92-93)   & 0.11 & 0.07  & 0.39 & 0.16 & 5 & 0.26 
   & 0.07 \\  \hline
   other non-leptonic (94-95)   & 0.11 & 0.07  & 0.37 & 0.16 & 3 & 0.16 
   & 0.02 \\  \hline
   $\phi \ell^+ \nu$ (92-93)    & 0.14 & 0.075 & 0.31 & 0.15 & 6 & 0.29 
   & 0.09 \\  \hline
   $\phi \ell^+ \nu$ (94-95)    & 0.14 & 0.075 & 0.31 & 0.15 & 6 & 0.21 
   & 0.07 \\
   \hline
\et
\vskip 0.5cm
\bt{|c|c|c|c|c|c|c|c|c|c|c|} \hline 
 \multicolumn{11}{|c|}{$\phi \ell h$ sample} \\ \hline
     $\sigma_{L1}(ps)$ & $\sigma_{P1}$ 
   & $\sigma_{L2}(ps)$ & $\sigma_{P2}$ 
   & $\sigma_{L3}(ps)$ & $\sigma_{P3}$ 
   & $\sigma_{L4}(ps)$ & $\sigma_{P4}$ 
   & $f_1$ & $f_2$ & $f_3$ \\ \hline \hline
     0.13 & 0.08 & 0.28 & 0.09 & 0.32 & 0.19 & 1.06 & 0.42 
   & 0.31  & 0.41  & 0.10 \\ \hline
\et
\ec
\caption{ \it Fitted values of the parameters of the resolution function 
${\cal R}_{bl}(t-t_{i})$ 
obtained, on simulated events, for the $\Dsl$ and $\phi \ell h$ samples.}
\label{tab:b}
\end{table}

Distortions on the reconstructed proper time can be due to a non-uniform
reconstruction efficiency as a function of the true proper time (acceptance).\\
Non-uniform efficiencies have been observed, on simulated events,
in the $\Ds$ decay modes $\phi \pi$, $\rm K^{*0} K$ and $\phi \ell \nu$
because of the selection criteria on $\rm L/\sigma(L)$.\\
This effect has been taken into account by inserting in the fitting function, 
for those channels, an acceptance function (${\cal A}(t)$) 
parametrized on simulated events.

\section{ Measurement of the $\rm B^0_s$ lifetime}
\label{sec:taubs}
The $\Bs$ meson lifetime has been studied using the signal 
sample (Section~\ref{sec:dsl_sel_3}) and a background sample
containing events selected in the sidebands of $\Ds$ ($\phi$) candidates.
Sidebands events are ``right'' sign events lying in the $\Ds$ mass interval
$[1.91-1.93]\cup[2.01-2.15]~GeV/c^2$ for the $\Ds$ hadronic decays and
``right'' sign events lying in the $\phi$ mass interval
$[0.990,1.005]\cup[1.035,1.060]~GeV/c^2$ for the $\Ds$ semileptonic decays.\\ 
In the $\Dsl$ analysis ``wrong'' sign candidates have been also included
in the background sample. 
This background sample is assumed to have the same proper
time distribution as the combinatorial background in the signal
sample. This assumption has been verified using the simulation.
The probability density function used for events in the signal region 
is given by:
\bes
P(t_i) =   f^{\Bs}_{b\ell}           P^{\Bs}_{b\ell}(t_i) 
         + f^{\rm B}_{b\ell}         P^{\rm B}_{b\ell}(t_i) 
         + f_{bc\ell}                P_{bc\ell}(t_i) 
         + f_{f\ell}                 P_{f\ell}(t_i) 
         + f_{bkg}                   P_{bkg}(t_i).  
\ees
where $t_i$ and $t$ are the measured and true 
proper times respectively.\\
The different probability densities are expressed as convolutions of the 
physical probability densities with the appropriate resolution (${\cal R}$) and acceptance (${\cal A}$)
functions:
\begin{itemize}
\item 
for the signal:
\bes
   P^{\mathrm \Bs}_{b\ell}(t_i) =  {1\over \tau_{\mathrm{\Bs}}}exp(-t/\tau_{\mathrm{\Bs}}) {\cal A}(t)\otimes 
   {\cal R}_{b\ell}(t-t_i)
\ees
\item for the background coming from non strange $B$ mesons: \\
\bes
   P^{\rm B}_{b\ell}(t_i) =  
   \sum_{q\neq s} f_{b \ell}^{\rm B_q}{1\over \tau_{B_q}}
   exp(-t/\tau_{B_q}) {\cal A}(t)\otimes {\cal R}_{b\ell}(t-t_i)
\ees
where $q$ runs over the various B-hadrons species contributing to this
background,
\item for the ``cascade'' background:\\
\bes
   P_{bc\ell}(t_i) =  
   \sum_{q} f_{bc \ell}^{\rm B_q}{1 \over \tau_{\rm B_q}}exp(-t/\tau_{\rm B_q}){\cal A}(t)\otimes 
   {\cal R}_{bc \ell}(t-t_i) 
\ees
\item for ``fake lepton'' candidates the function 
      $P_{f\ell}(t_i)$ has been parametrized using simulated events;
\item for the combinatorial background two different parametrizations 
      have been used:
  \begin{itemize}
    \item $\Dsl$ sample.

  \bes
  \ba{ll}
  P^{j}_{bkg}(t_i) = &  f^-   {1\over \tau^-}exp(-t/\tau^-)\otimes G(t-t_i,\sigma_j) +
                    f^+   {1\over \tau^+}exp(-t/\tau^+)\otimes G(t-t_i,\sigma_j) + \\
                 &   (1-f^--f^+)                G(t-t_i,\sigma_j)
  \ea
  \ees
 Three distributions have been used for each of the three classes of decay time
 resolution $\sigma_j~(j=1,3)$ (see Section \ref{sec:propt}). 
 A negative exponential for poorly measured 
 events (with negative lifetime $\tau^-$),
 an exponential distribution for the flying background 
 (with lifetime $\tau^+$) and a central Gaussian for the non-flying one. 
 The seven parameters ($f^-$, $f^+$, $\tau^+$, $\tau^-$ and $\sigma_j~(j=1,3)$)
 have been fitted independently for the 92-93 and 94-95 data samples.
 The parameter $\sigma_j~(j=1,3)$ are taken to be different for the three classes 
 of decay time resolution.
 \item $\phi \ell h$ sample.\\
   The combinatorial background shape has been described with a sum
   of four smeared exponentials ($exp(t_i,\tau)   \otimes G(t-t_i,\sigma)$).
 \end{itemize}
\end{itemize}

\begin{figure}[ph]
\begin{center}
\epsfig{figure=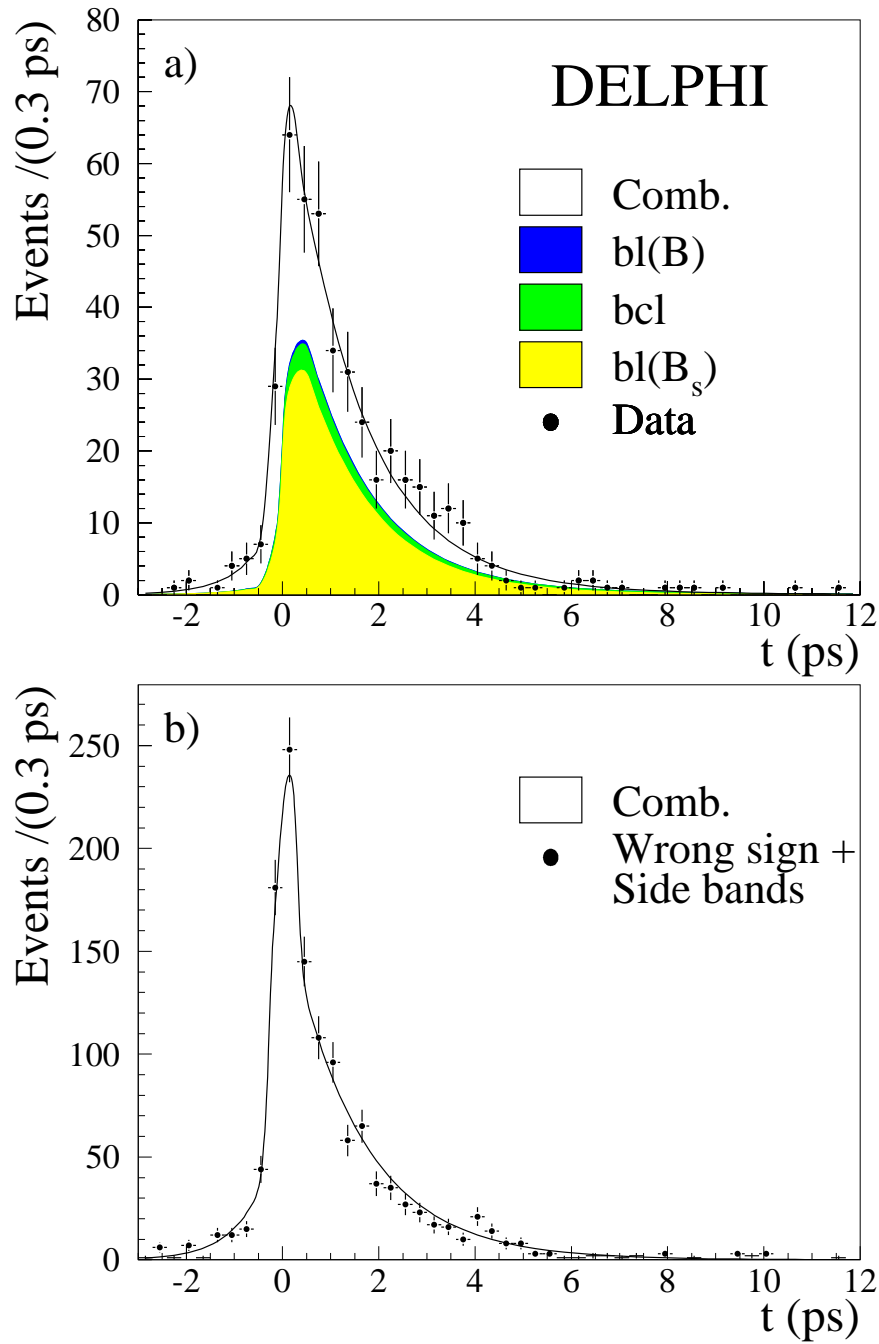,height=18cm}
\caption[]{  \it $\Dsl$ sample.
          a)  Likelihood fit for events in the signal mass region.
          The points show the data and the curves correspond to the different
          contributions to the selected events.
          b) The same as a) but for  ``wrong-sign'' events and for events 
          situated in the side band region.\\
          The labels of the signal components refers to Section~\ref{sec:dsl_comp}} 
\label{fig:dsl_tau}
\end{center}
\end{figure}

\begin{figure}[ph]
\begin{center}
\epsfig{figure=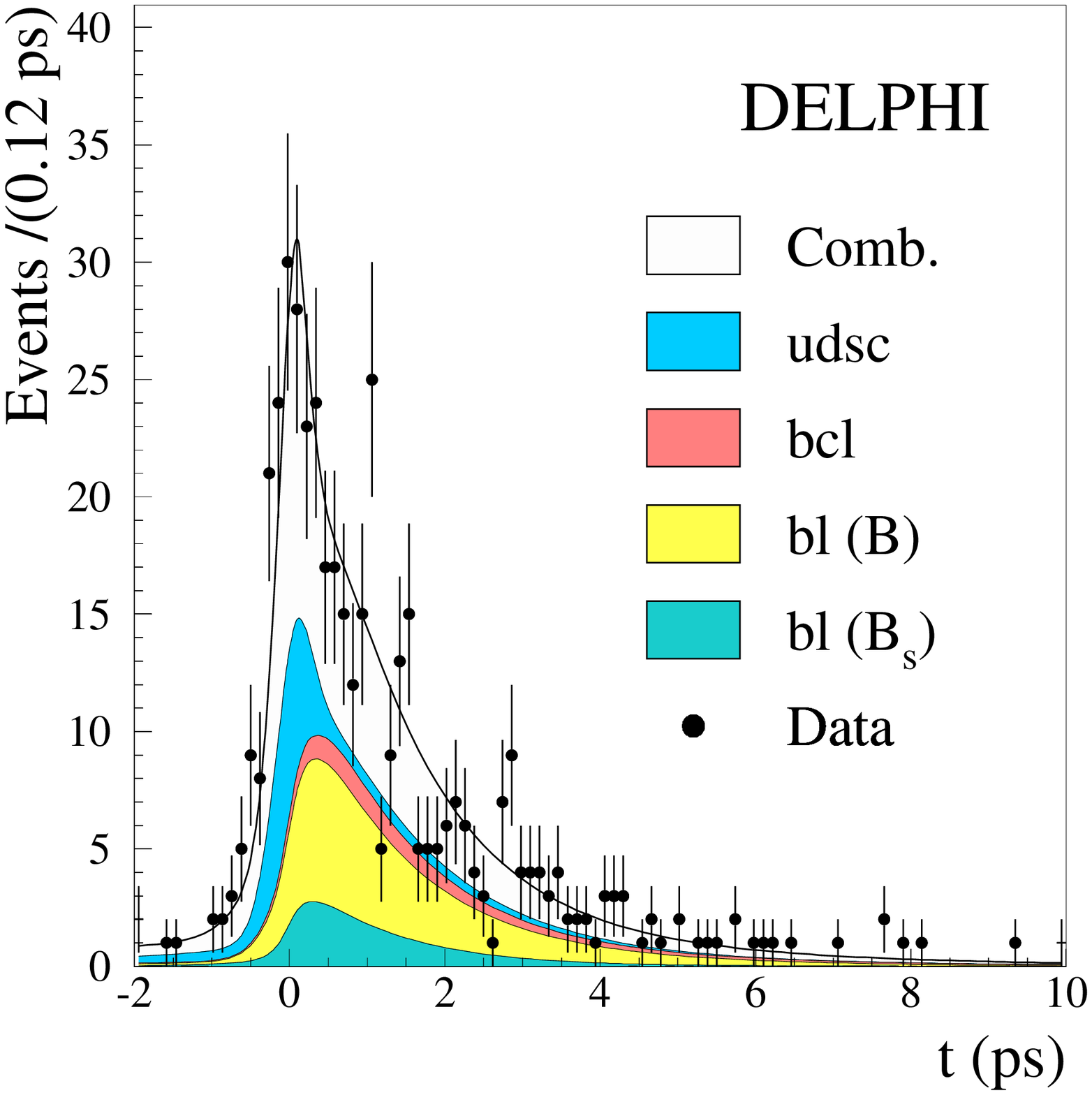,height=16cm}
\caption[]{  \it $\phi \ell h$ sample.\\
          Likelihood fit for events in the signal mass region.
          The points show the data and the curves correspond to the different
          contributions to the selected events.\\
          The labels of the signal components refers to Section~\ref{phi_comp}}
\label{fig:phil_tau}
\end{center}
\end{figure}

\noindent $\Bs$ lifetime fit has been performed simultaneously on the signal and 
background samples.
All parameters describing the shape of the background time distributions
in the $\Dsl$ and $\phi \ell h$ samples are left as free parameters.
Results of the fit are shown
in Figure~\ref{fig:dsl_tau}  ($\Dsl$ sample) and
in Figure~\ref{fig:phil_tau} ($\phi \ell h$ sample). 
Table~\ref{tab:sum_tbs} summarizes the different lifetimes measurements with their 
statistical errors.

\renewcommand\arraystretch{1.2}
\begin{table}[htb]
\bc
\begin{tabular}{|l|l|c|} \hline 
  \mbox{Decay mode} & \mbox{Data set} &$\tau_{\Bs}(ps)$ \\ \hline \hline
  $\Dsl; \Ds \rightarrow \phi\pi$               
  & \mbox(92-95)  & $1.44 ^{+0.26}_{-0.21}$ \\ \hline
  $\Dsl; \Ds \rightarrow {\overline {\mathrm{K}}}^{\star0} \mathrm{K}^+$ 
  & \mbox(92-95)  & $1.31 ^{+0.30}_{-0.25}$ \\ \hline
  $\Dsl; \Ds \rightarrow {\mathrm{K}^{0}_S} \mathrm{K}^+               $ 
  & \mbox(92-95)  & $1.43 ^{+0.61}_{-0.44}$ \\ \hline
  $\Dsl; \Ds \rightarrow {\overline {\mathrm{K}}}^{\star 0}{\mathrm{K}}^{\star +}$ 
  & \mbox(94-95)  & $1.00 ^{+0.50}_{-0.31}$ \\ \hline
  $\Dsl; \Ds \rightarrow \mathrm \phi \pi^{+} \pi^0          $ 
  & \mbox(94-95)  & $1.46 ^{+0.61}_{-0.42}$ \\ \hline
  $\Dsl; \Ds \rightarrow \mathrm \phi \pi^{+} \pi^{+} \pi^{-}$ 
  & \mbox(94-95)  & $1.96 ^{+1.16}_{-0.64}$ \\ \hline
  $\Dsl; \Ds \rightarrow \mathrm \phi \ell^{+} \nu         $ 
  & \mbox(92-95)  & $1.49 ^{+0.34}_{-0.27}$ \\ \hline
  $\phi \ell h$ 
  & \mbox(94-95)  & $1.41 \pm 0.68$ \\ \hline 
\end{tabular}
\ec
\caption{\it $\Bs$ lifetime determinations using the $\Dsl$ and $\phi \ell h$ events samples.}
\label{tab:sum_tbs}
\end{table}
\renewcommand\arraystretch{1.0}

\subsection{Systematic errors on the $\Bs$ lifetime}
Systematic uncertainties attached to the $\Bs$ lifetime determination
are summarized in Table~\ref{tab:syst_tau}.\\
\renewcommand\arraystretch{1.2}
\begin{table}[h]
\bc
  \bt{|c|c|} \hline 
       \mbox{Systematics} & \mbox{$\tau_{\Bs}$ variation in $ps$} \\ \hline \hline
      $f_{bkg.       }$              & $^{+0.0090}_{-0.0130}$ \\
      $f_{bc\ell  }$              & $^{-0.0100}_{+0.0110}$ \\
      $f^{\rm B}_{b\ell     }$              & $^{-0.0020}_{+0.0020}$ \\
     \mbox{$X_{\Ds}$ discrim. var.}   & ${+0.008}           $ \\
     \mbox{$p_T$    discrim. var.}   & $\pm0.004        $ \\
     \mbox{$\tau_{\rm B^+}(1.65\pm0.04~ps)$}   & $^{-0.0010}_{+0.0010}$ \\
     \mbox{$\tau_{\rm \Bd}(1.56\pm0.04~ps)$}   & $^{-0.0012}_{+0.0013}$ \\
      \hbox{$t$ resolution}          & $\pm 0.008$            \\ 
      \hbox{$t$ acceptance}          & $\pm 0.010$            \\ 
      \hbox{Simulated evts. statistics}  & $\pm 0.020$            \\ 
       \hline
       \hline
      \hbox{Total Syst. }  & $\pm0.03$      \\ 
       \hline
  \et
\ec
\caption{\it Different contributions to the systematic uncertainty attached
             to the $\Bs$ lifetime measurement.}
\label{tab:syst_tau}
\end{table}
\renewcommand\arraystretch{1.0}

\begin{figure}[ph]
\begin{center}
\epsfig{figure=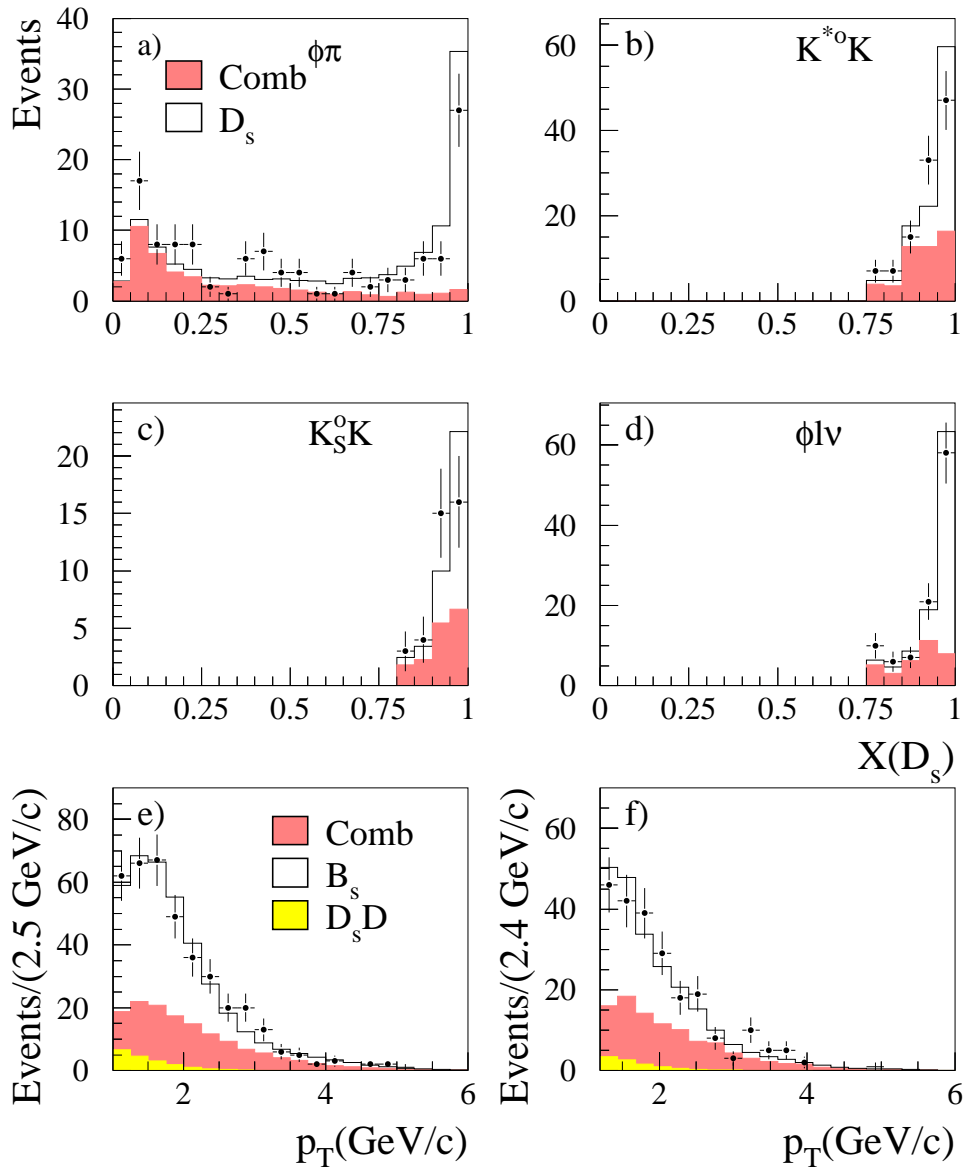,height=16cm}
\caption[]{ \it The plots illustrate
 the agreement between data and simulation 
 for the variables used in the estimate of the $\Bs$ purity
 on an event by event basis.\\
 a),b),c) and d) show the $X_{\Ds}$ distributions for the channels 
 $\phi \pi$, $\rm \overline{K^{0*}}K$, $\rm K^0_S K$ and $\phi \ell^+$ 
 respectively. White histograms and shaded histograms represent the
 signal and the background respectively. \\
 e) and f) show the $p_T$ distributions for the samples 
 selected with $p_T>1~GeV/c$ ($\phi \pi^+$ and $\phi \ell^+ \nu$)
 and $p_T>1.2~GeV/c$ (all the others) respectively.\\
 For the $p_T$ distribution the $\Ds\overline{D}$ and the combinatorial background
 are considered separately.}
\label{fig:plot_chk_bcl}
\end{center}
\end{figure}

The main contributions to the systematic uncertainties come from:
\begin{itemize}
\item { Systematics from the evaluation of the $\Bs$ purity.}\\
\begin{itemize}
\item $\Dsl$ sample:\\
The different fractions for signal and background events have been calculated 
on an event by event basis. The expressions 
defining the effective purities are given in Section~\ref{sec:dsl_comp}. The value of 
$f_{bkg.}$ has been varied according to the statistical 
uncertainties of the fitted combinatorial background fractions present in 
the  different, $\Ds$ or $\mbox{K}^+ \mbox{K}^-$, mass 
distributions.  The value of $f_{bc\ell}$ has been varied 
according to the errors given in Table~\ref{tab:comp} and in Table~\ref{tab:comdsl}, 
which takes into account both the statistical error from the simulation and the 
errors on measured branching ratios.\\
The evaluation of the systematics due to the procedure used to evaluate 
the $\Bs$ purity on a event by event basis has been evaluated in two steps.
The distributions of the variable $X_{\Ds}$ (Figure \ref{fig:plot_chk_bcl}) 
for signal and background events have been re-weighted with as linear 
function in order to maximize the Data-simulation agreement:
\bes
  {S(X_{\Ds}) \over B(X_{\Ds}) }_{\rm new} = 
  {S(X_{\Ds}) \over B(X_{\Ds}) }_{\rm old}(a+bX_{\Ds}) 
\ees
The linear behaviour of the correction has been chosen because of the limited 
statistics in the data: it has been verified that a quadratic correction
does not change the result significantly.\\ 
The fit has been redone with this new probability distribution and the 
variation of the fitted lifetime value (+0.008~$ps$)
has been taken as the systematic error.\\
Because of the agreement between data and simulation
(Figure \ref{fig:plot_chk_bcl}-e and \ref{fig:plot_chk_bcl}-f)
for the $p_T$ distribution, the systematic error associated to this variable 
has been evaluated varying its distributions by the uncertainties of the 
parametrization obtained from simulated events.

\item $\phi\ell h$ sample:\\
  In this analysis the fractions of signal and background events have not been calculated
  on an event by event basis. The systematic uncertainty due to the variation
  of the $f_{bc\ell},~f^{B}_{b\ell}~\mbox{and}~f_{bkg}$ fractions have been obtained by varying
  these parameters by the errors reported in Table~\ref{tab:comp}
  and in Table~\ref{tab1}. 
  The systematic uncertainty attached to the $f_{f\ell}$ fraction, affecting only 
  the $\phi \ell$ sample, has a negligible effect on the global result. 
\end{itemize}
\item{Validation of the fitting procedure
      using simulated events.}\\
The fitting method has been verified on pure $\Bs$ simulated events: 
the measured value on this sample has been 
$\tau_{\rm B^0_s}(\rm D^{\pm}_s \ell^{\mp})^{MC} = ( 1.605\pm 0.020 ) ps$ 
in agreement with the generated value
($\tau_{\rm B^0_s}=1.6~ps$). The statistical error of this verification
has been included in the systematic uncertainties.\\
A similar check has been performed on the $\phi \ell h$ sample giving
$\tau_{\rm B^0_s}(\rm \phi \ell h)^{MC} = ( 1.65\pm 0.04 ) ps$.
Since the statistical weight of the $\phi \ell h$ channel is small 
compared to the full sample, the error on the fitting procedure is 
dominated by the statistics of  $\Dsl$ simulated events.\\

\item{ Systematic from the proper time resolution.} \\
Uncertainties on the determination of the resolution on the proper time receive 
two contributions: one from errors on the decay distance evaluation and the other
from errors on the measurement of the $\Bs$ momentum.
The agreement between real and simulated events on the evaluation of the 
errors on the decay distance has been verified by comparing the widths of 
the negative part of the flight distance distributions, for events which are 
depleted in B-hadrons. The difference between the two widths has been found 
to be of the order of 10\%.\\
\noindent
The systematic on $\Bs$ momentum has been evaluated by comparing
the momentum distribution on simulated events with the distribution, 
background subtracted, obtained from the data sample (see Section~\ref{sec:dsl_comp}).
Effects from shift and width differences between the two distributions  
have been considered by changing the shape of the distribution of simulated events; 
it has been found that the main systematics comes from difference in width: 
the width on data has been estimated to be larger by a factor $1.07\pm0.04$.\\ 
\noindent
Taking into account these two effects the uncertainty on the time resolution 
has been, conservatively, evaluated by varying the parameters $\sigma_{L_i}$ and
$\sigma_{P_i}$ of the resolution functions (see Table~\ref{tab:b}) 
by $\pm$10$\%$.\\
Uncertainties on the acceptance determination have been also considered: 
the parameters entering in the definition of the acceptance function have 
been varied according to the errors given by the fit on simulated events.

\end{itemize}

The final result is:
\begin{equation}
   \tau_{\rm B^0_s} =  1.42^{+0.14}_{-0.13}(stat.)\pm 0.03(syst.) ~ps.
\end{equation}

\section{Lifetime difference between $\rm B^0_s$ mass eigenstates}
\label{sec:dg}

The $\Bs$ (or $\Bsb$) mesons are superpositions of the two mass eigenstates:
\bes
|\mathrm{B^0_s} \rangle = {\frac{1}{ \sqrt{2}}}(| \mathrm{B}^0_H \rangle + 
| \mathrm{B}^0_L \rangle)~~;~~
|\mathrm{\overline{B}^0_s} \rangle = {\frac{1}{ \sqrt{2}}}(| \mathrm{B}^0_H \rangle - 
| \mathrm{B}^0_L \rangle).
\ees
The probability density for N semileptonic $\Bs$ decays
is proportional to:
\begin{eqnarray}
  {dN \over dt} \propto ({\rm Br}({\rm B^0_H}\rightarrow \ell X)\Gamma_{\rm H}e^{-\Gamma_{\rm H}t}+
                         {\rm Br}({\rm B^0_L}\rightarrow \ell X)\Gamma_{\rm L}e^{-\Gamma_{\rm L}t})
\end{eqnarray}
where ${\rm Br}({\rm B^0_{H(L)} \rightarrow \ell X})$=
      ${\Gamma}({\rm B^0_{H(L)} \rightarrow \ell X})/\Gamma_{\rm H(L)}$.\\
The semileptonic partial widths for $\rm B^0_H$ and $\rm B^0_L$ are assumed to be
equal since only CP-eigenstates could generate a difference 
(semileptonic decays are not CP-eigenstates). \\
It follows that the two exponentials are multiplied by the same factor and
the probability density for the decay of a $\Bs$ or $\Bsb$ at time $t$ is
given, after normalization, by:
\begin{eqnarray}
   {\cal P}(t) = \frac{\Gamma_{\rm H}\Gamma_{\rm L}}
  {\Gamma_{\rm H}+\Gamma_{\rm L}} \; (e^{-\Gamma_{\rm H}t}+e^{-\Gamma_{\rm L}t}) 
\label{dg_form}
\end{eqnarray}
where
$\Gamma_{\rm L} = \Gamma_{\Bs} + \Delta \Gamma_{\Bs}/2$, 
$\Gamma_{\rm H} = \Gamma_{\Bs} - \Delta \Gamma_{\Bs}/2$.          \\

Two independent variables are then considered: $\tau \equiv {1/ \Gamma_{\Bs}}$\footnote{
$\tau$ does not coincide with the measured $\Bs$ lifetime if $\Delta \Gamma_{\Bs}$
is different from zero}
and $\delta \equiv \dgbs$. \\
As the statistics in the sample is not sufficient to fit
simultaneously $\tau$ and $\delta$,
the method used to evaluate $\delta$ consists in calculating 
the log-likelihood 
for the time distribution measured with the $\Dsl$ and $\phi \ell h$ samples
and deriving the probability density function for $\delta$ by
constraining $\tau$ to be equal to 
$1/\Gamma_{\Bd}\equiv \tau_{\mathrm{B_d}}=(1.56\pm0.04)~\mathrm{ps}$ \cite{ref:book}\footnote{
It has been assumed that $\Delta \Gamma_{\rm B_d}~=~0$.
}
($|\Gamma_{\mathrm{B_s}}/\Gamma_{\mathrm{B_d}}-1|~<~0.01$ is predicted in \cite{ref:bbd}).

\begin{figure}[ph]
\begin{center}
\epsfig{figure=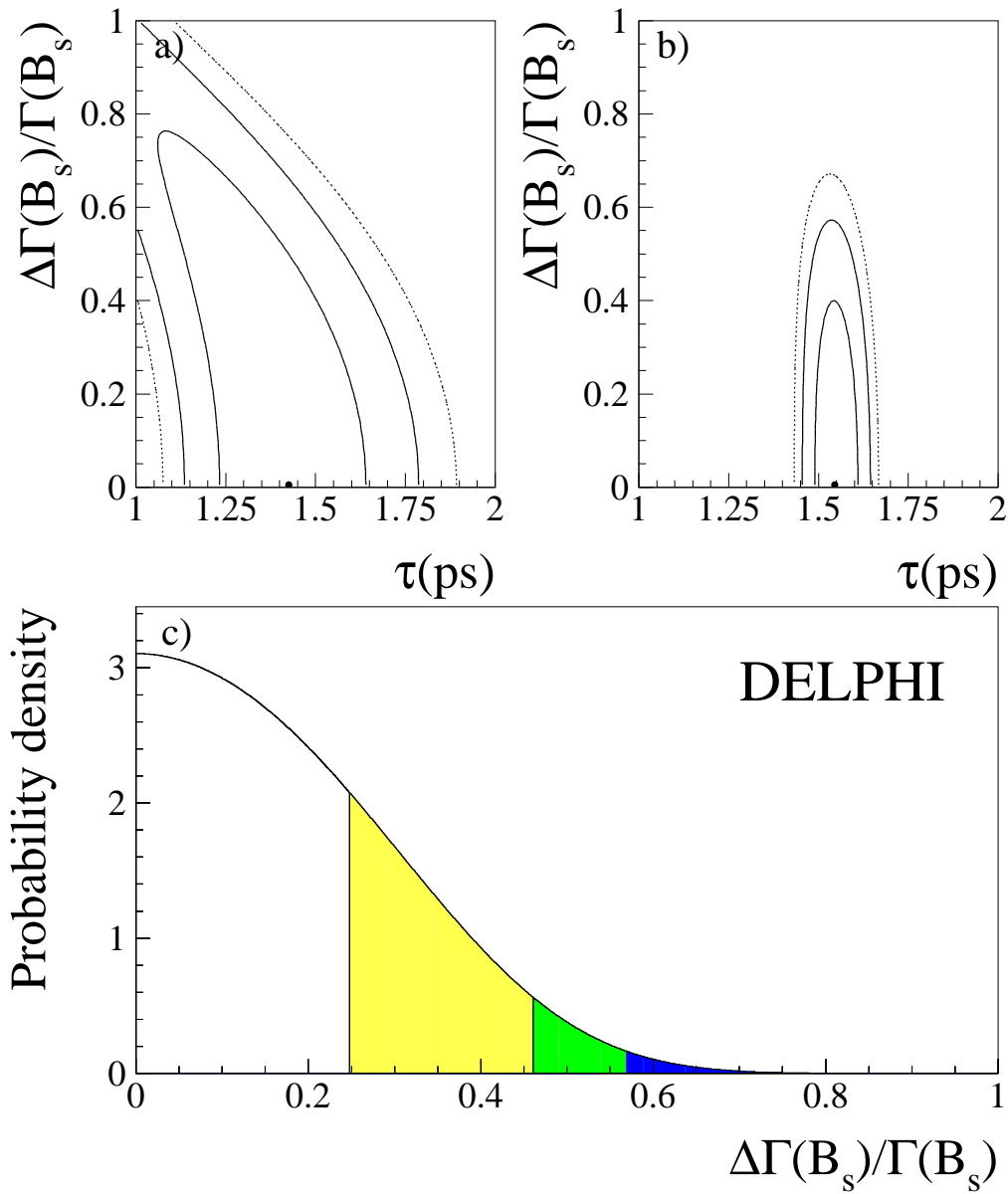,height=16cm}
\caption[]{ \it
          a) 68\%, 95\% and 99\% C.L. contours of the negative log-likelihood 
             in the plane $\tau(\equiv 1/\Gamma_{\rm B_s})$-$\dgbs$ evaluated on the $\Dsl$ sample.
          b) Same as a) but with the constraint $\tau=\tau_{\Bd}$.
          c) Probability density distribution for 
          $\dgbs$;
             the three shaded regions show the limits at 68 \%, 95\% and 99\% C.L. respectively.}
\label{fig:dg_va}
\end{center}
\end{figure}

The log-likelihoods function described in Section~\ref{sec:taubs}
have been modified by replacing the physical function 
$\Bs$ ($\exp(-t/\tau_{\mathrm{B^0_s}}$)) by Equation (\ref{dg_form}) and 
they have been added.\\
The log-likelihood sum has been minimized in the $(\tau,\delta)$ plane 
and the difference with respect to its minimum ($\Delta {\cal L}$) 
has been calculated (Figure~\ref{fig:dg_va}-a):
\bes
   \Delta{\cal L} = -\log{\cal L}^{\Dsl + \phi \ell h}_{tot}(\tau,\delta) + 
   \log{\cal L}^{\Dsl + \phi \ell h}_{tot}((\tau)^{min},(\delta)^{min}) \; .
\ees
The probability density function for the variables $\tau$ and $\delta$ is then proportional to:
\begin{eqnarray*}
   {\cal P}(\tau,\delta) \propto e^{-\Delta {\cal L}}
\end{eqnarray*}
The $\delta$ probability distribution is obtained by convoluting
${\cal P }(\tau,\delta)$ with the probability density function
$f_{\scriptsize({\tau=\tau_{\Bd}})}(\tau)$, expressing
the constraint $\tau = \tbd$, and normalizing the result:
\begin{eqnarray*}
   {\cal P}(\delta) = { \int {{\cal P}(\tau,\delta) f_{\scriptsize({\tau=\tau_{\Bd}})}(\tau) d\tau}  \over 
    \int {{\cal P}(\tau,\delta) f_{\scriptsize({\tau=\tau_{\Bd}})}(\tau) d\tau d\delta}}  
\end{eqnarray*}
\rm{where}
\begin{eqnarray*}
   f_{\scriptsize({\tau=\tau_{\Bd}})}(\tau)  = 
   {{1}/({\sqrt{2\pi}\sigma_{\tiny\tbd}})}\exp({-{{(\tau-\tbd)^2}/{2\sigma^2_{\tiny\tbd}}}}) 
\end{eqnarray*}
The upper limit on $\dgbs$, calculated from ${\cal P}(\delta)$, is:
\bes
   \dgbs < 0.45~\mbox{at the 95\% C.L.}
\ees
This limit takes into account both statistical uncertainties and the systematic 
coming from the uncertainty on the $\Bd$ lifetime.\\
The systematic uncertainty originating from other sources has been evaluated
by convoluting ${\cal P}(\tau,\delta)$
with the probability density functions of the corresponding parameters:
\bes
\small
  {\cal P}(\delta) = 
 {
  { \int {\cal P}(\tau,\delta,x^1_{sys}, ...,x^n_{sys})
   f_{\scriptsize({\tau=\tau_{\Bd}})}(\tau) 
   f({x^1_{sys}}) ...f_({x^n_{sys}})
   d\tau dx^1_{sys} ...dx^n_{sys}}
  \over
  { \int {\cal P}(\tau,\delta,x^1_{sys}, ...,x^n_{sys})
   f_{\scriptsize({\tau=\tau_{\Bd}})}(\tau) 
   f({x^1_{sys}}) ...f_({x^n_{sys}})
   d\tau dx^1_{sys} ...dx^n_{sys}} d\delta }
\ees
where $x^i_{sys}$ are the $n$ parameters considered in the systematic uncertainty and
$f(x^i_{sys})$ are the corresponding probability densities.\\
Since the method implies heavy numerical integrations over a $n$-dimensional grid
only two systematics have been considered here: the purity in $\Bs$
meson of the selected sample and the acceptance.
This approximation is justified since
systematic uncertainties are expected to be small 
(as they are in the lifetime measurement) and dominated by these two parameters.\\
The $\dgbs$ probability distribution, obtained with the inclusion of the systematics,
is shown in Figure~\ref{fig:dg_va}-c,
the most probable value for $\dgbs$ is $0$ and  the upper limit at 95\% confidence 
level is:
\bes
   \dgbs < 0.46~\mbox{at the 95\% C.L.}
\ees
It should be noted that the world average of the $\Bs$ lifetime 
cannot be used as constraint in such analysis,
since it depends on $\Delta\Gamma_{\rm B^0_s}$ and on
$\Gamma_{\rm B^0_s}$. Moreover, this dependence is also  different for
different decay channels.
In the $\Dsl$ case the expression of the average $\Bs$ lifetime
is given by:
\begin{equation}
   \tau_{\Bs}(\Dsl)  = 
    \frac{1 + (\frac{1}{2}\dgbs)^2}{\Gamma_{\Bs} (1 - (\frac{1}{2}\dgbs)^2)}
\end{equation}

\section{Study of $\rm B^0_s$-$\overline{\rm B^0_s}$ oscillations}
\label{sec:os}

The study of $\Bs$-$\Bsb$ oscillations requires the tagging of the sign 
of the $b$ quark in the $\rm B^0_s$ meson at the decay and production 
times.\\
The algorithm used for the $b (\overline b)$ tagging at production time 
has been tuned in order to have the best performances on the $\Dsl$ sample,
where all the charged particles from the $\Bs$ 
decays have been reconstructed.\\

\subsection{$b (\overline b)$ tagging at production time}
\label{sec:os1}
The signature of the initial production of a $b(\overline{b})$ quark
in the jet containing the $\Bs$ or $\Bsb$ 
candidate is determined using a combination of different variables which are
sensitive to the initial quark state following the same technique as
in Section~\ref{dsl_sel_0}. For each individual
variable $X_i$, the probability density functions
$f_b(X_i)$ $(f_{\overline {b}}(X_i))$ for $b$ ($\overline{b}$) quarks are 
obtained from the simulation
and the ratio $R_i = f_{\overline{b}}(X_i)/f_b(X_i)$ is computed. The combined
tagging variable is defined as:
\begin{equation}
\label{eq1}
x_{tag}  =  \frac{1 - R} {1 + R}, \: \mbox{where} \; \; R =  \prod R_i .
\end{equation}

\noindent The variable $x_{tag}$ varies between -1 and 1. High values of $x_{tag}$ 
correspond to a high probability that a given
hemisphere contains a $b$ quark in the initial state. If some of the
variables $X_i$ are not defined in a given event, the corresponding 
ratios $R_i$ are set to 1, corresponding to equal probabilities for the initial
state to be $b$ or $\overline{b}$. \\
An event is split into two hemispheres by the plane passing
through the beam interaction point and perpendicular to the direction 
of the $\Bs$ candidate;
then nine
discriminant variables have been selected for this
analysis.
Five variables are defined in the hemisphere opposite
to the $\Bs$ meson,
in which reconstructed charged particles have been used:
\begin{itemize}
\item
the mean hemisphere charge which is defined as :
\begin{equation}
\label{eq2}
 Q_{hem}= \frac{\sum_{i=1}^{n}q_i (|\vec{p_i} \cdot \vec{e_s}|)^{\kappa}}
{\sum_{i=1}^{n} (|\vec{p_i} \cdot \vec{e_s}|)^{\kappa}}.
\end{equation}
In this expression n is total number of charged particles in the hemisphere,
$q_i$  and  $\vec{p_i}$ are, respectively, 
the charge and the momentum of 
particle $i$ $\vec{e_s}$ is the unit vector along the thrust
axis and $\kappa$=0.6;
\item
the weighted sum of the charges of particles with tracks identified as kaon candidates: \\
\bc
 $Q_K = \sum q_i (|\vec{p_i} \cdot \vec{e_s}|)^{\kappa}$;
\ec
\item
the sum of the charges of tracks having significant impact parameters 
with respect to the event primary vertex;
\item 
the sum of the charges of the particles whose tracks are 
compatible with the event primary vertex;
\item 
the momentum transverse to the jet axis multiplied by the charge of the identified lepton 
candidate with the highest momentum.
\vskip 0.2cm
These variables have been combined to form the discriminant variable
$x^o_{tag}$.  

\end{itemize}

Another set of three variables are evaluated in the hemisphere
which contains the $\Bs$ meson candidate and 
only tracks not included in the $\Bs$ candidate decay products
have been used in their determination
\footnote { In the $\Ds \ell$ analysis all the $\Bs$ decay products are
identified and removed, for more inclusive analyses this is only partially possible}.
They are:
\begin{itemize}
\item
the mean hemisphere charge, computed using (\ref{eq2}) with 
$\vec{e_s}$ directed along the reconstructed momentum of 
the $\Bs$ candidate;
\item
the rapidity with respect to the direction of the thrust axis multiplied by the
charge of the identified kaon candidate 
with the highest momentum having a trajectory compatible with the primary 
vertex
(this algorithm aims at reconstructing the fragmentation kaon produced with
the $\Bsb$, this kaon has a sign opposite to the $b$ quark contained in 
the meson);
\item
the momentum of any reconstructed $\Lambda^0$ candidate multiplied by the 
charge
of the proton from its decay (same principle as in the previous item when
a baryon instead of meson is produced).
\end{itemize}
These variables have been combined to form the discriminant variable 
$x^s_{tag}$.  
In addition the distribution of the polar angle of the direction of the thrust 
axis, common to both hemispheres, is also used to benefit from the forward-backward
asymmetry of the $b$ quark production relative to the electron beam axis.\\

\subsection{Measurement of the tagging purity in events with an exclusively 
reconstructed $\mathrm D^*$}
\label{sec:os2}

The high statistics sample of exclusively reconstructed $\rm D^*$, 
accumulated in 1994-95, has been used to check the tagging procedure.
The purity of the tagging at production time, $\epst$, has been measured
on those events using the analysis of the $\Bd - \Bdb$ mixing. 

The $\Dstar$ candidates have been selected by reconstructing the  decay chain 
$\Dstarp \rightarrow \Do \pi^{+}$ followed by $\Do \rightarrow \Km \pi^+$ or 
$\Do \rightarrow \Km \pi^+\pi^0$.
The selection criteria rely mainly on the small mass difference between 
$\Dstarp$ and $\Do$ mesons \cite{ref:mixing}.
The measurement of the $\Bd - \Bdb$ mixing is performed by correlating
{\it a)} the sign of the $\Dstar$ charge, which 
tags the B flavour at decay time (since $\Dstarm$ in these events are mainly 
produced from $\Bd$ and
$\Dstarp$   from $\Bdb$), with {\it b)} the global tagging variable, 
$x^o_{tag}$, 
evaluated in the hemisphere opposite to the $\Dstar$ and
obtained by combining the five first quantities defined in the 
previous section. 
 If the $\rm \mbox{B}^0_d$ meson, decaying into a $\Dstar$, has oscillated, 
the $\Dstar$ 
 charge and the value of the variable $x^o_{tag}$ are expected to 
 be of unlike sign. 
The mass difference, $\dmd$, between the two physical states
of the $\Bd-\Bdb$ system
is obtained from the study of the $\Do$ decay distance distribution for unlike 
and like sign events. Details of the analysis can be found in  
\cite{ref:mixing}.
The amplitude of the time dependent oscillation is sensitive to the 
probability of correctly tagging events as unmixed and mixed $\Bd$ candidates. 
A fit has been performed, fixing the mass difference $\dmd$ to the world 
average \cite{ref:wav}, and leaving $\epst$ as a free parameter. 
The fit has been repeated for different minimum values of the global 
tagging variable $x^o_{tag}$. Results are reported in Table~\ref{tab:epstre},
together with the predictions from the simulation. The fraction
of events $ f_{events}$ remaining after the cut on the tagging variable 
is also reported.

\begin{table}[hbt]
\bc
\bt{|c|c|c|c|c|}
\hline 
        & \multicolumn{2}{c|}{Data} &  \multicolumn{2}{c|}{Simulation}  \\  
\hline
        &$\epst$& $f_{events}$ & $\epst$ &$f_{events}$ \\  
\hline \hline
 $|x^o_{tag}|>0.0$ &$0.68 \pm 0.02$& 1.0 & $0.69$ &1.0 \\  
\hline
 $|x^o_{tag}|>0.1$ &$0.69 \pm 0.02$& 0.88 & $0.71$ &0.89 \\  
\hline
 $|x^o_{tag}|>0.2$ &$0.71 \pm 0.02$& 0.77 & $0.74$ &0.78 \\  
\hline
\et
\caption []{\it Values of $\epst$ obtained from the 
analysis of exclusively reconstructed $\Dstar$ for different cuts on
the value of
the tagging variable $x^o_{tag}$. Also reported is the fraction of events 
remaining after the cut. Expectations from the simulation are also given.}
\label{tab:epstre}
\ec
\end{table}

The tagging efficiency, estimated using the $\Dstar$ sample,
is consistent within
its uncertainty with the expectation from the simulation. 

\begin{figure}[p]
\begin{center}
\begin{tabular}{c}
\epsfig{figure=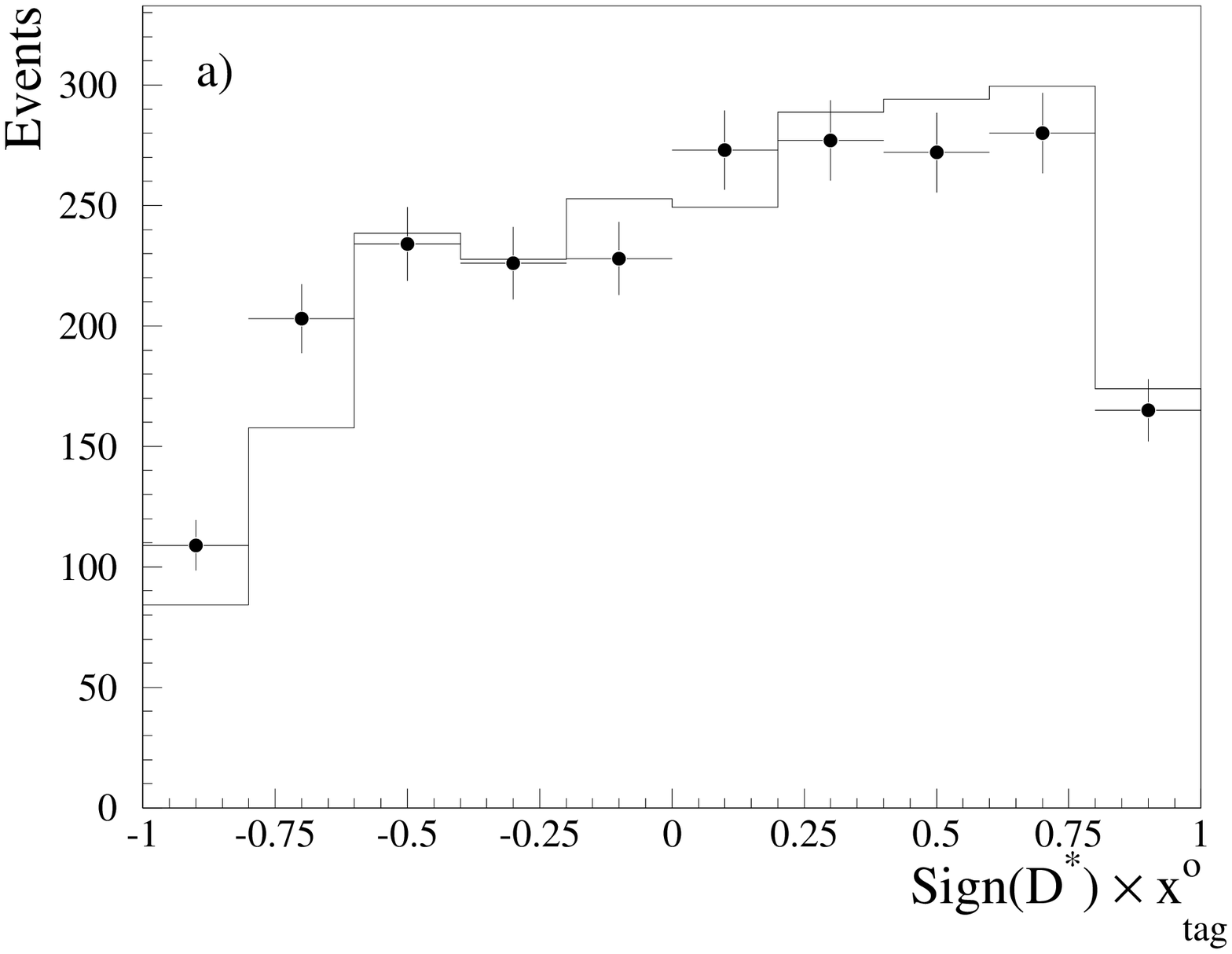,height=9cm} \\
\epsfig{figure=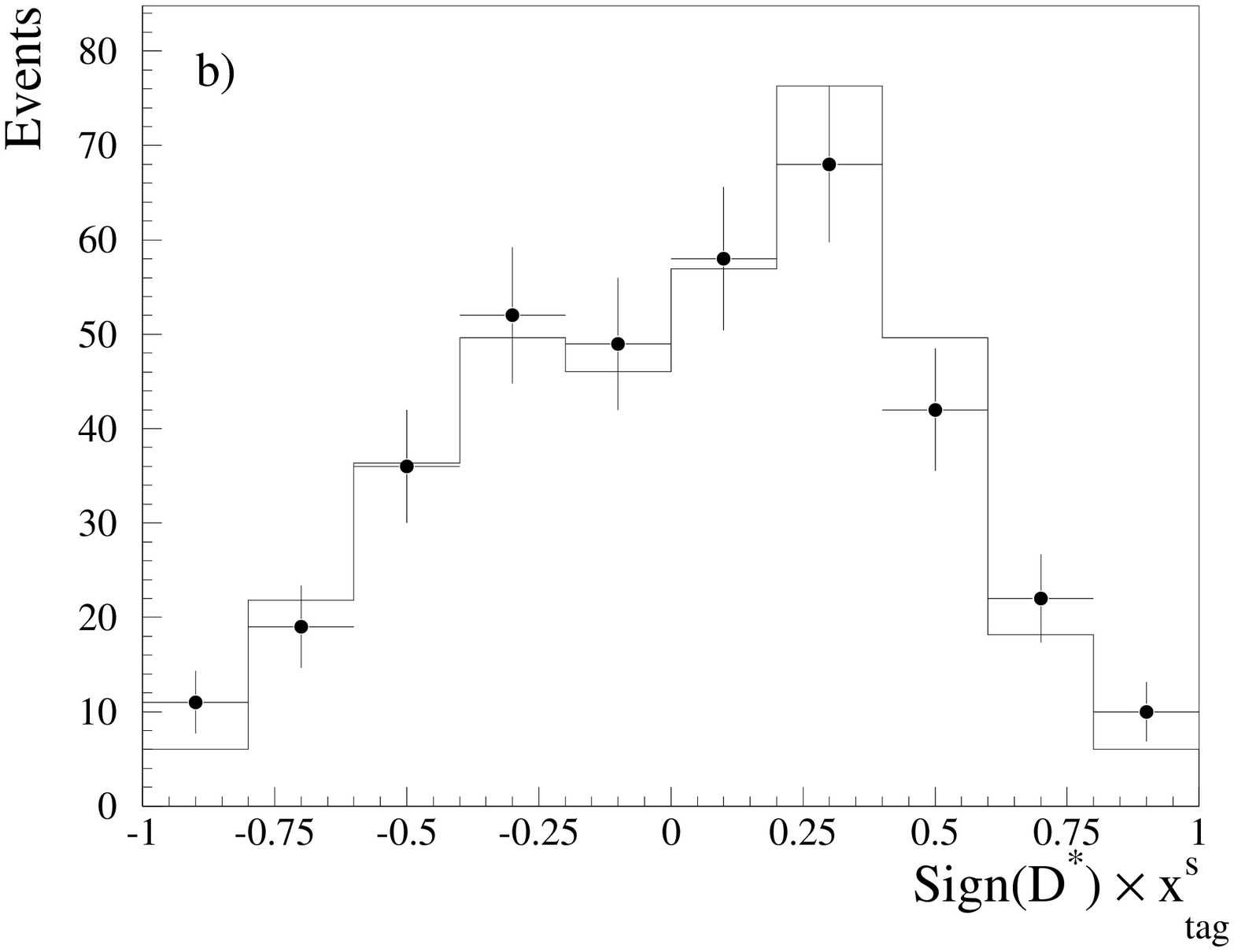,height=9cm}
\end{tabular}
\caption[]{ \it
  Check of the flavour tagging on the $\rm D^*$ sample. \\
  a) Distribution of the product between the global tagging variable $x^o_{tag}$ and the
            charge of the $\Dstar$ in 
            the hemisphere opposite to the $\Dstar$ candidate.\\
  b) Distribution of the product between the global tagging variable 
            $x^s_{tag}$ and the charge of the $\Dstar$ in the same 
            hemisphere as the $\Dstar$-lepton candidate.\\
            The full dots with the error bars represent the 
            data. The histogram is obtained in the simulation.\\
    The non perfect separation is due to the mistag fraction of $x_{tag}$
    but also to the $\Bd$-$\Bdb$ mixing. }
\label{fig:dstartag}
\end{center}
\end{figure}

 The selected sample of exclusively reconstructed $\Dstar$ still contains a 
significant fraction of events originating from charm and light quarks.
In order to study the distribution of the tagging variable $x^o_{tag}$, the 
b-tag probability for all tracks in the event has been required to be smaller 
than $10^{-3}$ \cite{btag}.
 The fraction of non-b events in the remaining 
sample is estimated to be $5\%$. The distribution of the product between 
the $\Dstar$ charge and the value of the tagging variable $x^o_{tag}$ 
is shown in Figure~\ref{fig:dstartag}--a, together with the expectation 
from the simulation. 

Another check has been performed by selecting events with an exclusively 
reconstructed $\Dstar$ accompanied by a lepton of opposite 
charge. This sample is highly enriched in $\Bd$, but has a limited 
statistics. However, it allows the study of the tagging 
variable $x^s_{tag}$ defined, in the same hemisphere as the $\Dstar$-lepton 
candidate, by combining the other three variables mentioned
in the previous section. 
The variable which quantifies the presence of an identified kaon 
of highest momentum compatible with the primary vertex has been removed from
the definition of $x^s_{tag}$. 
The distribution of the product between the $\Dstar$ charge
and the value of the tagging variable, $x^s_{tag}$, is shown in 
Figure~\ref{fig:dstartag}--b together with the expectation from 
the simulation.

\begin{figure}[ph]
\bc
\epsfig{figure=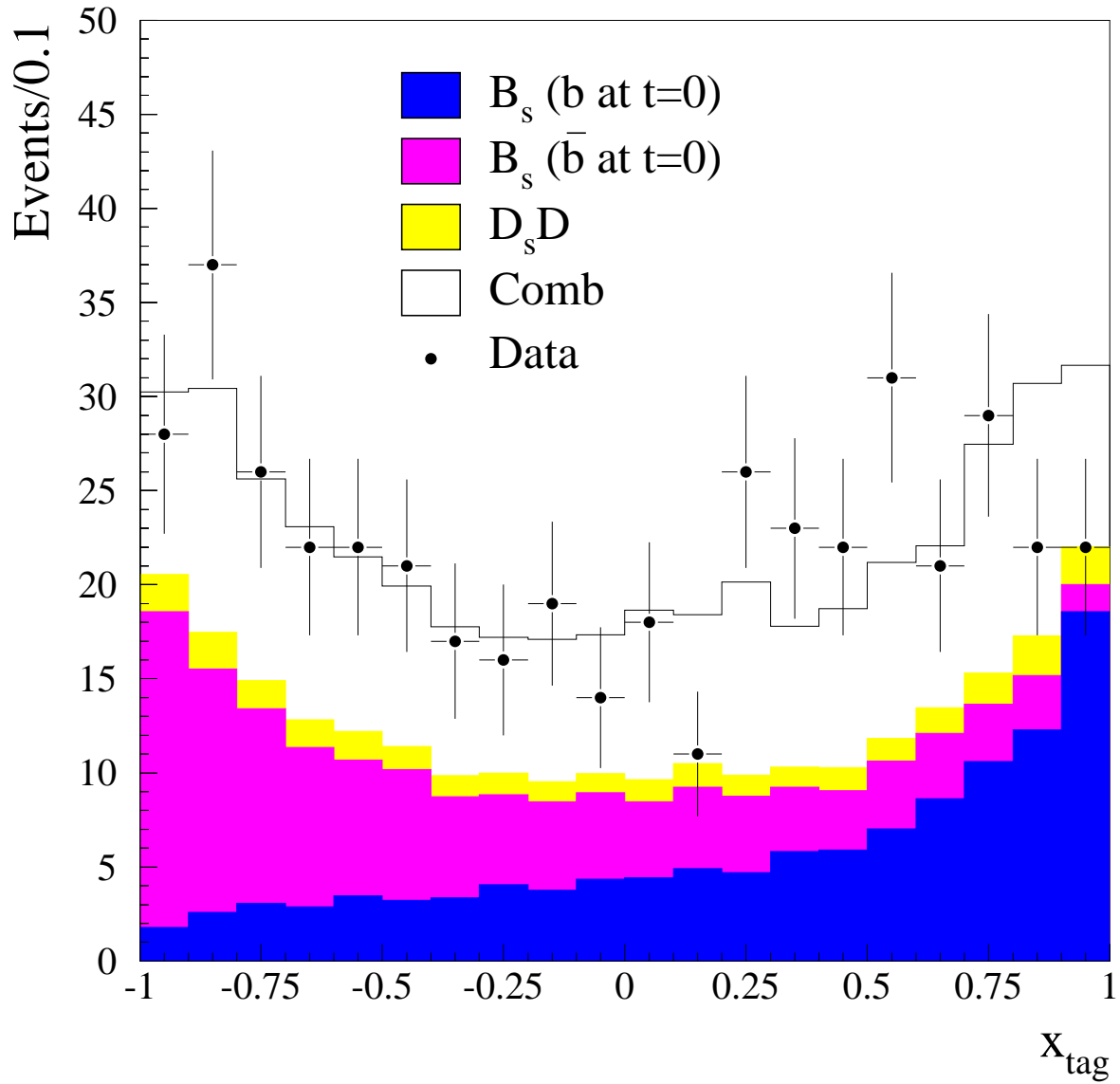,height=16cm}
\caption []{ \it The plot 
shows the distribution of the $x_{tag}$ discriminant variable in the $\Dsl$ sample. 
The points with the error bars represent the data, 
the white histogram shows the contribution from combinatorial background,
the lighter histogram the contribution from $\Ds \rm D$ events and
the darker histograms 
the contribution from the $\Bs$ signal in which $\Bs$ mesons produced 
from $b$ or $\bar b$ quarks have been distinguished.\\
The degree of separation between the $b$ and $\bar b$ histograms quantify the tagging 
purity of $x_{tag}$. }
\label{fig:tag_plot}
\ec
\end{figure}

The selected $\Dsl$ sample do not have enough statistics to perform 
a quantitative check. The $x_{tag}$ distributions
expected from the simulation and measured data, using the $\Dsl$ 
sample, are found to be compatible within statistics (Figure~\ref{fig:tag_plot}).

\subsection{Tagging procedure}
\label{sec:os3}

An event is classified as a mixed or an unmixed candidate according to 
the relative signs  of the ${\mbox{D}_{\mathrm s}}$ electric charge, $Q_{D}$, 
and of the $x_{tag}$ variable.
Mixed candidates have $x_{tag} \times \mbox{Q}_D < 0$, and unmixed
ones $\mbox{x}_{tag} \times \mbox{Q}_D > 0$. \\
The probability, ${\epsilon}_b$, of tagging the $b$ or the 
$\overline{b}$ quark correctly from the measurement of $x_{tag}$ has been 
evaluated using a dedicated simulated event sample and has been found 
to be, in the $\Dsl$ sample, $74.5\pm0.5$\% in 94-95 data and 
$71.5\pm1.2$\% in 92-93 data.\\
In the $\phi \ell h$ sample the tracks
from the $\rm B$ decay have not been all reconstructed.
The tagging purity is lower with respect to the one estimated in the $\Dsl$ sample
due to some possible misidentification between primary and secondary tracks 
present in the same hemisphere as the $\phi$ meson. The value found in simulated events is 
($\epsilon_b=0.69\pm 0.01$).

To improve the tagging purity further, the shape of the $x_{tag}$ 
distribution can be included in the analysis.\\
Four purities enter in the analysis:
  \begin{itemize}
     \item[-] $\epsilon_{b\ell}$: tagging purity for the direct $b\rightarrow \ell$ decays; 
     \item[-] $\epsilon_{bc\ell}$: tagging purity for $b\rightarrow c \rightarrow \ell$ ``cascade'' decays;
     \item[-] $\epsilon_{bkg}^{mix(unmix)}$ probability of 
              classifying  background candidates as mixed or as unmixed
              (computed on sidebands events);
     \item[-] $\epsilon_{f\ell}^{mix(unmix)}$ probability of 
              classifying fake lepton candidates as mixed or as unmixed.
  \end{itemize}
using $x_{tag}$ as a discriminant variable each of these purities
is replaced by the function $\epsilon{X}(x_{tag})$,
where ${X}$ is the $x_{tag}$ probability density function.\\
The global probability density function has been divided by the sum
$\epsilon{X}^{r}_{b\ell}(x_{tag}) + (1-\epsilon){X}^{w}_{b\ell}(x_{tag})$
($r\equiv\mbox{right tag}$ and $w\equiv\mbox{wrong tag}$) in order
to keep, for the signal part, the relation $\epsilon^{w} = 1-\epsilon^{r}$.\\
The functions entering in the final likelihood are then re-defined as:
\renewcommand\arraystretch{1.5}
\bes
  \begin{array}{ccc}
    X^{r}_{b\ell} =  
           { \textstyle \epsilon_{b\ell}{X}^{r}_{b\ell}(x_{tag}) 
             \over
             \textstyle \epsilon_{b\ell}{X}^{r}_{b\ell}(x_{tag}) +  (1-\epsilon_{b\ell}){X}^{w}_{b\ell}(x_{tag}) } & 
    X^{w}_{b\ell} = 1-X^{r}_{b\ell}  & \\
    X^{r}_{bc\ell} = {\textstyle \epsilon_{bc\ell}{X}^{r}_{bc\ell} 
                      \over
                      \textstyle \epsilon_{b\ell}{X}^{r}_{b\ell}(x_{tag}) +  (1-\epsilon_{b\ell}){X}^{w}_{b\ell}(x_{tag}) } & 
    X^{w}_{bc\ell} = {\textstyle (1-\epsilon_{bc\ell}){X}^{w}_{bc\ell} 
                      \over
                      \textstyle \epsilon_{b\ell}{X}^{r}_{b\ell}(x_{tag}) +  (1-\epsilon_{b\ell}){X}^{w}_{b\ell}(x_{tag}) } &  \\
    X^{mix}_{bkg}    = { \textstyle \epsilon_{bkg}^{mix}X^{mix}_{bkg}
                         \over
                         \textstyle \epsilon_{b\ell}{X}^{r}_{b\ell}(x_{tag}) +  (1-\epsilon_{b\ell}){X}^{w}_{b\ell}(x_{tag}) } &  
    X^{unmix}_{bkg}  = { \textstyle \epsilon_{bkg}^{unmix}X^{unmix}_{bkg}
                         \over
                         \textstyle \epsilon_{b\ell}{X}^{r}_{b\ell}(x_{tag}) +  (1-\epsilon_{b\ell}){X}^{w}_{b\ell}(x_{tag}) } & \\  
    X^{mix}_{f\ell}    = { \textstyle \epsilon_{f\ell}^{mix}X^{mix}_{f\ell}
                         \over
                         \textstyle \epsilon_{b\ell}{X}^{r}_{b\ell}(x_{tag}) +  (1-\epsilon_{b\ell}){X}^{w}_{b\ell}(x_{tag}) } &  
    X^{unmix}_{f\ell}  = { \textstyle \epsilon_{f\ell}^{unmix}X^{unmix}_{f\ell}
                         \over
                         \textstyle \epsilon_{b\ell}{X}^{r}_{b\ell}(x_{tag}) +  (1-\epsilon_{b\ell}){X}^{w}_{b\ell}(x_{tag}) } & \\  

  \end{array}
\ees
\renewcommand\arraystretch{1.0}
The effective tagging purities obtained, in the $\Dsl$ sample, with this method correspond to $78\pm0.5$\%
for 94-95 data and to $74\pm1.2$\% for 92-93 data.

\subsection{Fitting procedure}
\label{sec:os4}

From the expected proper time distributions and the tagging probabilities,
the probability functions for mixed and unmixed events candidates have been 
computed
\footnote{In the following, only the probability function for mixed events
is written explicitly; the corresponding probability for unmixed events can be
obtained by changing $r\rightarrow w$.}:
\begin{eqnarray}
P^{mix}(t_i) = f^{\Bs}_{b\ell} P_{\Bs}^{mix}(t_i) + 
        f^{\rm B}_{b\ell}P_{\rm B}^{mix}(t_i) + 
        f_{bc\ell}P_{bc\ell}^{mix}(t_i) + 
        f_{f\ell}P_{f\ell}^{mix}(t_i) +  
        f_{bkg}P_{bkg}^{mix}(t_i).  
\end{eqnarray}
where $t_i$ is the reconstructed proper time. 
The analytical probability densities are as follows, with $t$ being the true 
proper time:
\begin{itemize}
\item $\Bs$ mixing probability.
 \begin{eqnarray}
   \begin{array}{ll}
    {P_{\Bs}^{mix}}(t_i) &= \{~~X^{r}_{b\ell}
    {\cal P}_{\Bs}^{mix}(t) +
            X^{w}_{b\ell}{\cal P}_{\Bs}^{unmix}(t)~~\} {\cal A}(t)\otimes 
   {\cal R}_{b\ell}(t-t_i)
   \end{array}
 \end{eqnarray}
\item ``cascade'' background mixing probability.\\

\renewcommand\arraystretch{1.1}
 \begin{eqnarray}
   \begin{array}{ll}
  P_{bc\ell}^{mix}(t_i) = \{ & 
  f^{\rm B_d}_{bc\ell}
  (~X^{r}_{bc\ell}{\cal P}_{{\mathrm B_d}}^{unmix}(t) + 
  (~X^{w}_{bc\ell}{\cal P}_{{\mathrm B_d}}^{mix}(t)~) + \\ 
& {(f^{\Bs}_{bc\ell}/2)}
  (~X^{r}_{bc\ell}{\cal P}_{{\Bs}}^{unmix}(t) + 
    X^{w}_{bc\ell}{\cal P}_{{\Bs}} ^{mix}(t)~) + \\ 
& {(f^{\Bs}_{bc\ell}/2)} (
   ~X^{w}_{bc\ell}{\cal P}_{{\Bs}}^{unmix}(t) + 
    X^{r}_{bc\ell}{\cal P}_{{\Bs}}^{mix}(t)~) + \\ 
&    f^{\rm B^+}_{bc\ell}X^{r}_{bc\ell}/\tau_{\rm B^+}\exp(-t/\tau_{\rm B^+}) \\
&    f^{\Lambda_b}_{bc\ell}X^{r}_{bc\ell}/\tau_{\Lambda_b}
     \exp(-t/\tau_{\rm \Lambda_b}) \hspace{1cm} \}
   {\cal A}(t)\otimes {\cal R}_{bc\ell}(t-t_i)    
   \end{array}
 \end{eqnarray}
\renewcommand\arraystretch{1.0}
  Note that the two terms contributing to the $\rm B^0_s$ are due to the 
  fact that, in the decay
  $\rm B^0_s \rightarrow D^+_s D^-_s X$, the lepton can originate from either 
  $\rm D_s$ mesons. The $\rm B^0_s$ contribution can then be simplified in the
  expression and becomes, mixing independent:
  \bes
      f_{\Bs}/\tau_{\Bs}{\rm exp(-t/\tau_{\Bs})}
  \ees
\item non-strange $\rm B$-hadrons mixing probability.

\renewcommand\arraystretch{1.1}
 \begin{eqnarray}
   \begin{array}{ll}
  P^{B~mix}_{b\ell}(t_i) = \{ & 
  f^{{\mathrm B_d}}_{b \ell} (~X^{r}_{b\ell}
  {\cal P}_{{\mathrm B_d}}^{mix}(t) + 
   X^{w}_{b\ell}{\cal P}_{{\mathrm B_d}}^{unmix}(t)~) + \\ 
&    f^{\rm B^+}_{b\ell}X^{w}_{b\ell}/\tau_{\rm B^+}
     \exp(-t/\tau_{\mathrm B^+}) \\
&    f^{\Lambda_b}_{b\ell}X^{w}_{b\ell}/\tau_{\Lambda_b}
     \exp(-t/\tau_{\mathrm \Lambda_b}) \hspace{1cm} \}
{\cal A}(t)\otimes {\cal R}_{b \ell}(t-t_i)   
   \end{array}
 \end{eqnarray}
\renewcommand\arraystretch{1.0}

\item mixing probability for candidates from light quark events or fake leptons:
 \begin{eqnarray}
     {\mathrm P}_{f\ell}^{mix}(t_i) = X^{mix}_{f\ell}  
  {\cal P}_{f\ell}(t_i) 
 \end{eqnarray}
\item combinatorial background mixing probability:
 \begin{eqnarray}
     {\mathrm P}_{bkg}^{mix}(t_i) = X^{mix}_{bkg}  
  {\cal P}_{bkg}(t_i) 
 \end{eqnarray}
  The parameters entering in the proper time distribution for this 
  background have been determined in the lifetime fit.
\end{itemize}

The oscillation analysis has been performed in the framework of the amplitude
method \cite{ref:amplitude} which consists in measuring, for each value of
the frequency $\dms$, an amplitude $A$ and its error
$\sigma(A)$. The parameter $A$ is introduced in the time evolution of pure 
$\mbox{B}^0_s$
or $\overline{\mbox{B}^0_s}$ states so that the value $A=1$ corresponds to
a genuine signal for oscillation:
$$
{\cal P}(\mbox{B}^0_s\rightarrow (\mbox{B}^0_s,~\overline{\mbox{B}^0_s}))~
=~\frac{1}{2 \tau_{s}} e^{- \frac{t}{\tau_{s}}} \times
 ( 1 \pm A~cos ({\dms t} ) )
$$
The 95$\%$ C.L. excluded region for $\dms$ is obtained by evaluating the 
probability that, in at
most 5$\%$ of the cases, a real signal having an amplitude equal to unity
would give an observed amplitude smaller than the one measured. 
This corresponds to the condition:
$$A(\dms) + 1.645~\sigma(A(\dms))~<~1.$$\\
In the amplitude approach it is possible to define the exclusion 
probability, that is the probability that a certain $\dms$ value
lies in an excluded region if the generated $\dms$ was very
large ($\dms \rightarrow \infty$). 
The sensitivity is the value of $\dms$ corresponding to 50 \% of 
exclusion probability.

\begin{figure}[ph]
\begin{center}
\epsfig{figure=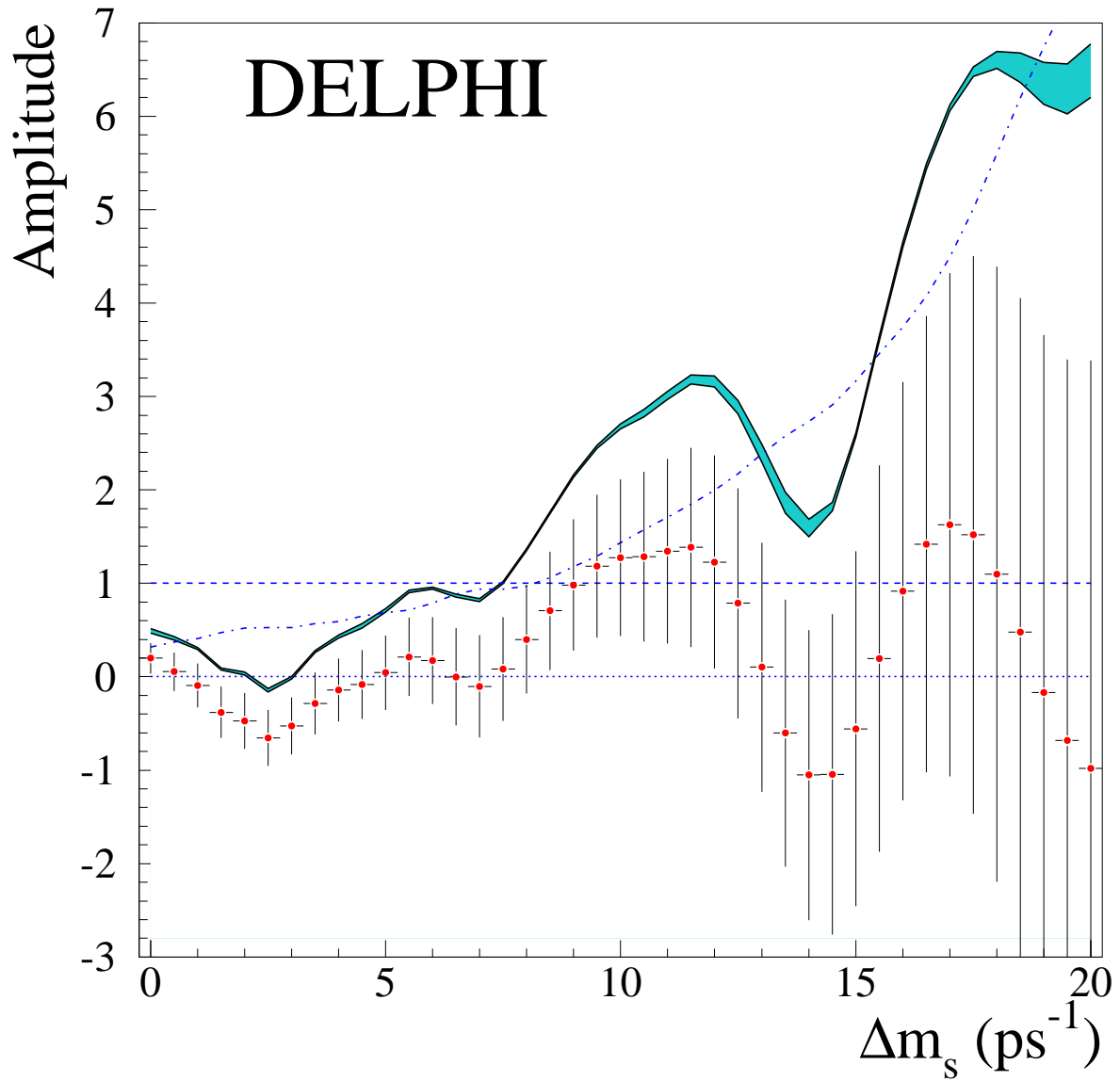,height=16cm}
\caption []{\it Variation of the oscillation  amplitude $A$ as a function 
of $\Delta m_{s}$.
The lower continuous line corresponds to $ A + 1.645~\sigma_A$ where $\sigma_A$ includes 
statistical uncertainties only, while the shaded area includes the contribution from systematics. The dashed-dotted line corresponds 
to the sensitivity curve. The lines at A=0 and A=1 are also given. }
\label{fig:plot_dms_va}
\end{center}
\end{figure}

Using the amplitude approach (Figure~\ref{fig:plot_dms_va}), and 
considering only statistical uncertainties, 
a limit has been obtained: 
\begin{eqnarray}
   \begin{array}{ll}
~~~~~~~~~~\dms > 7.4 ~ \mbox{ps}^{-1}~\mbox{at 95\% C.L.} & \\
   \end{array}
 \end{eqnarray}
with a corresponding sensitivity at $~\dms = 8.3 ~ \mbox{ps}^{-1}$.
At $\dms$ = 10 $ps^{-1}$, the error on the amplitude is 0.85.

\begin{figure}[ph]
\begin{center}
\epsfig{figure=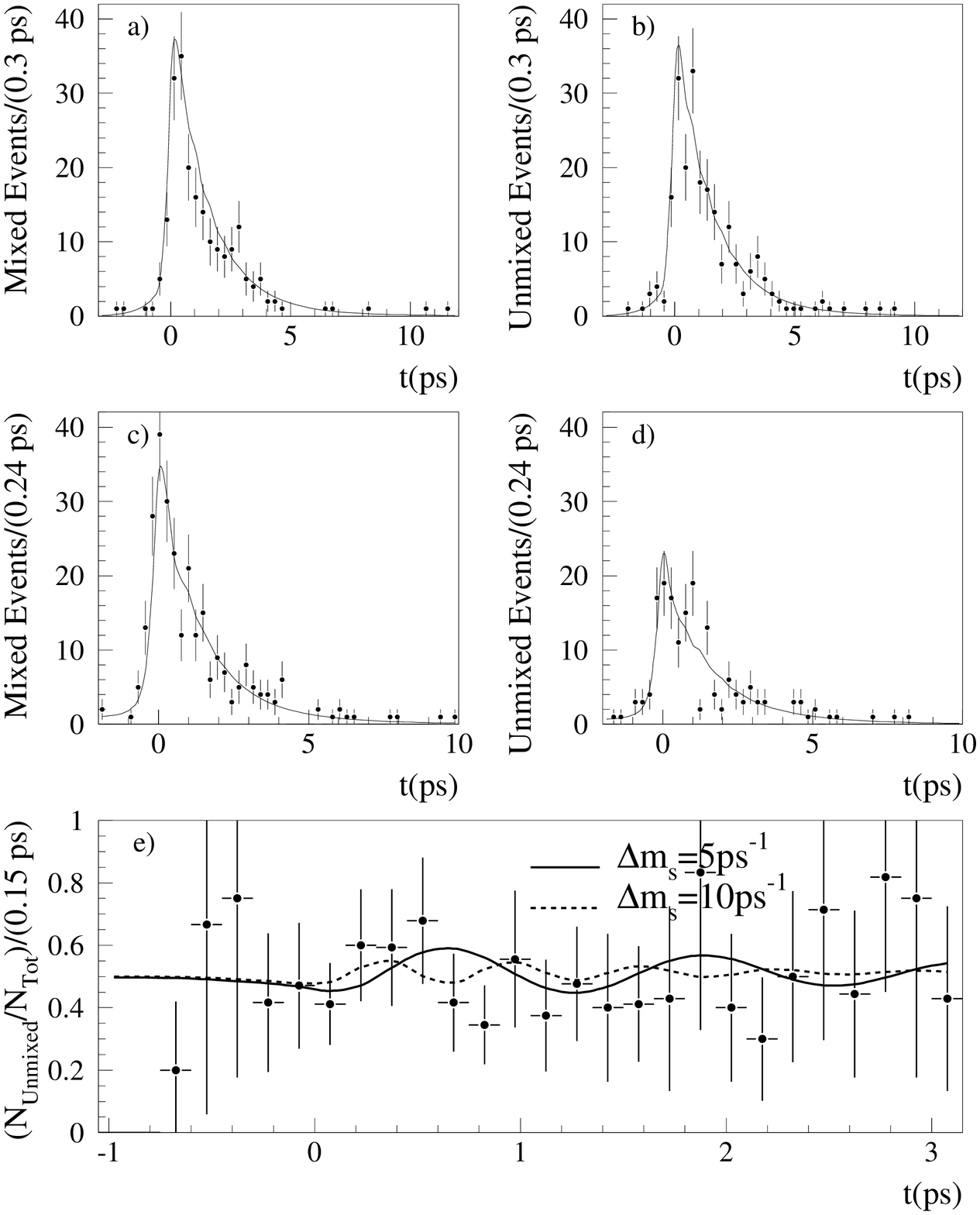,height=16cm}
\caption []{\it 
 Proper time distribution of mixed and unmixed events in $\Dsl$ sample (a-b) and
 in the $\phi \ell$ sample (c-d); the full dots with error bars represent the data, the
 curves are the corresponding distributions for $\dms=10~ps^{-1}$.\\
 c) Ratio between the mixed events and the total number of events
    in bins of proper time in the $\Dsl$ sample. 
    The full (dashed) line represents
    the prediction for an oscillation ($A=1$) with $\dms=5~ps^{-1}(10~ps^{-1})$.}
\label{fig:do_plot_osc}
\end{center}
\end{figure}

Several checks have been done to verify the reliability of the amplitude fit:
the proper time distributions for mixed and unmixed events have been verified
to be well reproduced by the fit (Figure~\ref{fig:do_plot_osc}-a,-b) and the ratio between
mixed events and the total number of events in bins of the proper time
has been compared with the expected distribution for $\dms=5~ps^{-1}$
and $\dms=10~ps^{-1}$. These values have been chosen to illustrate
the behaviour of the expected oscillation curve for $A=0$ ($\dms=5~ps^{-1}$) and
$A=1$ ($\dms=10~ps^{-1}$) (Figure~\ref{fig:do_plot_osc}-e). 
It could be seen that the oscillation curve at $\dms=10~ps^{-1}$
(where $A$ is close to 1) fits the data better than the corresponding curve
at $\dms=5~ps^{-1}$ (where $A$ is compatible with 0), as expected from the definition of $A$.

\subsection{Study of systematic uncertainties}
\label{sec:os5}

Systematic uncertainties have been evaluated by varying the parameters which 
have been  kept constant in the fit, according to their measured or 
expected errors, using the 
formula \cite{ref:amplitude}:

\begin{eqnarray*}
\sigma \left [
    A(\nu) \right ]_{sys} =
   \Delta A(\nu) + (1-A){ {\Delta \sigma(A)} 
                 \over    {\sigma\left[ A(\nu) \right]}}.
\end{eqnarray*}
$\Delta A(\nu)$ and $\Delta \sigma(A)$
indicate the variations of the amplitude, in the central value and 
in the error, due to the considered systematics.\\
Three main sources of systematic uncertainties have been identified:
\begin{itemize}

\item { Systematics from the tagging purity. } 
\begin{itemize}
\item $\Dsl$ sample.\\
The studies done in Section~\ref{sec:os2} show that, using the tagging variables 
in the opposite hemisphere and
requiring 
$|x^o_{tag}|>0.$, the difference between the values of the tagging
purity measured in real and simulated events is 
$\epst(DATA) - \epst(MC)$ = -0.01 $\pm$ 0.02. It has been verified 
that the real and the simulated distributions for the 
tagging purities agree in both hemispheres. \\
The systematics coming from the control of the tagging purity has been  
evaluated by varying the probability distributions of the discriminant 
variable for $b$ and $\overline{b}$ quarks in a way to induce 
an absolute variation on the effective value of the tagging purity of $\pm 3.0 \%$.\\
\item $\phi \ell h$ sample.
The agreement between data and simulation has not been checked for this sample;
a conservative absolute variation of 5\% in the tagging purity has been assumed.
\end{itemize}

\item { Systematics from the $\Bs$ purity.}\\
The same procedure already applied for the lifetime measurement has been used.

\item { Systematics from the resolution on the B decay proper time. } \\
The same procedure already applied  for the lifetime measurement has been used.
In addition, the systematic error due to the variation of the proper 
time distribution of the combinatorial background,
has been considered: the parameters used to define the 
background shape,
in the lifetime fit, have been varied according to their fitted errors.
\end{itemize}

The inclusion of systematic uncertainties lowers the sensitivity 
to $8.1~ps^{-1}$ without affecting the $95\%$~C.L limit. In Table~\ref{tabdmssys}
the amplitude values are reported, together with their statistical and systematical errors, 
for five different values of $\dms$.

\begin{table}
\begin{center}
\begin{tabular}{|c|c|c|c|c|c|}
\hline 
 $\dms(ps^{-1})$ & $A$ & $\sigma_A(stat)$ 
& $\sigma_A(\mbox{total $syst.$ (but $t$ resolution)})$
& $\sigma_A(\mbox{$t$ resolution $syst.$})$ \\ \hline
 $2.5  $    & -0.638  & 0.304 & 0.112 & 0.033 \\ \hline
 $5    $    &  0.037  & 0.400 & 0.118 & 0.060 \\ \hline
 $7.5  $    &  0.182  & 0.561 & 0.069 & 0.098 \\ \hline
 $10   $    & 1.343   & 0.846 & 0.160 & 0.180 \\ \hline
 $12.5 $    & 0.867   & 1.241 & 0.285 & 0.389 \\ \hline
\end{tabular}
\end{center}
\caption{\it Amplitude values with statistical and systematic errors 
for three different values of $\dms$}
\label{tabdmssys}
\end{table}

\begin{figure}[ph]
\begin{center}
\epsfig{figure=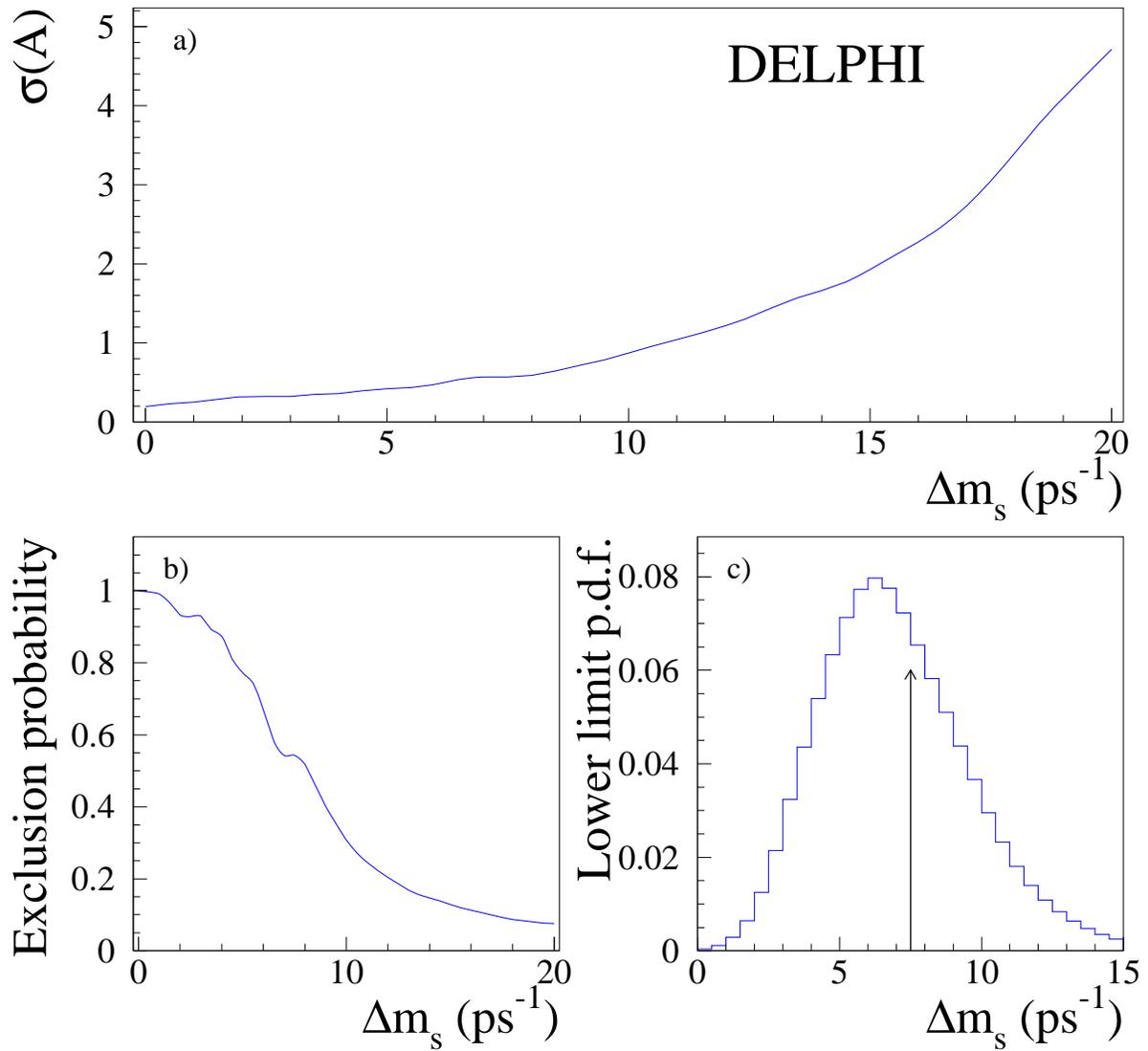,height=16cm}
\caption []{\it 
        a) Variation  of the error on the amplitude as a function of $\Delta m_{s}$.
        b) Exclusion probability vs. $\Delta m_{s}$.
        c) Lower limit probability density function vs. $\Delta m_{s}$.}
\label{fig:pl_dms_prob}
\end{center}
\end{figure}

The exclusion probability  of $\dms=7.4~ps^{-1}$ is 54\% while
the probability of obtaining a limit on $\dms$ higher than the actual one is
38\% (Figure~\ref{fig:pl_dms_prob}-c). \\
Figure~\ref{fig:pl_dms_prob}-a and Figure~\ref{fig:pl_dms_prob}-b
represent, respectively, the error on the amplitude and the exclusion probability 
as a function of $\dms$.

\section{ Conclusion }
\label{sec:conc}
A sample of 436 $\mathrm{D_s^{\pm}}\ell^\mp$ candidate events has been 
selected from about 3.6 million hadronic $\Zz$ decays accumulated by
DELPHI between 1992 and 1995, using seven different $\Ds$ decay modes.
The number of events coming from $\Bs$ semileptonic decays has been estimated
to be $230\pm18$ in this sample. In addition, a sample of 441 $\phi \ell h$,
containing $41 \pm 12$ $\Bs$ semileptonic decays, has been also used.
Events contained in the $\Dsl$ sample, with a reconstructed $\phi$ and
have been removed from this last sample. \\
Using these samples, three analyses have been performed.
The $\Bs$ lifetime has been measured and a limit on the fractional width
difference between the two physical $\Bs$ states has been set:
$$
\tau (\Bs)  = ( 1.42^{+0.14}_{-0.13}(stat.)\pm 0.03(syst.) )~ps
$$
$$
\dgbs < 0.46 ~\mbox{at the 95\% C.L.}
$$
This last result has been obtained under the hypothesis that 
$\tau_{\Bs}=\tau_{\rm B_d}$.\\
The study of $\Bs-\Bsb$ oscillations sets a limit $\mbox{at 95\% C.L.}$ on the mass 
difference between the physical $\mbox{B}^0_{\mathrm s}$ states:
\begin{eqnarray}
   \begin{array}{ll}
~~~~~~~~~~\dms > 7.4 ~ \mbox{ps}^{-1}~\mbox{at 95\% C.L.} & \\
   \end{array}
\end{eqnarray}
with a corresponding sensitivity equal to $~~ 8.1 ~ \mbox{ps}^{-1}$.\\
Previous DELPHI results obtained with $\Dsl$ and $\phi \ell$ samples 
(\cite{ref:oldbs},\cite{ref:bspaper}) are superseded by the analyses 
presented in this paper.

\subsection*{Acknowledgements}
\vskip 3 mm
 We are greatly indebted to our technical 
collaborators, to the members of the CERN-SL Division for the excellent 
performance of the LEP collider, and to the funding agencies for their
support in building and operating the DELPHI detector.\\
We acknowledge in particular the support of \\
Austrian Federal Ministry of Science and Traffics, GZ 616.364/2-III/2a/98, \\
FNRS--FWO, Belgium,  \\
FINEP, CNPq, CAPES, FUJB and FAPERJ, Brazil, \\
Czech Ministry of Industry and Trade, GA CR 202/96/0450 and GA AVCR A1010521,\\
Danish Natural Research Council, \\
Commission of the European Communities (DG XII), \\
Direction des Sciences de la Mati$\grave{\mbox{\rm e}}$re, CEA, France, \\
Bundesministerium f$\ddot{\mbox{\rm u}}$r Bildung, Wissenschaft, Forschung 
und Technologie, Germany,\\
General Secretariat for Research and Technology, Greece, \\
National Science Foundation (NWO) and Foundation for Research on Matter (FOM),
The Netherlands, \\
Norwegian Research Council,  \\
State Committee for Scientific Research, Poland, 2P03B06015, 2P03B1116 and
SPUB/P03/178/98, \\
JNICT--Junta Nacional de Investiga\c{c}\~{a}o Cient\'{\i}fica 
e Tecnol$\acute{\mbox{\rm o}}$gica, Portugal, \\
Vedecka grantova agentura MS SR, Slovakia, Nr. 95/5195/134, \\
Ministry of Science and Technology of the Republic of Slovenia, \\
CICYT, Spain, AEN96--1661 and AEN96-1681,  \\
The Swedish Natural Science Research Council,      \\
Particle Physics and Astronomy Research Council, UK, \\
Department of Energy, USA, DE--FG02--94ER40817. \\

\newpage

\newpage

\newpage

\newpage

\newpage

\newpage

\newpage

\end{document}